\documentclass[letterpaper,12pt]{article}
\pdfoutput=1

\usepackage{silence}
\usepackage{jheppub}
\usepackage{epsfig}
\usepackage{amsmath}
\usepackage{graphicx}
\usepackage{capt-of}
\usepackage{tikz-cd}
\usepackage[missing=]{gitinfo2}

\DeclareSymbolFont{AMSa}{U}{msa}{m}{n}
\DeclareSymbolFont{AMSb}{U}{msb}{m}{n}
\let\Box\relax
\DeclareMathSymbol{\Box}{\mathord}{AMSa}{"03}

\allowdisplaybreaks

\newcommand{\ti}[1]{\textit{#1}}
\newcommand{\widet}[1]{\widetilde{#1}}

\newcommand{\half}{{\frac12}}
\newcommand{\eps}{{\epsilon}}

\newcommand{\bbS}{{\mathbb S}}

\newcommand{\cH}{{\mathcal H}}

\newcommand{\N}{{\mathcal N}}
\newcommand{\cN}{{\mathcal N}}
\newcommand{\cK}{{\mathcal K}}

\newcommand{\cF}{{\mathcal F}}

\newcommand{\fg}{{\mathfrak g}}

\newcommand{\C}{{\mathbb C}}

\newcommand{\bbZ}{{\mathbb Z}}

\newcommand{\bbR}{{\mathbb R}}
\newcommand{\Z}{{\bbZ}}
\newcommand{\R}{{\bbR}}

\newcommand{\fro}{{\overline{\underline\Omega}}}

\newcommand{\abs}[1]{\lvert#1\rvert}

\newcommand{\de}{{\mathrm d}}
\newcommand{\e}{{\mathrm e}}
\newcommand{\HOMFLY}{{\mathrm {HOMFLY}}}
\newcommand{\odd}{{\mathrm {odd}}}

\newcommand{\fX}{{\mathfrak X}}

\newcommand{\tM}{\widetilde M}
\newcommand{\tbbS}{\widetilde {\mathbb S}}
\newcommand{\tL}{\widetilde L}
\newcommand{\tl}{\tilde l}
\newcommand{\tC}{\widetilde C}
\newcommand{\tK}{\widetilde K}

\newcommand{\IP}[1]{\langle#1\rangle}

\newcommand{\here}{\mathrm{{here}}}
\newcommand{\there}{\mathrm{{there}}}

\newcommand{\even}{\mathrm{{even}}}

\DeclareMathOperator{\Tr}{Tr}
\DeclareMathOperator{\Sk}{Sk}
\DeclareMathOperator{\GL}{GL}

\DeclareMathOperator{\SU}{SU}
\DeclareMathOperator{\U}{U}
\DeclareMathOperator{\fgl}{{\mathfrak{gl}}}
\DeclareMathOperator{\fsl}{{\mathfrak{sl}}}

\newcommand{\insfigsvg}[3]{

\medskip
\noindent
\begin{minipage}{\linewidth}

\makebox[\linewidth]{\includegraphics[keepaspectratio=true,scale=#2]{figures/#1.pdf}}

\captionof{figure}{#3}

\label{fig:#1}
\end{minipage}
\medskip

}

\newcommand{\insfigsvgs}[2]{
	
	\medskip
	\noindent
	\begin{minipage}{\linewidth}
		
		\makebox[\linewidth]{\includegraphics[keepaspectratio=true,scale=#2]{figures/#1.pdf}}
		
	\end{minipage}
	\medskip
	
}

\title{$q$-nonabelianization for line defects}

\author[1]{Andrew Neitzke}
\author[2]{and Fei Yan}
\affiliation[1]{Department of Mathematics, Yale University}
\affiliation[2]{Department of Physics, The University of Texas at Austin\\
NHETC and Department of Physics and Astronomy, Rutgers University}
\emailAdd{andrew.neitzke@yale.edu}
\emailAdd{fyan.hepth@gmail.com}

\abstract{We consider the \textit{$q$-nonabelianization}
map, which maps links $L$ in a 3-manifold $M$ to combinations of 
links $\tL$ in a branched $N$-fold cover $\tM$. In quantum field theory terms, $q$-nonabelianization is the UV-IR
map relating two different sorts of defect:
in the UV we have the six-dimensional $(2,0)$ superconformal field theory 
of type $\fgl(N)$ on $M \times \R^{2,1}$, and we consider surface defects placed on
$L \times \{x^4 = x^5 = 0\}$; in the IR we have 
the $(2,0)$ theory 
of type $\fgl(1)$ on $\tM \times \R^{2,1}$, and put the defects on
$\tL \times \{x^4 = x^5 = 0\}$.
In the case $M = \R^3$, $q$-nonabelianization 
computes the Jones polynomial of a link,
or its analogue associated to the group $U(N)$.
In the case $M = C \times \R$, when the projection of $L$ to $C$ is a simple non-contractible loop, $q$-nonabelianization 
computes the protected spin character
for framed BPS states in 4d $\cN=2$ theories of class $S$.
In the case $N=2$ and $M = C \times \R$, we give a concrete 
construction of the $q$-nonabelianization map.
The construction uses the data of the WKB foliations
associated to a holomorphic covering $\tC \to C$.}

\begin{document}

%\tracingall

\noindent{{\tiny \color{gray} \tt \gitAuthorIsoDate \gitAbbrevHash}}

\maketitle
\flushbottom

%%%%%%%
\section{Introduction}

This paper concerns a geometric construction which 
we call \ti{$q$-nonabelianization}. In short,
$q$-nonabelianization is an operation which maps
links on a 3-manifold $M$ to links on a branched
$N$-fold cover $\tM$. A bit more precisely, $q$-nonabelianization
is a map of
linear combinations of links modulo certain skein 
relations, encoded in \ti{skein modules} associated 
to $M$ and $\tM$.

In this introduction we formulate the notion of $q$-nonabelianization
rather generally: in particular, we discuss arbitrary $N$,
and a general 3-manifold $M$.
In the body of the paper, we describe a concrete way to construct
$q$-nonabelianization in some detail, but only when $N=2$ and $M = C \times \R$,
with $C$ a surface (which could be noncompact, e.g. $C = \R^2$).
The extensions to higher $N$ and general 3-manifolds $M$
involve similar ideas but various new difficulties, and
will appear in upcoming work.

The $q$-nonabelianization map we describe is related to many previous constructions
in the literature, as we discuss in the rest of this introduction,
and particularly close to the works \cite{Bonahon2010,Galakhov:2014xba,Gabella:2016zxu} which provided important inspirations for our approach;
indeed this paper was motivated by the problem of understanding their constructions in a more covariant and local way.
Our description of $q$-nonabelianization 
can also be viewed as an extension 
of an approach to the Jones polynomial
described in \cite{Gaiotto:2011nm}, particularly Section 6.7
of that paper; from that point of view, what we are doing
in this paper is explaining a way to replace $\R^3$
by $C \times \R$.

\subsection{Physical setup} \label{sec:physics-setup}

Our starting point is the six-dimensional $(2,0)$ superconformal field theory
of type $\fgl(N)$, which we call $\fX[\fgl(N)]$.
We will consider the theory $\fX[\fgl(N)]$ 
in six-dimensional spacetimes of the form
\begin{equation}
	M \times \R^{2,1}
\end{equation}
where $M$ is a Riemannian manifold.
We adopt coordinates as follows: $M$ is coordinatized by
$x^{1,2,3}$, $\R^{2,1}$ by $x^{4,5,0}$.

The theory $\fX[\fgl(N)]$ admits supersymmetric surface defects labeled
by representations of $\fgl(N)$; in this paper we only consider
the defect labeled by the fundamental representation.
Given a 1-manifold (link) $L \subset M$,
we insert this surface defect on a locus
\begin{equation}
	L \times \{x^4 = x^5 = 0\} \subset M \times \R^{2,1},
\end{equation}
and call the resulting insertion $\bbS[L]$.

This kind of setup has been discussed frequently in the physics 
literature. As a tool for studying link invariants,
it appears explicitly in \cite{Witten:2011zz}, and its
various dimensional reductions or M/string theory realizations 
have appeared in many other places,
such as \cite{Ooguri1999,Gukov:2004hz,Chun2015}.
The case of $M=C\times \R$ with $L$ at a fixed $x^3$-coordinate has also been studied in the context of line defects in class $S$ theories, e.g.
\cite{Drukker:2009tz,Drukker:2009id,Gaiotto:2010be,Gaiotto:2012rg},
while for more general $M$ and $L$ see e.g. \cite{Dimofte:2011py}.

\subsection{The UV-IR map} \label{sec:uv-ir}

Our approach to studying this setup
is to pass to the ``Coulomb branch'' of the theory.
The quickest way to understand what this means is to use the
M-theory construction of the theory $\fX[\fgl(N)]$ on $M$. 
This construction involves $N$ fivebranes wrapped on the zero section 
$M \subset T^* M$. To go to the Coulomb branch we consider
instead one fivebrane wrapped on an $N$-fold cover $\tM \subset T^* M$
(possibly branched). This construction has been used frequently in the study
of class $S$ and class $R$ theories, e.g.
\cite{Gaiotto:2009hg,Gaiotto:2011nm,Dimofte2011,Cecotti:2011iy,Gaiotto:2012rg}.

Away from its branch locus the covering $\tM$ is represented
by $N$ 1-forms $\lambda_i$ on $M$.
In order to preserve supersymmetry, the $\lambda_i$ should be harmonic.
In this paper, we will only consider two
kinds of example:
\begin{enumerate}
\item $M = \R^3$, and the 1-forms $\lambda_i$ are constants,
\item $M = C \times \R$ for a Riemann surface $C$, 
and the 1-forms $\lambda_i$ are the real parts of
meromorphic 1-forms on $C$.
\end{enumerate}
(Case 1 is actually a special case of case 2 where
we take $C = \C$, but it is interesting enough to merit
mention on its own.)

Now we ask the question: 
\ti{what does $\bbS[L]$ look like in the IR?}
In this limit the bulk theory
is well approximated by the theory $\fX[\fgl(1)]$ on
$\tM \times \R^{2,1}$ \cite{Cecotti:2011iy}. 
In that theory we could consider surface defects
$\tbbS[\tL]$ on the loci
\begin{equation}
	\tL \times \{x^4 = x^5 = 0\} \subset \tM \times \R^{2,1},
\end{equation}
for links $\tL \subset \tM$.
The picture we propose, similar to e.g. \cite{Gaiotto:2010be,Coman:2015lna}, 
is that \ti{$\bbS[L]$ 
decomposes into a sum of defects $\tbbS[\tL]$},
in the form
\begin{equation} \label{eq:uv-ir-surface}
	\bbS[L] \rightsquigarrow \sum_{\tL} a(\tL) \tbbS[\tL].
\end{equation}
The meaning of the ``coefficients'' $a(\tL)$ 
is a bit subtle since we are not talking about local operators 
but rather extended ones. Nevertheless, we want to compute
something concrete, and we proceed as follows.
\eqref{eq:uv-ir-surface}
implies a decomposition of the Hilbert space
in the presence of the defect $\bbS[L]$ as a direct sum over
Hilbert spaces associated to the IR defects, of the form
\begin{equation}
	\cH_L = \bigoplus_{\tL} V_{a(\tL)} \otimes \cH_{\tL}.
\end{equation}
What we will really compute is the
Laurent polynomials $\alpha(\tL) = \Tr_{V_{a(\tL)}} (-q)^{2 J_3} q^{2 I_3}
\in \Z[q,q^{-1}]$ (see below).

\subsection{Computing the UV-IR map}

The construction of the UV-IR map \eqref{eq:uv-ir-surface} which we propose
goes as follows.
The links $\tL \subset \tM$ are
built out of two sorts of local pieces:
\begin{itemize}
	\item Lifts of segments of $L$ to one of the sheets $i$ of the 
	covering $\tM\to M$.
	\insfigsvgs{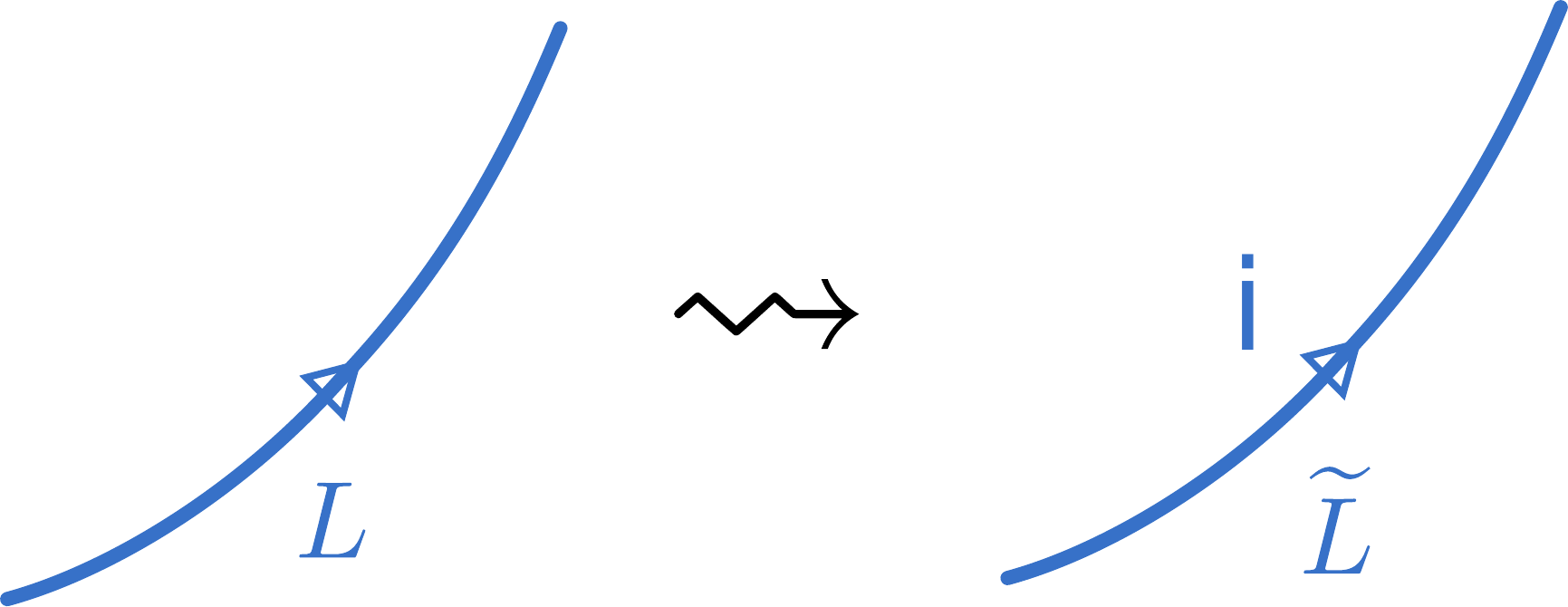}{0.25}
\item Lifts to $\tM$ of webs of strings, 
labeled by pairs $ij$ of sheets of the covering
$\tM \to M$, ending on branch points of the covering or on the link
$L$.
\insfigsvgs{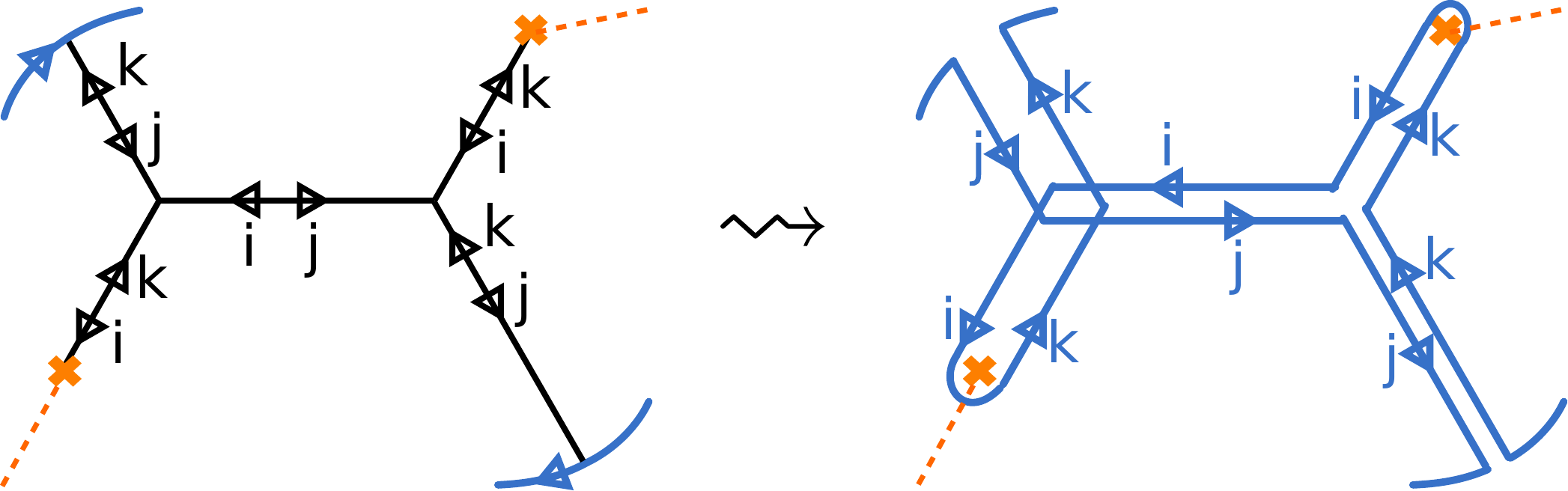}{0.35}
\end{itemize}

To each link $\tL$ built from these pieces we assign a corresponding weight $\alpha(\tL)\in\Z[q^{\pm1}]$, built as a product of elementary local factors.
Most of the difficulty in constructing the UV-IR map is to get these factors
correct. 

In this paper we develop this scheme in detail only for $N=2$.
In the $N=2$ case there are no trivalent string junctions, which
enormously simplifies the situation:
the only kinds of webs we have to deal with are 
the two shown below.
\insfigsvgs{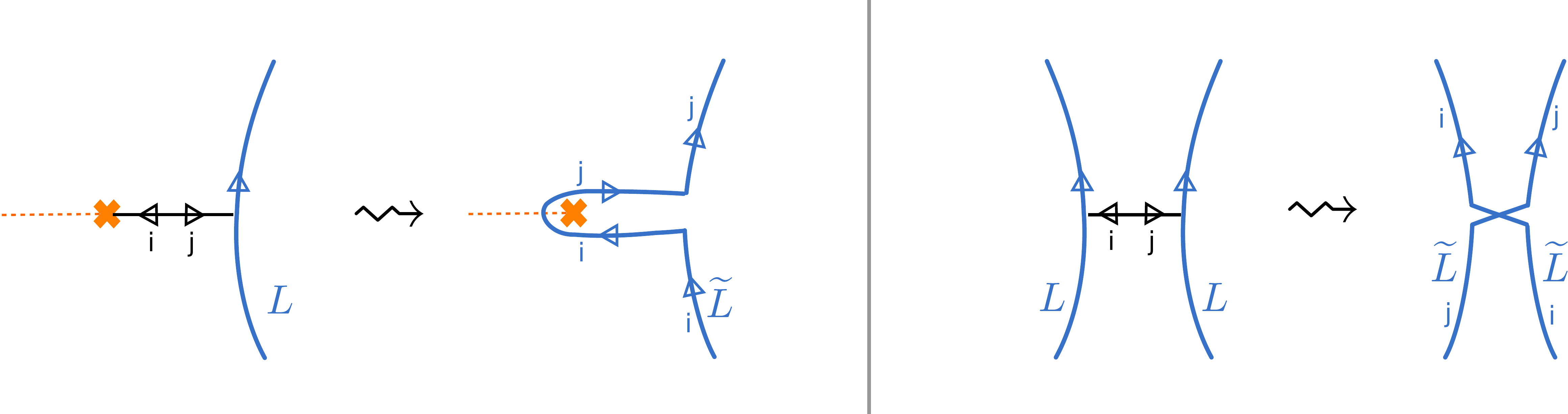}{0.27}
On the left is a string connecting the 
link $L$ to a branch point of the covering $\tM \to M$;
the lift $\tL$ takes a \ti{detour} from the lift of $L$ to the
branch point along sheet $i$ and back again along sheet $j$.
On the right is a string connecting two different strands of $L$;
we call this an \ti{exchange}, because
the two strands of the lift $\tL$ exchange sheets
along the lifted string.

There is an important difference between the detours and the exchanges.
As we discuss in \autoref{sec:qab-N=2}, each exchange
comes with a factor of $(q - q^{-1})$, so the contribution from exchanges
vanishes at $q=1$. (More generally, for $N>2$, a web
with $k$ ends on $L$ will come with a factor of $(q-q^{-1})^{k-1}$.)
When we set $q = 1$, therefore, the only web contributions
which survive are those from webs with exactly one end on $L$.
Then our description of the UV-IR map reduces to
a construction which already appeared in \cite{Gaiotto:2012rg},
for line defects in class $S$ theories. In that context
the webs attach to $L$ only at places where $L$ 
crosses the WKB spectral network.
In contrast, when we allow $q \neq 1$, the webs can 
attach anywhere on $L$.

In \cite{Gabella:2016zxu} a version of the UV-IR map is described, in
the language of spectral networks,
which uses the detours but not the exchanges: in lieu of the exchanges 
one isotopes $L$
to a specific profile relative to the spectral network,
and then inserts by hand additional $R$-matrix factors.
In our approach we do not make such an isotopy and we do not 
insert these $R$-matrix factors.
The relation between the two approaches is roughly that,
if we apply our approach to an $L$ in this specific profile,
then exchanges appear and produce automatically the off-diagonal parts
of the $R$-matrix. 

\subsection{Skein relations}

BPS quantities in the theory with surface defects
$\bbS[L]$ inserted, such as supersymmetric indices, 
obey some relations.
First, they depend only on the isotopy class of the link $L$, because 
deforming by an isotopy is a $Q$-exact deformation.
Second, and more interestingly, there are also some 
$Q$-exact deformations which relate
$\bbS[L]$ for non-isotopic links $L$.

We do not have a first-principles derivation 
of what these additional relations are, but to get a hint,
we can use a picture from \cite{Witten:2011zz}:
Euclideanize and compactify the $x^0$ direction,
replace the $x^4$-$x^5$ directions by a cigar and
compactify on its circular direction. The resulting effective
theory is $\N=4$ super Yang-Mills on $M \times \R_+$,
with a boundary condition at the finite end.
The operators $\bbS[L]$ reduce to
supersymmetric Wilson lines in the boundary $M$.

From this point of view one should expect $\bbS[L]$
to obey the skein relations of analytically 
continued Chern-Simons theory.
We formulate this as the conjecture that
up to $Q$-exact deformations, $\bbS[L]$
is determined by the class of $L$ in the \ti{skein module}
$\Sk(M, \fgl(N))$, described 
explicitly in \autoref{sec:gln-skein} below.
This sort of skein relation in supersymmetric field theory
has been discussed in many different contexts,
e.g. \cite{Drukker:2009id,Gaiotto:2010be,Coman:2015lna,Gabella:2016zxu,Tachikawa2015,Longhi:2016wtv}.
Similarly we propose that the IR defect $\tbbS[\tL]$
depends only on the class of $\tL$ in another skein module 
$\Sk(\tM, \fgl(1))$, described in
\autoref{sec:gl1-skein}.

The UV-IR decomposition takes $Q$-exact deformations
in the UV to $Q$-exact deformations in the IR. 
It follows that it should descend to a map of skein modules,
\begin{equation}
	F: \Sk(M, \fgl(N)) \to \Sk(\tM, \fgl(1)).
\end{equation}
This is a strong constraint, which in particular is strong
enough to determine all the weight factors $\alpha(\tL)$.
In \autoref{sec:isotopyproof} we verify that our rules indeed 
satisfy this constraint in the $N=2$ case.
In case $M = C \times \R$, the skein modules are actually
algebras, because of the operation of ``stacking'' links in the 
$\R$ direction (see \autoref{sec:skein-algebra}); 
this algebra structure gets related to the OPEs
of the operators $\bbS[L]$ or $\bbS[\tL]$, and $F$ is then
a homomorphism of algebras.

We should mention one subtlety we have been ignoring: the 
skein modules we consider involve \ti{framed} links.
The need to frame links, though familiar for Wilson lines in Chern-Simons
theory, is not immediately
obvious from the point of view of the theory $\fX[\fgl(N)]$.
Nevertheless, it seems that they do need framing, and that 
shifting the framing in the $\fgl(N)$ theory by $1$ unit
is equivalent to changing the angular momentum of the defect 
in the $x^4$-$x^5$ direction by $N$ units.
It would be desirable to understand this in a more fundamental way.

\subsection{Framed BPS state counting}

In this subsection we discuss one application of the UV-IR map:
it can be used to compute the spectrum of ground states of
the bulk-defect system. 
This point of view unifies the computation
of link polynomials and that of framed BPS states in 4d 
class $S$ theories. When we say ``framed BPS state'' we also include 
states associated to a slightly
unfamiliar sort of line defect which breaks rotation invariance;
we call these ``fat line defects'' and discuss them
in \autoref{sec:non-flat-links} below.

\subsubsection{Links in \texorpdfstring{$\R^3$}{R3}} \label{sec:links-in-r3}

The simplest case arises when we take 
\begin{equation}
	M = \R^3,
\end{equation}
i.e. we consider $\fX[\fgl(N)]$ 
in the spacetime $\R^{5,1}$.
This theory has $16$ supercharges.
For generic $L$, the defect $\bbS[L]$ preserves $2$ of
these supercharges \cite{Zarembo2002}.
The setup also preserves a rotational 
$\U(1)_P$ in the $x^4$-$x^5$ plane and a $\U(1)_R$ symmetry.

Now we make a ``Coulomb branch'' perturbation $\tM$ as mentioned above.
The ground states of the system form a vector
space depending on the link $L$ and 
the perturbation $\tM$, which we call $\cH_L(\tM)$.
$\cH_L(\tM)$ is a representation of $\U(1)_P \times \U(1)_R$.

It was proposed in \cite{Witten:2011zz,Gaiotto:2011nm} following
\cite{Gukov:2004hz} that $\cH_L(\tM)$ is a homological invariant of the link $L$,
likely closely related to the  \ti{Khovanov-Rozansky homology}.
The action of $\U(1)_P \times \U(1)_R$ is responsible for the bigrading
of the link homology.
In particular, suppose we let $J_3$ denote the generator of
$\U(1)_P$, and $I_3$ the generator of $\U(1)_R$. Then the
proposal of \cite{Witten:2011zz,Gaiotto:2011nm} implies that
the generating function 
\begin{equation} \label{eq:psc-homfly}
	\fro(L,\tM) := \Tr_{\cH_{L}(\tM)} (-q)^{2 J_3} q^{2 I_3} \, \in \, \Z[q,q^{-1}],
\end{equation}
is related to the (un-normalized) 
HOMFLY polynomial $P_\HOMFLY(L)$. In our conventions the precise relation 
is 
\begin{equation} \label{eq:homfly-relation-intro}
\fro(L,\tM) = q^{N w(L)} P_\HOMFLY(L, a = q^N, z = q - q^{-1})	
\end{equation}
where $w(L)$ is the self-linking number of $L$.
(In particular, this generating function is actually
independent of the chosen perturbation $\tM$, although \ti{a priori}
it might have depended on it.)

The generating function $\fro(L,\tM)$ 
can also be computed using the IR
description of the theory. Indeed the IR skein module 
$\Sk(\tM,\fgl(1))$ is easy to understand: every $\tL$ is equivalent
to the class of some multiple of the empty link $[\cdot]$.
It is not completely obvious what the contribution from the empty link 
to $\fro(L,\tM)$ should be (this amounts to counting the ground states
of the system without a defect), but we propose that this contribution
is just $1$, and thus that
\begin{equation}
F([L]) = \fro(L,\tM) [\cdot].
\end{equation}
Thus, in this particular case, the UV-IR map 
described in \autoref{sec:uv-ir} must reduce to a method of computing 
the knot polynomial \eqref{eq:homfly-relation-intro}.
We discuss this more in \autoref{sec:state-sum} below.

\subsubsection{Flat links in \texorpdfstring{$C \times \R$}{C x R}} \label{sec:flat-links}

Now we consider the case
\begin{equation}
	M = C \times \R
\end{equation}
where $C$ is a Riemann surface (perhaps with marked points.)
In this case we take a twisted version of $\fX[\fgl(N)]$
in the spacetime $C \times \R^{3,1}$.
This kind of twist was used in
\cite{Gaiotto:2009we,Gaiotto:2009hg}; it preserves $8$ supercharges.
If $C$ is compact, then on flowing to the IR
one arrives at an $\cN=2$ theory of class $S$ in $\R^{3,1}$; we
denote this theory $\fX[C,\fgl(N)]$.

We are going to consider a link $L \subset M$ and a surface defect
$\bbS[L]$.
We fix a
branched holomorphic $N$-fold covering $\tC \to C$, where $\tC \subset T^*C$.
(Such a covering $\tC$ is also a Seiberg-Witten curve, associated to 
a point of the Coulomb branch of the theory $\fX[C, \fgl(N)]$,
as described in \cite{Gaiotto:2009hg}.)
Then we define
\begin{equation}
\tM = \tC \times \R.
\end{equation}

The symmetries preserved by the defect $\bbS[L]$ depend on how $L$ is placed.
In this section we consider the following: pick a simple closed curve $\wp \subset C$
and then take 
\begin{equation}
	L = \wp \times \{ x^3 = 0 \}.
\end{equation}
We call such an $L$ ``flat'' since it explores only 2 of the 3 dimensions of $M$.
The remaining $x^3$-direction combines with the $x^4$-$x^5$ plane to give an
$\R^3$, in which $\bbS[L]$ sits at the origin;
thus in this setup we have 3-dimensional rotation invariance $\SU(2)_P$,
and it turns out to also preserve $R$-symmetry $\SU(2)_R$, and $4$ supercharges.

For $C$ compact and $L$ flat, from the point of view of the reduced theory $\fX[C,\fgl(N)]$,
$\bbS[L]$ is a $\half$-BPS line defect. These line defects
have been studied extensively, beginning with
\cite{Drukker:2009tz,Drukker:2009id,Gaiotto:2010be}.\footnote{More exactly, those
	works considered the case where the simple closed curve $\wp$ is not contactible. 
	The case of contractible $\wp$
	turns out to be a bit special, as we will see below.}
In particular, the vector space $\cH_L(\tM)$ of ground states has been studied,
e.g. in
\cite{Gaiotto:2010be,Gaiotto:2012rg,Galakhov:2014xba,Gabella:2016zxu}; it 
is called the space of
\ti{framed BPS states} of the line defect.\footnote{The ``framed''
	in ``framed BPS states'' is not directly related to the framing of links;
	we will discuss the role of framing of links below.}

$\cH_L(\tM)$ admits a grading:
\begin{equation}\label{eq:HLdecomposition}
	\cH_L(\tM) = \bigoplus_{\gamma \in \Gamma(\tM)} \cH_{L,\gamma}(\tM)
\end{equation}
where $\Gamma(\tM) = H_1(\tM, \Z)$ is the IR charge lattice of the four-dimensional theory
$\fX[C, \fgl(N)]$, in the vacuum labeled
by the covering $\tM$.
The grading by $\Gamma(\tM)$ keeps track of electromagnetic and 
flavor charges of the framed BPS states.
Each $\cH_{L,\gamma}(\tM)$ is a representation of $\SU(2)_P \times \SU(2)_R$.

Now fix Cartan generators $J_3$, $I_3$ of $\SU(2)_P$ and $\SU(2)_R$ respectively.
Then we can consider the \ti{protected spin character}\footnote{In comparing to \cite{Gaiotto:2010be} we have $q_\here = -y_\there$.}
\begin{equation} \label{eq:psc-first}
	\fro(L,\tM,\gamma) := \Tr_{\cH_{L,\gamma}(\tM)} (-q)^{2 J_3} q^{2 I_3} \, \in \, {\bbZ}[q,q^{-1}].
\end{equation}
Unlike the case of $M=\R^3$, here $\cH_L(\tM)$ does depend
strongly on $\tM$; as we vary $\tM$ (moving in the Coulomb branch)
the invariants $\fro(L,\tM,\gamma)$ can change.
This is the phenomenon of \ti{framed wall-crossing}. 
(We discuss this phenomenon in more detail in \autoref{sec:wall-crossing}, where we show that our map $F$
obeys the expected framed wall-crossing formulas
associated to BPS hypermultiplets and vector multiplets.)

Once again, we can compute the invariants $\fro(L, \tM, \gamma)$
using the IR description of the theory on its Coulomb branch.
In this case the IR skein module $\Sk(\tM,\fgl(1))$ 
is more interesting:
it has one generator $X_\gamma$ for each class $\gamma \in H_1(\tC, \Z)$.
Roughly speaking $X_\gamma$ is just represented by a loop on $\tC$ in class $\gamma$; 
see \autoref{sec:quantumtorus} for
the precise statement. A link in class $X_\gamma$ gives an IR 
line defect carrying charge $\gamma$, which contributes
one state to $\cH_{L,\gamma}(\tM)$. This leads to the proposal
\begin{equation} \label{eq:nonab-psc}
	F([L]) = \sum_{\gamma \in H_1(\tC,\Z)} \fro(L, \tM, \gamma) X_\gamma,
\end{equation}
i.e., $q$-nonabelianization gives the generating function of
the protected spin characters.

\subsubsection{Non-flat links in \texorpdfstring{$C \times \R$}{C x R}} \label{sec:non-flat-links}

Continuing with the case $M = C \times \R$, now we consider
a more general link $L \subset M$. In this case we
only have 2-dimensional rotation invariance $\U(1)_P$ (in the $x^4$-$x^5$ plane),
and $\U(1)_R$, and 2 supercharges. From the point of view of symmetries,
this is the same as the case of general links in $M = \R^3$ which we considered
in \autoref{sec:links-in-r3}.

What does the defect $\bbS[L]$ look like from the point of view of
the reduced theory $\fX[C,\fgl(N)]$?
In the IR, the position variations of the defect in the $x^3$ direction are
suppressed, so the effective support of the defect in the spatial $\R^3$ 
is a point. Thus we obtain a $\frac14$-BPS
line defect in the theory $\fX[C, \fgl(N)]$. 

If $L$ is isotopic to a flat link, then
we expect that this defect is actually $\frac12$-BPS in the IR, and 
all the IR physics
should be the same as in \autoref{sec:flat-links}.
If $L$ is not isotopic to a flat link, though, then the 
line defect we get is really only $\frac14$-BPS.
Moreover, in the UV the $\frac14$-BPS defect breaks the rotational 
symmetry $\SU(2)_P$ to the $\U(1)_P$ rotation in the $x^4$-$x^5$ plane, 
and the full $\SU(2)_P$ need not be restored 
in the IR.
Thus we expect that a general link $L \subset C \times \R$ corresponds to 
an unconventional sort of line defect in theory $\fX[C,\fgl(N)]$,
which partially breaks rotation invariance in the spatial $\R^3$.
We call these \ti{fat line defects}.

As for a $\half$-BPS line defect, we can consider a protected spin character
$\fro(L, \tM, \gamma) \in \Z[q,q^{-1}]$
for a fat line defect; it is defined by exactly the same equation \eqref{eq:psc-first}
which we used before,
with $J_3$ and $I_3$ the generators of $\U(1)_P$ and $\U(1)_R$.
We propose that this protected spin character is also computed
by $q$-nonabelianization, just as in \eqref{eq:nonab-psc} above.

\subsubsection{Positivity} \label{sec:positivity}

It was conjectured in \cite{Gaiotto:2010be} that, when $L$ is flat and not contractible,
the action of $\SU(2)_R$ on $\cH_L$ is trivial; this is the ``framed no-exotics''
conjecture\footnote{There is an analogous no-exotics conjecture for bulk BPS states. Important progress towards fully proving the no-exotics conjectures via physical arguments has been made by Clay C\'{o}rdova and Thomas Dumitrescu \cite{CD}.}.
If the framed no-exotics conjecture is true, then when $L$ is flat,
\eqref{eq:psc-first} reduces to the simpler
\begin{equation} \label{eq:psc-reduced}
	\fro(L,\tM,\gamma) := \Tr_{\cH_{L,\gamma}(\tM)} (-q)^{2 J_3}.
\end{equation}
This would imply that the coefficients in the $(-q)$-expansion
of $\fro(L,\tM,\gamma)$
are all positive, and moreover they have symmetry and monotonicity
properties following from the fact that they give a character
of the full $\SU(2)_P$, not only of $\U(1)_P$.

For $N=2$ the expected 
positivity has been established in \cite{Allegretti:2016jec,Cho:2017ymn},
for the version of $F$ given in \cite{Bonahon2010}.
The expected monotonicity property has not been proven as
far as we know.\footnote{We thank Dylan Allegretti for several
useful explanations about this.}
We will see in various examples below that the coefficients 
in $\fro(L,\tM,\gamma)$ computed from our $F$ do
have the expected properties; it would be very interesting
to give a proof directly from our construction of $F$.

In contrast, for non-flat links $L$ there is no reason to expect any
kind of positivity property (and indeed,
for a link contained in a ball in $M$ we get the 
polynomial \eqref{eq:homfly-relation-intro}, which is 
in general not sign-definite.)
Likewise when $L$ is contractible we do not necessarily expect positivity,
and indeed for a small unknot in $M$ (even a flat one) 
we will get $F(L) = q + q^{-1}$, which does not have a positive
expansion in $-q$.

\subsection{Connections and future problems}

\begin{enumerate}

\item In the language of theory $\fX[\fgl(N)]$,
our UV-IR map can be interpreted roughly as follows.
The direct lifts of $L$ to $\tM$ come from 
the usual symmetry-breaking
phenomenon: moving to the Coulomb branch breaks 
the symmetry locally from $\fgl(N)$ to
its Cartan subalgebra of diagonal matrices, 
and correspondingly decomposes the fundamental representation
of $\fgl(N)$ into $N$ one-dimensional weight spaces.
The terms involving webs are contributions
from massive BPS strings of theory $\fX[\fgl(N)]$
on its Coulomb branch. In particular, a physical interpretation of the exchange factor for $N=2$ has been given in \cite{Gaiotto:2011nm}. Moreover its categorification in terms of knot homology was studied in \cite{Galakhov:2016cji}. We hope that similar interpretations also exist for generic web factors.

\item
In the language of the M-theory construction of 
$\fX[\fgl(N)]$, 
compactifying the time direction to $S^1$ to compute the
BPS indices 
could optimistically be understood as computing 
a partition function in Type IIA string theory on $T^* M$,
with $N$ D6-branes inserted on the zero section $M \subset T^* M$,
and D4-branes placed on the conormal bundle to $L$ in $T^* M$.
Adding the parameter $q$ would be implemented by making a rotation
in the $x^4$-$x^5$ plane as we go around the $S^1$;
from the Type IIA point of view this corresponds to 
activating a graviphoton background.
Such a partition function is computed by the
$A$ model topological string on $T^* M$, with Lagrangian boundary
conditions at the D-brane insertions,
with string coupling $g_s$ where $q = \e^{-g_s}$ \cite{Ooguri1999}.

In this language the webs are interpreted as 
M2-branes in $T^*M$ whose boundary lies partly on $\tM$ and partly on 
the conormal to $L$. For $N=2$ this is indicated in \autoref{fig:branefig}.

\insfigsvg{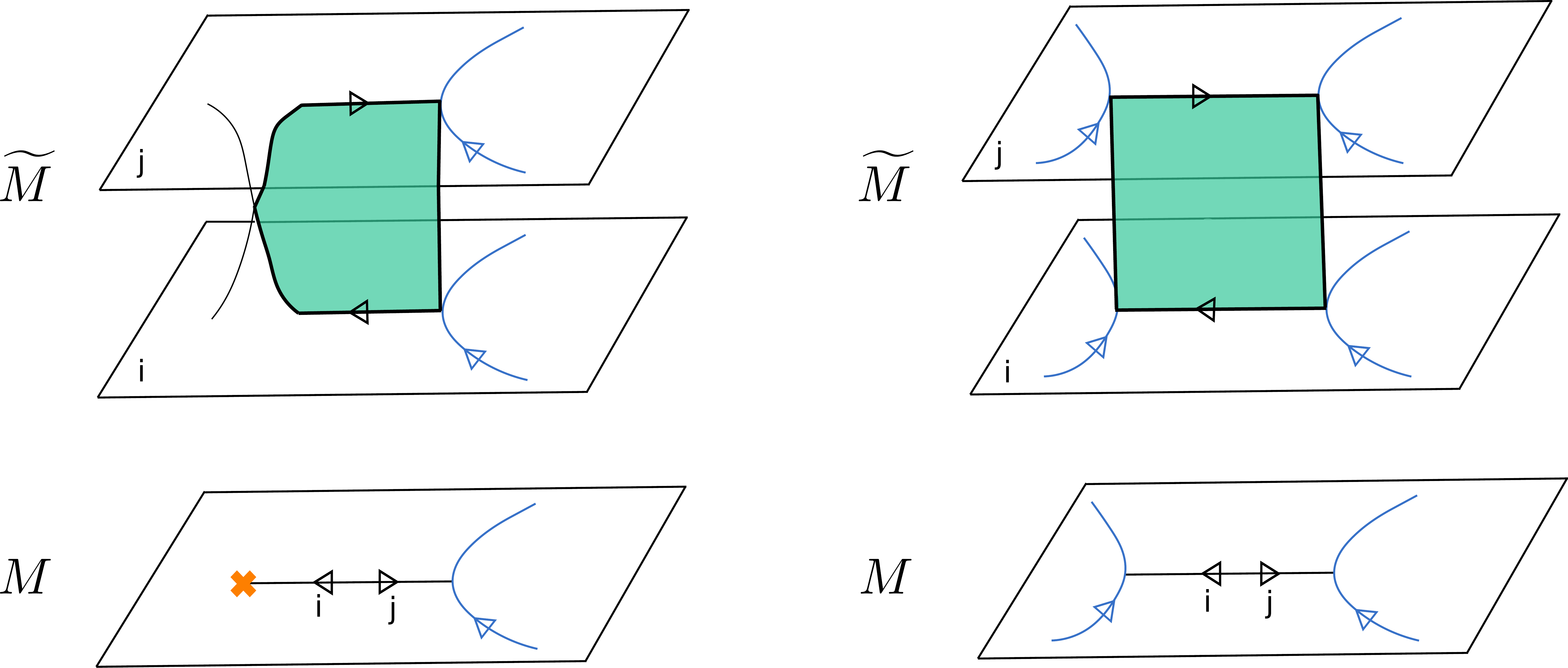}{0.18}{The interpretation of detour (left) and exchange (right) as M2-branes in $T^*M$ bounded by $\tM$ and the conormal to $L$. }

\item
In the topological string approach to link invariants
pioneered in \cite{Ooguri1999},
one takes the above setup with $M=S^3$, but then passes 
through the conifold transition.
After the conifold transition we do not have the $N$ D6-branes anymore,
but we do have a compact holomorphic 2-cycle with volume $N g_s$.
In consequence, this sort of computation involves the variable
$N$ only through the combination $a = \e^{- N g_s} = q^N$; so for example
one gets the HOMFLY polynomial directly as a function of $a$, 
rather than its specialization \eqref{eq:homfly-relation-intro}
to a particular $N$. Recently this open topological
string computation has been interpreted in the language
of skein modules \cite{Ekholm2019}.
We want to emphasize that our computation is on the other
side, ``before'' the conifold transition.

\item We discuss in this paper mainly the cases $M = \R^3$ and $M = C \times \R$.
Having come this far, it is natural to consider the case of a general 3-manifold
$M$. There is a twist of $\fX[\fgl(N)]$ on $M \times \R^{2,1}$ which preserves
$4$ supercharges. If $M$ is compact, then on flowing to the IR
one arrives at a theory $\fX[M, \fgl(N)]$ in $\R^{2,1}$;
examples of these theories have been studied e.g. in \cite{Dimofte2011,Dimofte:2011py,Dimofte2013}.
The surface defect $\bbS[L]$ gives in the IR a $\half$-BPS line defect in theory
$\fX[M, \fgl(N)]$.

As before we could imagine perturbing the theory to reach a ``Coulomb branch''
using an $N$-fold covering $\tM \to M$.
Unlike the cases we have discussed up to now, though,
here we do not have a good understanding of the IR physics.
In particular, it is not clear to us
that we can define a meaningful ground state Hilbert
space $\cH_L(\tM)$ in this case.
Nevertheless, we could still think about
the IR decomposition of line defects, and the corresponding
map of skein modules.
We believe that the same rules we use in this paper
(in their covariant incarnation, \autoref{sec:covariant})
will work for more general $M$, but there is a subtlety in formulating
the skein module $\Sk(\tM, \fgl(1))$ in this case: one needs to include 
corrections from boundaries of holomorphic discs in $T^* \tM$.
The $q = 1$ version of this problem will be treated in \cite{FreedNtoappear};
the case of general $q$ is ongoing work.

\item One way of thinking of the skein modules is that they describe
relations obeyed by Wilson lines in Chern-Simons theory. With that in mind,
our $q$-nonabelianization map could be interpreted as part of a general
relation between $\GL(N)$ Chern-Simons theory on $M$ and $\GL(1)$
Chern-Simons theory on $\tM$, with the latter corrected by holomorphic discs
--- or more succinctly, as a relation between the A model topological string
on $T^*M$ with $N$ D-branes on $M$ and the A model topological string 
on $T^*M$ with
$1$ D-brane on $\tM$. The possibility of such a relation was
proposed in \cite{Cecotti:2011iy,Galakhov:2014xba}.
Its avatar in classical Chern-Simons theory will appear in \cite{FreedNtoappear}.
The existence of a relation between local systems on $M$
and A branes in $T^*M$ is also known in the mathematics literature, e.g.
\cite{nadler2006constructible,nadler2006microlocal,Jin_2015}.

\item Quantization of character varieties, skein modules and skein algebras
have recently been very fruitfully investigated 
from the point of view of topological field theory
 \cite{BenZvi2015,BenZvi2016,Ganev2019,Gunningham2019}. Roughly speaking,
 the underlying idea is that an appropriate version of the
 $\fgl(N)$ skein module of $M$ can be identified with the space
 of states of a topologically 
 twisted version of $\cN=4$ super Yang-Mills theory
 on $M \times \R$ with gauge group $\U(N)$. 
 It seems natural to ask whether $q$-nonabelianization
 can be understood profitably in this language.\footnote{This perspective
 was emphasized to us by Davide Gaiotto and David Jordan.}

\item As we have mentioned, the map $F$ of skein modules which we
construct for $N=2$ in this paper
closely resembles the ``quantum trace'' map constructed first in \cite{Bonahon2010}
and revisited in \cite{Gabella:2016zxu}. Our map cannot 
be precisely the same
as the map considered there, if only because the relevant skein modules are
different: ultimately this is related to the fact that we consider
the $\fgl(2)$ theory while those references consider $\fsl(2)$.
We believe that after appropriately decoupling the central
$\fgl(1)$ factor, the maps are likely the same;
we discuss this point a bit more in \autoref{sec:sl2}.

Our map is also closely related
to the computation of framed BPS states for certain
interfaces between surface defects given in \cite{Galakhov:2014xba};
in particular, the writhe used in \cite{Galakhov:2014xba} is essentially
the same as the power of $q$ which appears in relating
a class $[\tL] \in \Sk(\tM, \fgl(1))$ 
to a quantum torus generator $X_\gamma$,
as described in \autoref{sec:quantumtorus}.

Nevertheless, our way of producing $F$ looks quite different from the
constructions in \cite{Bonahon2010,Gabella:2016zxu,Galakhov:2014xba}, 
and has some advantages: 
it is more local, it can be covariantly formulated on a general $M$,
it does not require us to make a large isotopy of the link
$L$ to put it in some special position,
and it makes the connections to the strings of the theory $\fX[\fgl(N)]$
or to holomorphic discs more manifest.
We expect that these advantages will be useful for further developments.
(In particular, for $N>2$ our approach here is well suited for 
dealing with the case of theories $\fX[C,\fgl(N)]$
at general points of their Coulomb branch, involving more general spectral networks
than the ``Fock-Goncharov'' type considered in \cite{Gabella:2016zxu}.)

\item In this paper we compute indexed dimensions of spaces 
$\cH_L(\tM)$ of framed BPS
states. It would be very interesting to try to promote our construction to 
get the actual vector spaces $\cH_L(\tM)$, with their bigrading
by $\U(1)_P \times \U(1)_R$. When $M = \R^3$ this is expected
to give some relative of the Khovanov-Rozansky homology as discussed e.g.
in \cite{Gukov:2004hz,Witten:2011zz,Gaiotto:2011nm,Gaiotto:2015aoa,Galakhov:2016cji}.

\item At least for flat links in $C\times \R$, there is also an algebraic approach to studying the framed BPS states. In \cite{Cordova:2013bza, Cirafici:2013bha, Chuang:2013wt} it was proposed that for theories of {\it quiver type} as studied in \cite{Douglas:1996sw,Douglas:2000ah,Douglas:2000qw,Cecotti:2010fi, Alim:2011kw, Cecotti:2011gu, DelZotto:2011an}, framed BPS spectra of line defects could be computed by methods of quiver quantum mechanics. More recently this idea has been further developed in \cite{Cirafici:2017iju,Cirafici:2017wlw,Cirafici:2018jor,Cirafici:2019otj}. It would be nice to understand better the connection between this algebraic method and our geometric approach. One possibility might be to compare the method of {\it BPS graphs} as in \cite{Longhi:2016wtv,Gabella:2017hpz,Gang:2017ojg} to the framed BPS quivers.

\item As we have mentioned, when $M = C \times \R$ our $q$-nonabelianization map is closely connected
to the quantum trace of \cite{Bonahon2010}. For other aspects of the quantum trace see e.g. 
\cite{Le2015,Allegretti:2015nxa,Allegretti:2016jec, Cho:2017ymn,Kim:2018dux,Korinman-Quesney}.
It is also closely related to the
quantum cluster structure on moduli spaces of flat connections,
discussed in \cite{Fock-Goncharov,MR2567745,alex2019quantum};
for some choices of the covering $\tC \to C$,
we expect that the formula \eqref{eq:nonab-psc} 
gives the expansion of a distinguished element $F([L])$
of the quantum cluster algebra, relative to the variables $X_\gamma$
of a particular cluster determined by $\tC$.

\end{enumerate}

\subsection*{Acknowledgements}
We thank Dylan Allegretti, J\o rgen Andersen, Sungbong Chun, Clay C\'ordova, Davide Gaiotto, Po-Shen Hsin, Saebyeok Jeong, David Jordan, Pietro Longhi, Rafe Mazzeo, Gregory Moore, Du Pei, Pavel Putrov, Shu-Heng Shao, and Masahito Yamazaki for extremely helpful discussions. We thank Dylan Allegretti, Pietro Longhi, Gregory Moore, and Du Pei for extremely helpful comments on a draft. AN is supported 
in part by NSF grant DMS-1711692.
FY is supported by DOE grant DE-SC0010008. This work benefited from the 2019 Pollica summer workshop, which was supported in part by the Simons Collaboration on the Non-Perturbative Bootstrap and in part by the INFN.

\section{The state sum model revisited} \label{sec:state-sum}

We begin with the simplest case of our story.
As we have discussed above, when $M = \R^3$, $q$-nonabelianization should 
boil down to a way of computing the specialization \eqref{eq:homfly-relation-intro} 
of the HOMFLY polynomial.

The precise way in which this works depends on the shape of the covering $\tM \to M$.
If $\tM$ is given by $N$ constant 1-forms $\lambda_i$ which are all nearly
parallel, then $q$-nonabelianization
is essentially equivalent to a known method of computing
\eqref{eq:homfly-relation-intro}, known as the \ti{state sum model}.
In this section we first review the state sum model and then describe the ways
in which $q$-nonabelianization generalizes it.

\subsection{The state sum model} \label{sec:state-sum-details}

Choose a distinguished axis in $\R^3$, say the $x^3$-axis. This induces a projection $\R^3 \to \R^2$.
We assume that $L$ is in general position relative to this projection; if not, make a small
isotopy so that it is.
Then the projection induces an oriented knot diagram in $\R^2$.

A \ti{labeling} $\ell$
of the knot diagram
consists of an assignment of a label $i \in \{1, \dots, N\}$ to each arc. 
The state sum is a sum over all labelings.
Each labeling $\ell$ is assigned a weight 
$\alpha(\ell) \in \Z[q,q^{-1}]$
according to the following rules:

\insfigsvg{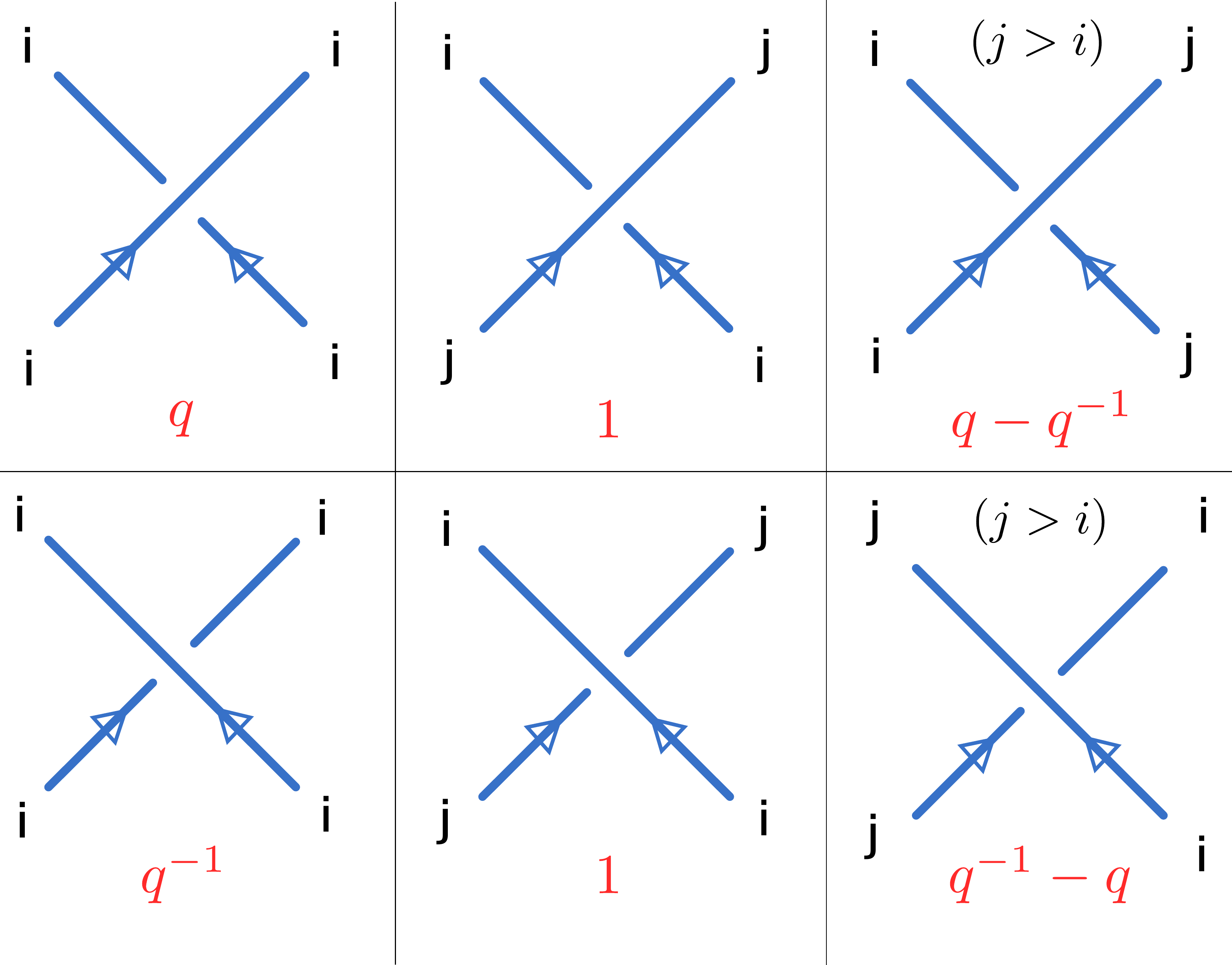}{0.24}{Multiplicative weights 
in the state sum model, associated to labeled crossings in a knot diagram.}

\begin{itemize}
\item 
Each crossing gives a factor depending on the labels of the four involved arcs, 
as indicated in \autoref{fig:state-sum-formulas}.
If the labels at any crossing are not of one of the types shown in the figure, then
the factor for that crossing is $0$ (and thus $\alpha(\ell) = 0$.)
\item 
If the diagram includes crossings where the labels change, 
we ``resolve'' the crossings as shown in \autoref{fig:state-sum-formulas-2}.
After so doing, the diagram consists of loops, each carrying a fixed
label $i$. We assign each such loop a factor $q^{w(N+1-2i)}$, where $w$
is the winding number of the projection of the loop to $\R^2$ (so e.g.
for a small counterclockwise loop the winding number is $+1$.)\footnote{Our
description of this factor is a bit different from what usually appears
in the literature. One could
equivalently define this factor by first making the stipulation that whenever 
there is a crossing, the orientations of both arcs in the 
crossing should have the same sign for their $y$-component (both up or both down), 
and then assigning factors
$q^{\pm \frac12(N-1-2i)}$ to local maxima and minima of $y$ along arcs;
these locations are sometimes called ``cups'' and ``caps'' in the knot diagram.}
\end{itemize}

\insfigsvg{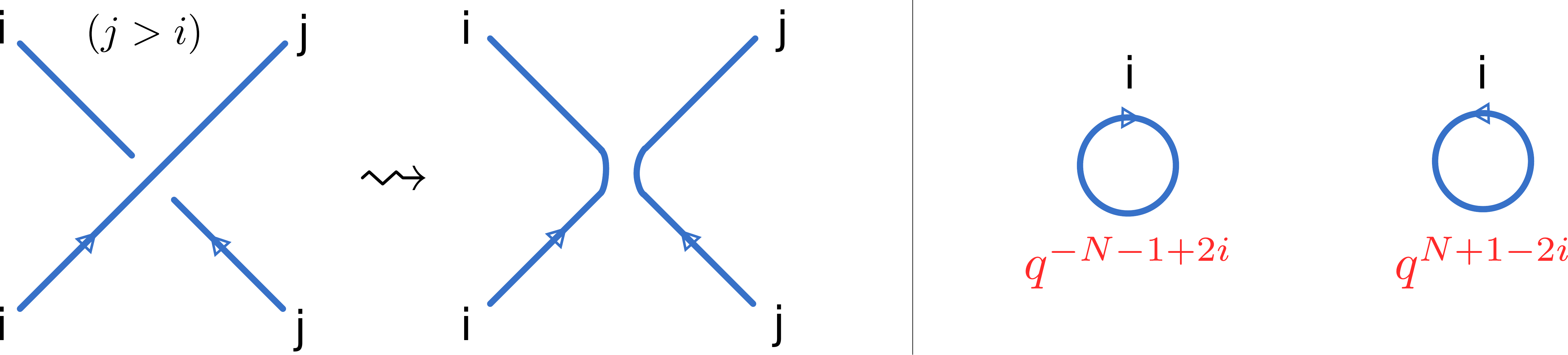}{0.19}{Left: resolving a labeled knot diagram. Right: 
multiplicative weights associated to winding of loops.}

Then, the state sum formula is
\begin{equation} \label{eq:state-sum-formula}
	\sum_\ell \alpha(\ell) = q^{N w(L)} P_\HOMFLY(L, a = q^N, z = q - q^{-1}),
\end{equation}
where $w(L)$ is the writhe of the knot diagram (ie the number of overcrossings
minus the number of undercrossings.)

Let us illustrate \eqref{eq:state-sum-formula} with a few examples:

\begin{itemize}
	\item 
The most trivial example is the unknot, placed in $\R^3$ so that its
projection to $\R^2$ is a circle.
The HOMFLY polynomial for this knot is
\begin{equation}
	P_\HOMFLY(L, a, z) = z^{-1} (a - a^{-1}).
\end{equation}
Since the diagram has only one arc, the state sum model just sums over the $N$ possible labels
for that arc. Since there are no crossings, the weight $\alpha(\ell)$ reduces to the winding factor, giving
\begin{equation}
	\sum_\ell \alpha(\ell) = \sum_{i=1}^N q^{N+1-2i} = \frac{q^N - q^{-N}}{q - q^{-1}} = P_\HOMFLY(L, a=q^N, z = q-q^{-1})
\end{equation}
as desired.

\item
As a more interesting example, suppose we take $L$ to be the left-handed trefoil, placed in $\R^3$
so that its projection to $\R^2$ is the diagram in \autoref{fig:trefoil}.

\insfigsvg{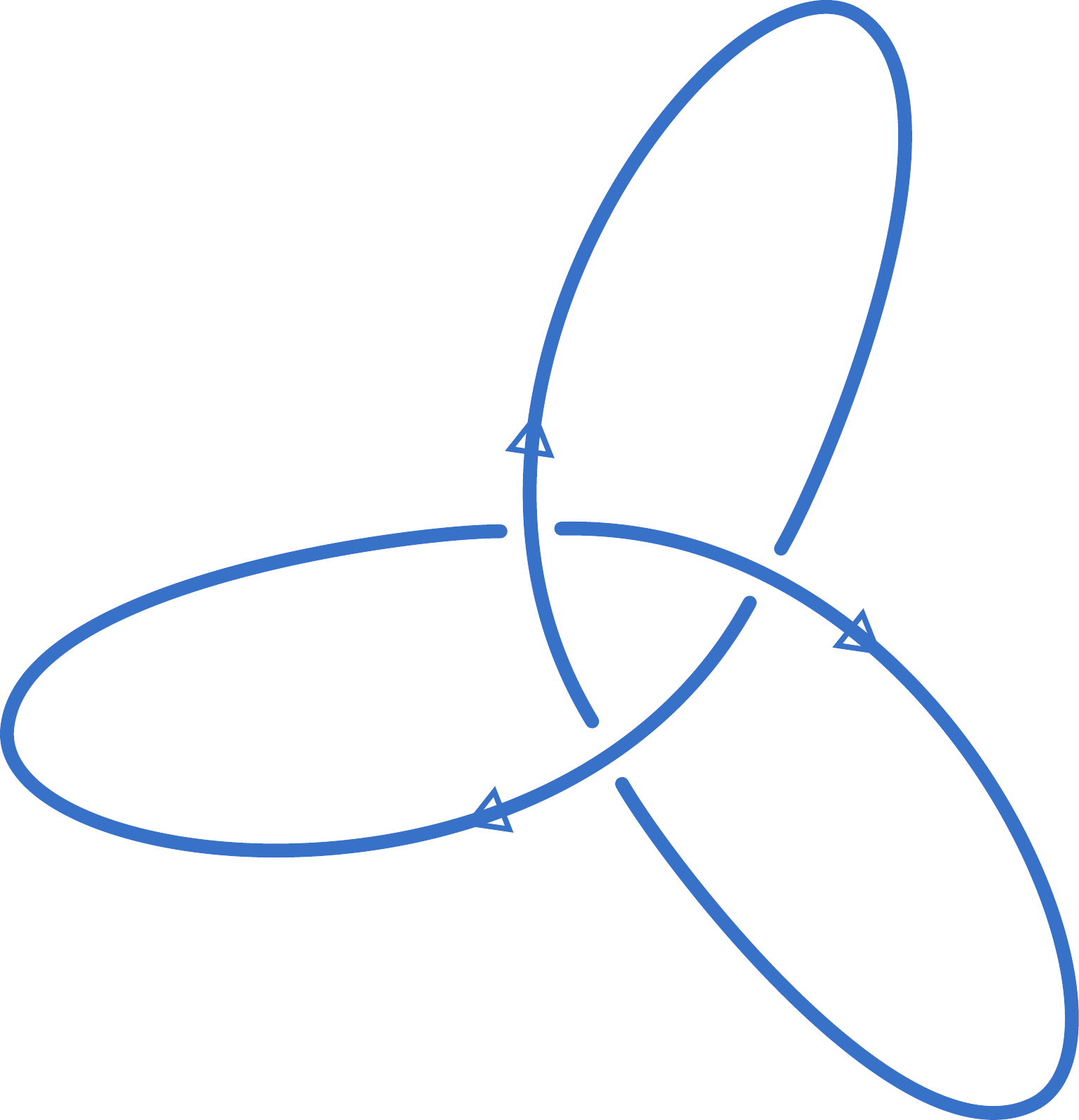}{0.18}{A diagram for the left-handed trefoil.}

The HOMFLY polynomial for this knot is
\begin{equation}
	P_\HOMFLY(L, a, z) = z^{-1} (a - a^{-1}) (  -a^4 + a^2 z^2 + 2 a^2 ).
\end{equation}

Let us see how the state sum model reproduces this formula.
The knot diagram in \autoref{fig:trefoil} has $6$ arcs; thus the state sum model
involves a sum over $N^6$ different arc label assignments $\ell$.
The $\ell$ for which $\alpha(\ell) \neq 0$ fall into five classes, as shown below:

\insfigsvg{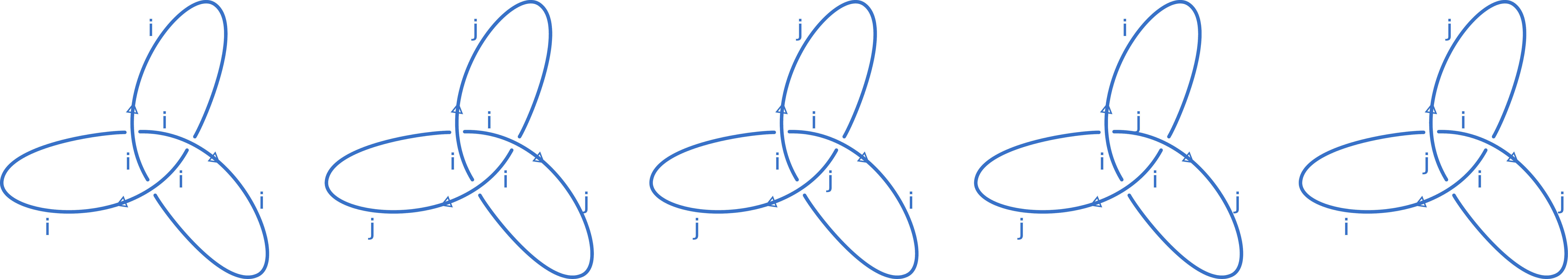}{0.16}{Arc labelings $\ell$ for which $\alpha(\ell) \neq 0$.
The labels $i,j \in \{1,\dots,N\}$ and $j > i$.}

Applying the state sum model rules, we obtain
\begin{align*}
\sum_\ell \alpha(\ell) &= q^{-3}\sum\limits_{i=1}^Nq^{-2N-2+4i}+\Big((q^{-1}-q)^3+3(q^{-1}-q)\Big)\sum\limits_{1\leq i<j\leq N}q^{-2N-2+2i+2j}\\
&=q^{-3N}\frac{q^N-q^{-N}}{q-q^{-1}}\Big(-q^{4N}+q^{-2+2N}+q^{2+2N}\Big)\\
&=q^{-3N}P_\HOMFLY(L, a=q^N, z = q-q^{-1})
\end{align*}
This matches \eqref{eq:state-sum-formula} as desired, since the writhe of the diagram in \autoref{fig:trefoil} is $w = -3$.
\end{itemize}

\subsection{Reinterpreting the state sum model}

With an eye toward generalization, we now slightly reinterpret the state sum rules.

We think of the index $i$ as labeling the $i$-th sheet of a trivial $N$-fold covering
\begin{equation} \label{eq:trivial-covering}
	\tM \to M, \qquad M = \R^3, \qquad \tM = \sqcup_{i=1}^N \R^3.
\end{equation}
Each arc labeling $\ell$ 
with $\alpha(\ell) \neq 0$ 
gets interpreted as representing some link $\tL$ in $\tM$.
If an arc in $L$ is labeled $i$, it means $\tL$ contains 
the lift of that arc to the $i$-th sheet in $\tM$.

The simplest situation arises when all arcs in $L$ carry the same label $i$: 
then $\tL$ is just the lift of $L$ to the $i$-th sheet.
More generally, if the labels of the arcs change at the crossings, simply taking the 
lift of each arc would not give a closed link on $\tM$. At such a crossing involving
labels $i$ and $j$, we 
insert two segments, traveling along the $z$-direction on sheets $i$ and $j$
in opposite directions, to
close up the link, as shown in \autoref{fig:state-sum-formulas-2}. We call
this pair of segments an \ti{exchange}.
The resulting $\tL$ is a disjoint union of closed links on various sheets of $\tM$.

Now we reinterpret the weight factor $\alpha(\ell)$ in terms of the link $\tL$:
\begin{itemize}
	\item The weights $q^{\pm 1}$ assigned to crossings where all arc labels are the same give altogether
	$q^{\sum_{i=1}^N n_i}$, where $n_i$ is the self-linking number of the part of $\tL$ on sheet $i$. This 
	factor can be understood as using the relations in $\Sk(\tM,\fgl(1))$
	to express $\tL$ as a multiple of the empty link: $[\tL] = q^{\sum_{i=1}^N n_i} [ \cdot ]$.
	\item The weights $\pm (q - q^{-1})$ assigned to crossings where the arc labels change 
	are interpreted as universal factors associated to exchanges in $\tL$.
	\item The winding factors $q^{w(N+1-2i)}$ can be interpreted as follows: for 
	a loop of $\tL$ on sheet $i$, we include a factor $q^w$ for each $j > i$, 
	and a factor $q^{-w}$ for each $j < i$.
\end{itemize}

\subsection{\texorpdfstring{$q$}{q}-nonabelianization as a generalization}
The $q$-nonabelianization map can be thought of
as a generalization of the state sum model. Some of the key elements are:
\begin{itemize}
	\item The trivial covering of $\R^3$ given in \eqref{eq:trivial-covering} above is replaced by the (in general nontrivial) branched covering $\tM \to M$,
	with $\tM \subset T^* M$. The 
	labels $i = 1, \dots, N$ which we used above are replaced by local choices of a 
	sheet of this covering over a patch of $M$.
	\item The single projection $\R^3 \to \R^2$ along the $z$-axis 
	is replaced by many different projections, 
	along the leaves of $\binom{N}{2}$ different locally defined foliations of $M$, labeled by
	pairs of sheets $ij$. 
	The directions of the different projections are determined
	by the 1-forms $\lambda_i - \lambda_j$.
	To define the winding factor for
	an arc on sheet $i$, we sum the winding of $N-1$ different projections of the arc 
	to the leaf spaces of the $N-1$ $ij$-foliations. (In the case of the state sum model
	all of the foliations coincide, and so all of the projections also coincide, but the 
	leaf space $\R^2$ gets a different orientation depending on whether $i > j$ or $i < j$;
	this recovers the recipe above.)
	\item Instead of simple exchanges built from segments traveling along the $z$-axis, we  have to consider more general webs built out of segments of leaves of 
	the $\binom{N}{2}$ foliations. Each end of each segment lies either on the link $L$,
	on the branch locus of $\tM \to M$, or at a trivalent junction between three segments. 
	Each such web comes with a weight factor
	generalizing the $\pm (q - q^{-1})$ we had above.
\end{itemize}

Even when $M = \R^3$, we can take $\tM$ to be a general branched cover,
and then $q$-nonabelianization looks quite different from the state sum model,
involving sums over various sorts of webs;
nevertheless it still computes \eqref{eq:homfly-relation-intro} in
the end. We will 
show how this works in a few examples in \autoref{sec:unknotexample}
and \autoref{sec:Jonespolynomials} below.

\section{Skein modules} \label{sec:skeins}

\subsection{The \texorpdfstring{$\fgl(N)$}{gl(N)} skein module} \label{sec:gln-skein}

Fix an oriented 3-manifold $M$.
The $\fgl(N)$ skein module $\Sk(M,\fgl(N))$\footnote{The literature contains a number of 
variants of this
skein module; the one we consider here was also considered in \cite{Ekholm2019} where it is
called the ``$\U(N)$ HOMFLYPT skein,'' except that $q_{\here} = q^{\frac12}_{\there}$.} is the free $\mathbb{Z}[q^{\pm 1}]$-module generated by ambient isotopy classes of framed oriented links in $M$, modulo the submodule generated
by the following relations:

\insfigsvg{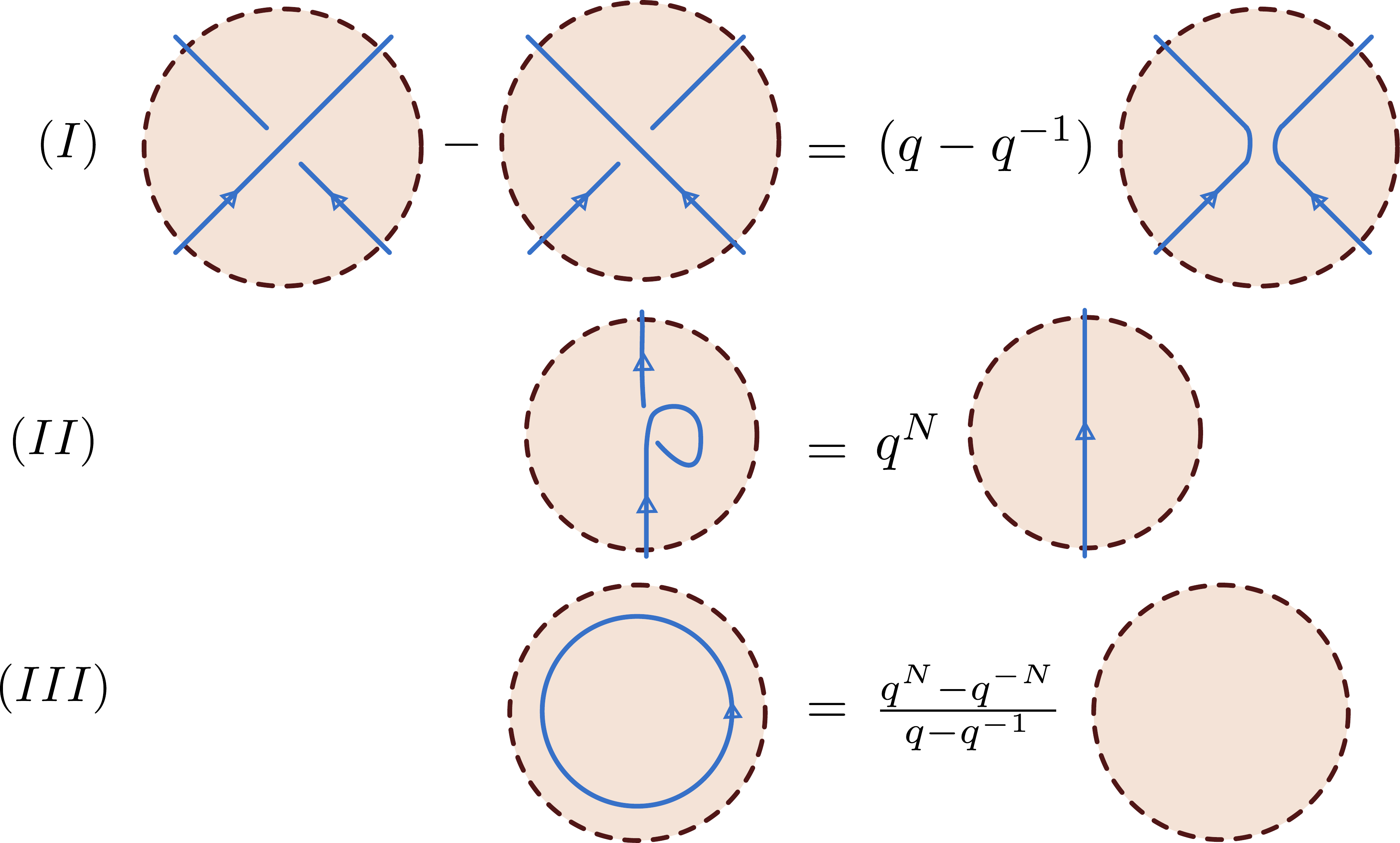}{0.25}{The skein relations defining $\Sk(M,\fgl(N))$.}

\noindent
In each skein relation, all the terms represent links which are the same outside a ball in $M$, 
and are as pictured inside that ball, with blackboard framing.
Relation (II) can be thought of as a ``change of framing'' relation: the two links are 
isotopic as unframed links, but as framed links with blackboard framing, 
they differ by one unit of framing.

\subsection{The \texorpdfstring{$\fgl(1)$}{gl(1)} skein module with branch locus} \label{sec:gl1-skein}

Now consider an oriented 3-manifold $\tM$ decorated by a codimension-2 locus $\cF$. (We sometimes
call $\cF$ the \ti{branch locus} since in our application below, $\tM$ will be a covering
of $M$, branched along $\cF$.)
The $\fgl(1)$ skein module with branch locus, 
$\Sk(\tM,\fgl(1))$, is the free $\mathbb{Z}[q^{\pm 1}]$-module generated by ambient isotopy classes of framed oriented links in $\tM \setminus \cF$, modulo the submodule generated by the following skein relations:

\insfigsvg{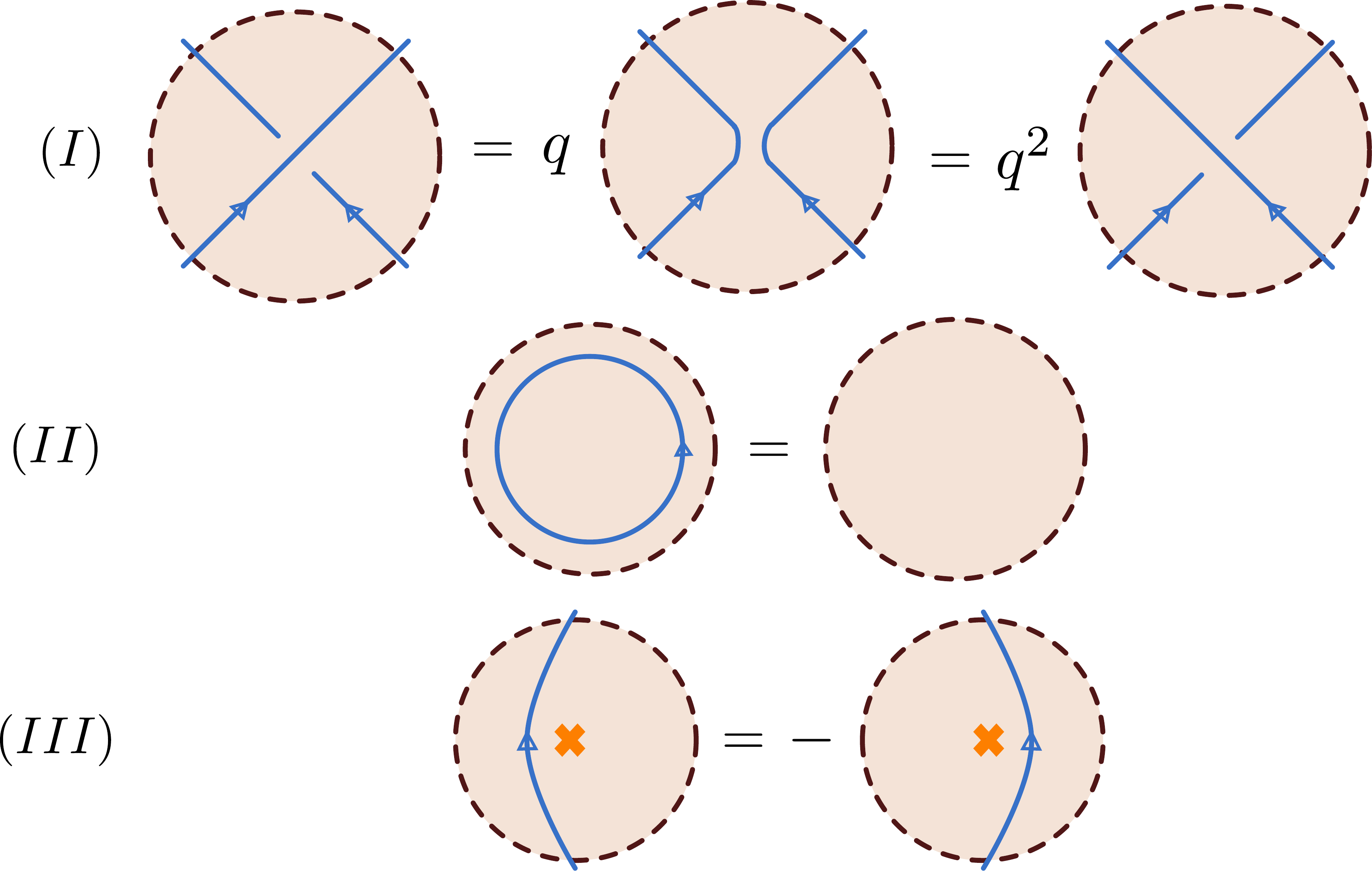}{0.25}{The skein relations defining $\Sk(\tM,\fgl(1))$.}

In the skein relation (III) the orange cross represents the codimension-2 locus $\cF \subset \tM$. This skein relation says that we can isotope a link segment
across $\cF$ at the cost of a factor $-1$.

We remark that relations (I) and (II) imply a simple relation between links 
whose framings differ by one unit, parallel to what we had for $\fgl(N)$ above:
\insfigsvg{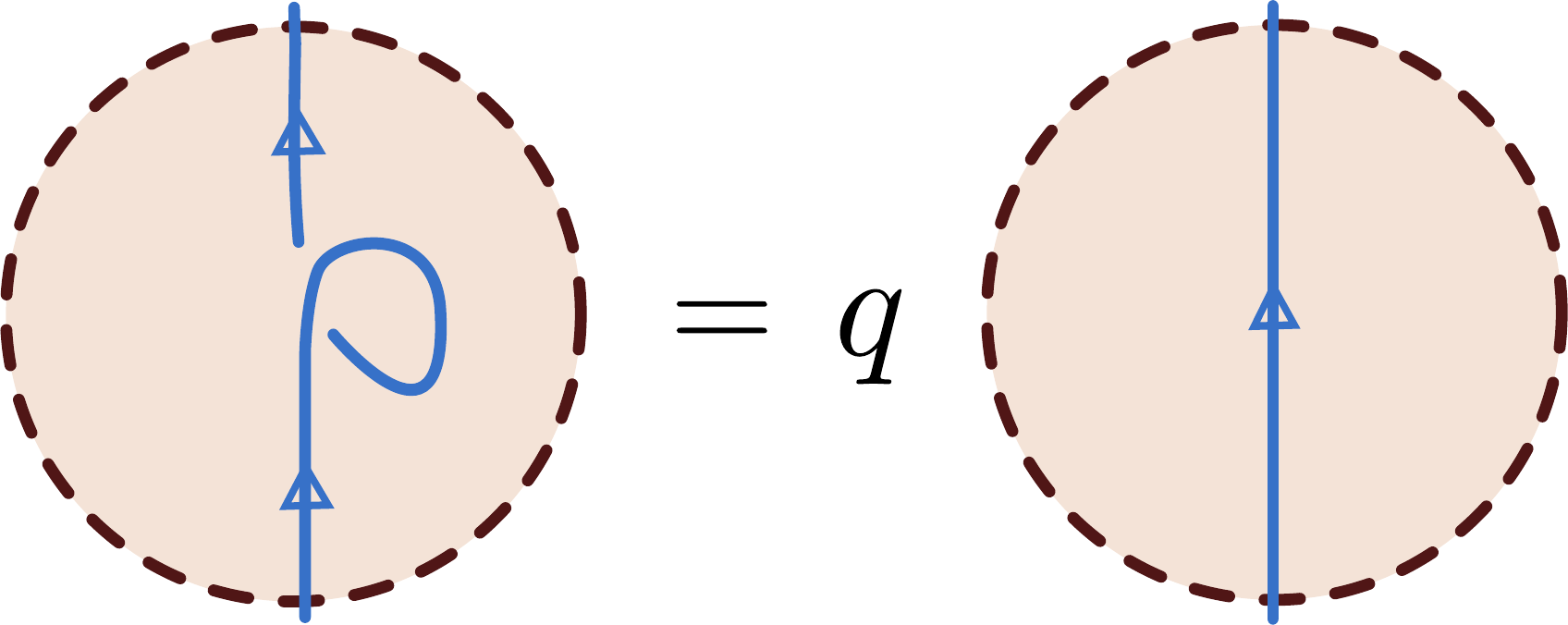}{0.25}{The change-of-framing relation in $\Sk(\tM,\fgl(1))$.}

\subsection{Skein algebras and their twists} \label{sec:skein-algebra}

Now we take $M=C\times\R$ and $\tM=\tC\times\R$,\footnote{For later convenience we sometimes call the $\R$-direction the ``height'' direction.} where $C$ and $\tC$ are oriented surfaces.
We take the orientation of $M$ (resp. $\tM$) to be the one induced from the orientation of $C$
(resp. $\tC$)
and the standard orientation of $\R$.

In this case $\Sk(M,\fgl(N))$ and $\Sk(\tM,\fgl(1))$ 
are algebras
over $\Z[q^{\pm 1}]$, where the multiplication is given by ``stacking'' links along 
the $\R$ direction:
$[L][L'] = [LL']$, where $LL'$ is the link defined by superposing $L'$ with the
translation $L^t$ of $L$ in the positive $x^3$-direction, so that all points of $L^t$ have larger $\R$ coordinate than all points of $L'$.
We emphasize that this algebra structure comes from the $\R$ factor:
for a general 3-manifold $M$, there is no algebra structure on $\Sk(M, \fgl(N))$.

We have to mention a little subtlety: the product 
structure that is most
convenient for our purposes below is a \ti{twisted} version of the 
usual one. Given two links $L$, $L'$ in $C \times \R$, let
$\eps(L,L') = \pm 1$ be the mod 2 intersection number of their projections to
 $C$.
Then we define the twisted product in $\Sk(M,\fgl(N))$ by the rule
\begin{equation}
	[L] [L'] = \eps(L,L') [LL']
\end{equation}
and in $\Sk(\tM,\fgl(1))$ by
\begin{equation} \label{eq:gl1-twist}
	[\tL] [\tL'] = \eps(\pi(\tL), \pi(\tL')) [\tL \tL'].
\end{equation}
In what follows we will only use the twisted products, not the
untwisted ones. (The twisted
and untwisted versions of the skein algebra are actually isomorphic,
but not canonically so; to get an isomorphism between them one needs
to choose a spin structure on $C$. For our purposes it will be more
convenient not to make this choice.)

\subsection{Standard framing}

Another extra convenience in the case $M = C \times \R$ is that
we have a distinguished framing
available: as long as the projection of a link 
$L \subset M$ (resp. $\tL \subset \tM$) to $C$ (resp. $\tC$) 
is an immersion, we can equip the link with a framing vector pointing 
along the positive $x^3$-direction. 
We call this \ti{standard framing} and will use it frequently.

\subsection{The \texorpdfstring{$\fgl(1)$}{gl(1)} skein algebra is a quantum torus} \label{sec:quantumtorus}

When $\tM = \tC \times \R$ we can describe the skein 
algebra $\Sk(\tM,\fgl(1))$ explicitly as follows.
(Descriptions of the $\fgl(1)$ skein algebra similar to what follows
have been used before in connection with $\fgl(1)$ Chern-Simons theory 
and BPS state counting, e.g.
\cite{Cecotti:2010fi,Galakhov:2014xba,Dimofte2016}.)

Given any lattice $\Gamma$ with a skew bilinear pairing $\IP{\cdot,\cdot}$, the 
\ti{quantum torus} $Q_\Gamma$ is a $\Z[q,q^{-1}]$-algebra
with basis $\{X_\gamma\}_{\gamma \in \Gamma}$ and the product law
\begin{equation}\label{eq:quantumtorus}
X_{\gamma_1} X_{\gamma_2} = (-q)^{\IP{\gamma_1,\gamma_2}} X_{\gamma_1+\gamma_2}.
\end{equation}
Now let $\Gamma$ be $H_1(\tC,\Z)$, with $\IP{\cdot,\cdot}$ the 
intersection pairing. Then, there is a canonical isomorphism
\begin{equation} \label{eq:skein-iso-quantumtorus}
	\iota: \Sk(\tM,\fgl(1)) \quad \simeq \quad Q_\Gamma.
\end{equation}
The construction of $\iota$ is as follows.
Suppose given a link $\tL$ on $\tM$ with standard framing.
Let $n(\tL)$ be the writhe (number of overcrossings minus
undercrossings) in
the projection of $\tL$ to $\tC$, and let $n'(\tL)$ be
the number of ``non-local crossings'' in 
the projection of $\tL$ to $\tC$: these are places which are
not crossings on $\tC$, but become crossings after further 
projecting from $\tC$ to $C$.
Then, we let
\begin{equation}
	\iota([\tL]) \ = \ (-1)^{n'(\tL)} q^{n(\tL)} X_{\gamma(\tL)}
\end{equation}
where $\gamma(\tL) \in H_1(\tM,\Z)$ is the homology class of $\tL$.

To see that $\iota$ is really well defined, we must check
that it respects the $\fgl(1)$ skein relations. For this the key point 
is that when we perturb the link $\tL$ across a branch point we shift
$n'(\tL)$ by $1$, compatibly with relation (III) in \autoref{fig:gl1-skein-relations}.

We also need to check that $\iota$ respects the algebra structures.
For this consider two loops $\tL$, $\tL'$. Let $k(\tL,\tL')$ be the signed
number of crossings between $\tL$ and $\tL'$, and $k'(\tL,\tL')$
the number of non-local crossings. Then using the definition of $\iota$
we get directly
\begin{align}
\iota([\tL \tL']) &= (-1)^{n'(\tL)} q^{n(\tL)} (-1)^{n'(\tL')} q^{n(\tL')} (-1)^{k'(\tL,\tL')} q^{k(\tL,\tL')} X_{\gamma(\tL) + \gamma(\tL')} \\
&= (-1)^{n'(\tL)} q^{n(\tL)} (-1)^{n'(\tL')} q^{n(\tL')} (-1)^{k'(\tL,\tL')-k(\tL,\tL')} X_{\gamma(\tL)} X_{\gamma(\tL')} \\
&= (-1)^{k'(\tL,\tL')-k(\tL,\tL')} \iota([\tL]) \iota([\tL']) \\
&= \eps(\pi(\tL),\pi(\tL')) \, \iota([\tL]) \iota([\tL']),
\end{align}
so $\iota$ is indeed a homomorphism (recall the twisted algebra
structure \eqref{eq:gl1-twist}.)

For an example of a quantum torus relation see 
\autoref{fig:quantum-torus-example} below.
The relation in $\Sk(\tM,\fgl(1))$ shown 
there is
\begin{equation}
	[p_1] [p_2] = (-q^{-1}) [p_3],
\end{equation}
or equivalently
\begin{equation}
X_{\gamma_1} X_{\gamma_2} = (-q^{-1}) X_{\gamma_1+\gamma_2},
\end{equation}
matching \eqref{eq:quantumtorus}.

\insfigsvg{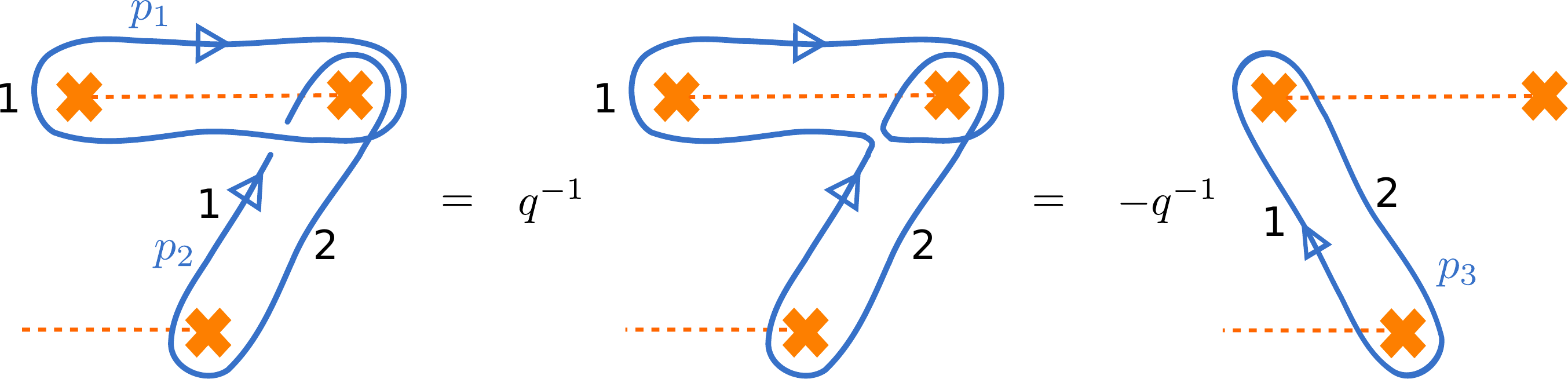}{0.38}{A sample quantum torus relation. We show a patch of
$\tC$, presented as a branched double cover of a patch of 
$C$. The covering is branched at the orange crosses, and branch cuts
are shown as dashed orange segments. Labels $1$ and $2$ next
to loop segments indicate which sheet of $\tC$ they lie on. At a crossing between
segments on the same sheet, the $x^3$-direction is
represented as the ``height'' (loops which are closer to the eye are
ones with a larger $x^3$ coordinate.) At a ``non-local crossing'' where 
two segments on different sheets of $\tC \to C$ cross, 
we do not indicate the relative height.}

\section{\texorpdfstring{$q$}{q}-nonabelianization for \texorpdfstring{$N=2$}{N=2}} \label{sec:qab-N=2}

For the rest of this paper, we will focus on the case $N=2$ and $M = C \times \R$. In this section we spell out the concrete $q$-nonabelianization map in this case.

\subsection{WKB foliations} \label{sec:foliation-data}
 
As we have discussed, we fix a complex structure on $C$ and a holomorphic 
branched double cover $\tC \to C$. 
Locally on $C$ we then have $2$ holomorphic 
$1$-forms $\lambda_i$, corresponding to the two sheets
of $\tC$. 
We define a foliation of $C$
using these one-forms: the leaves are the paths along which
$\lambda_i - \lambda_j$ is real.
The leaves are not naturally oriented, but if we choose one of the two
sheets (say sheet $i$) then we get an orientation: 
the positive direction is the direction in which $\lambda_i - \lambda_j$
is positive; see \autoref{fig:ij-leaf}.
Thus the lift of a leaf to either sheet of $\tC$ is naturally oriented.
\insfigsvg{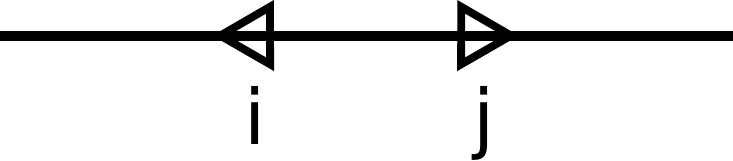}{0.42}{A leaf with its two orientations labeled.}
 
 At branch points of $\tC \to C$ there is a three-pronged singularity 
 as shown in \autoref{fig:foliation-zero}.\footnote{To understand this three-pronged structure
 	note that around a branch point at $z=0$ we have $\lambda^{(i)} - \lambda^{(j)} \sim c z^{\frac12} \, \de z$, so $w^{(ij)} \sim c z^{\frac32}$.}
 The three leaves ending on each branch point are called \ti{critical}.

  \insfigsvg{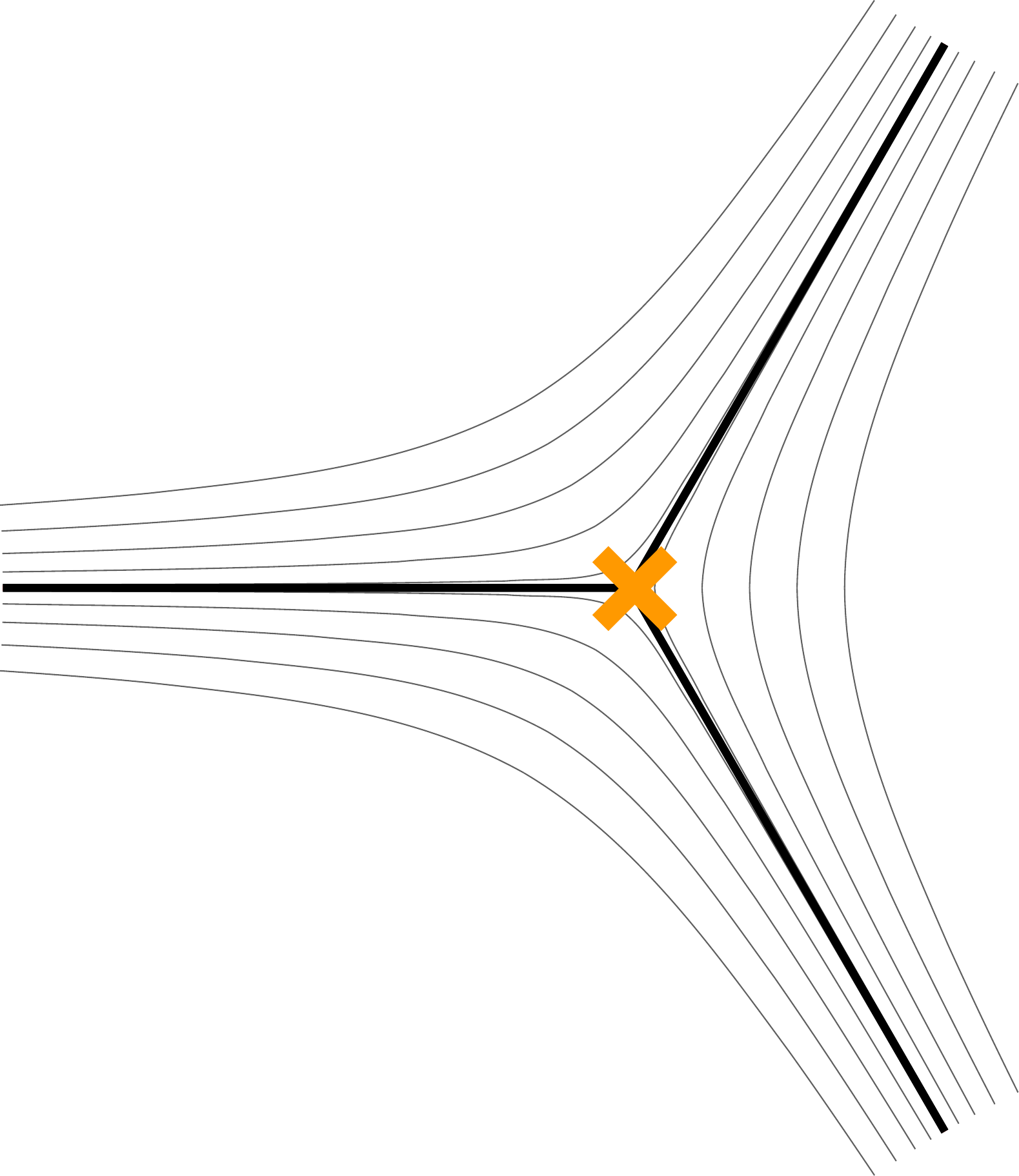}{0.2}{The local structure of the WKB foliation for $N=2$ around a branch point. The branch point is represented by an orange cross. 
 	The dark lines represent critical leaves, while the lighter curves are generic leaves.}

 Dividing $C$ by the equivalence relation that identifies points lying on the same leaf, one obtains the \ti{leaf space} of the foliation.
 This space is a trivalent tree, as 
 indicated in \autoref{fig:leaf-space-C}. It will be convenient later to 
 equip it (arbitrarily) with a Euclidean structure.

 \insfigsvg{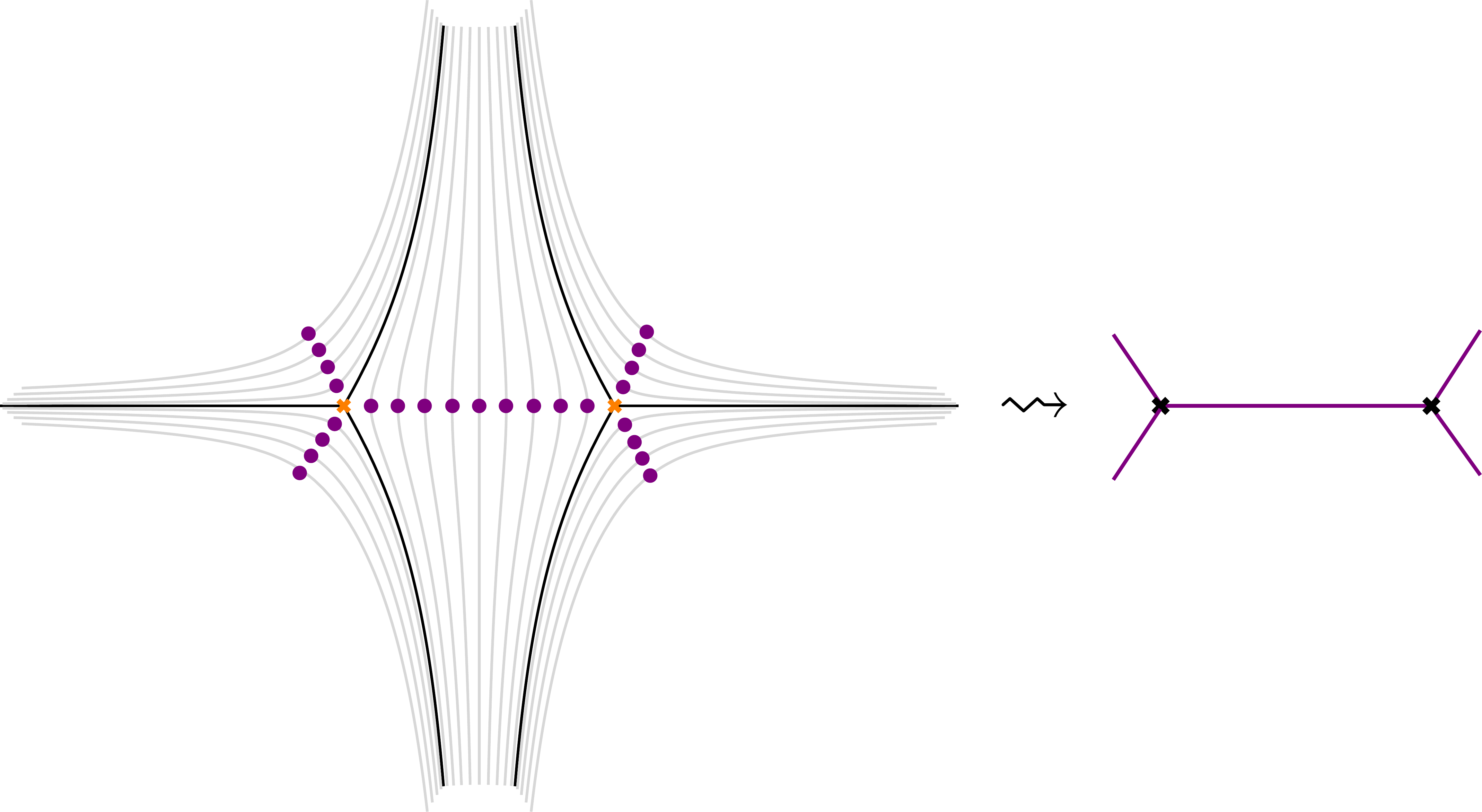}{0.16}{Left: A portion of the foliation 
 	on $C$. One point on each leaf is marked. Right: The corresponding
 	portion of the leaf space, which is a trivalent tree.}
 
 \insfigsvg{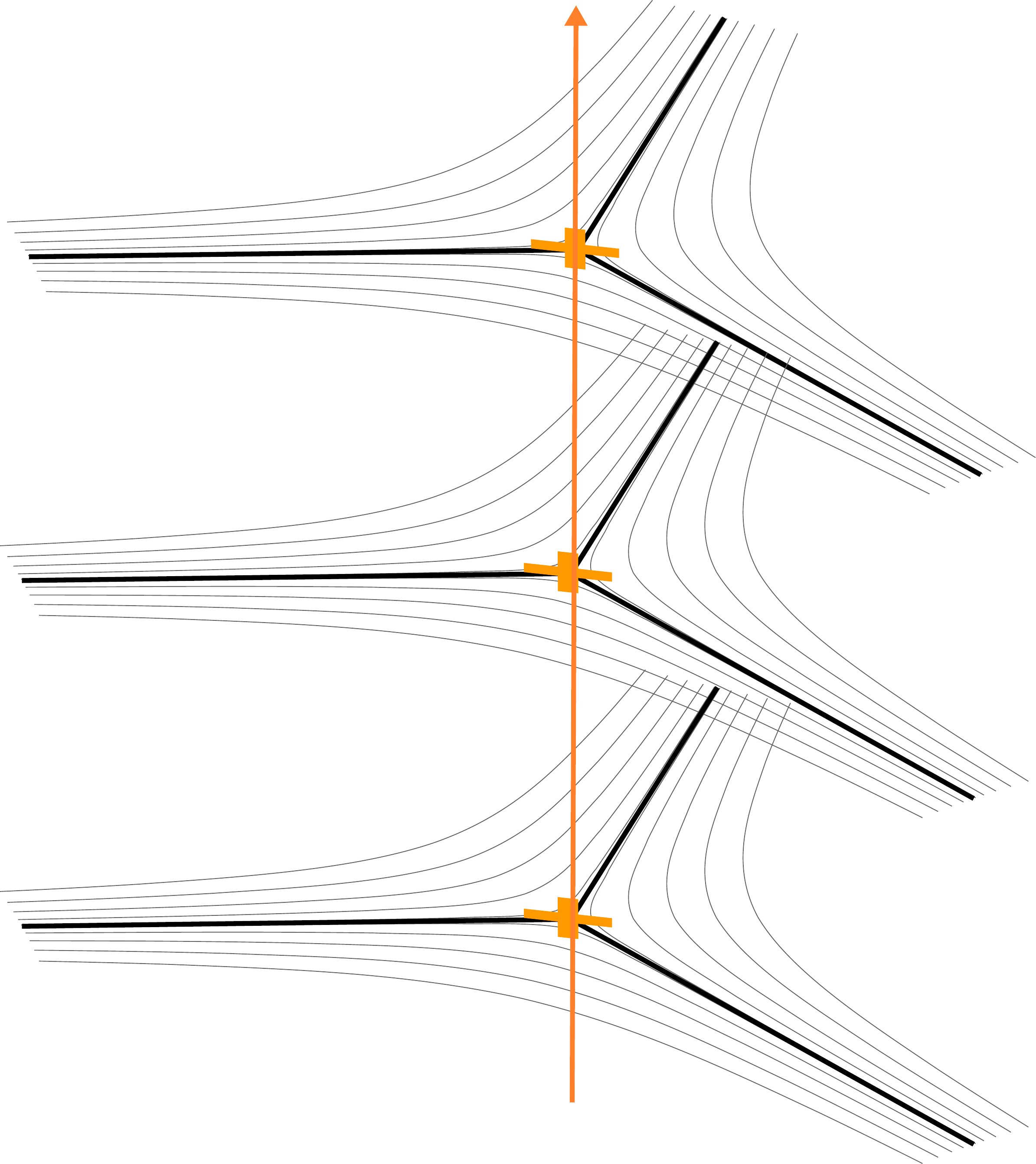}{0.16}{The local structure of the WKB foliation of $M=C\times \R$ for $N=2$ around the branch locus. We explicitly show leaves at three discrete positions in the $x^3$-direction.}
 
 The WKB foliation of $C$ also induces a 
 foliation of $M = C \times \R$: the leaves on $M$
 are of the form $\ell \times \{x^3 = c\}$, where $\ell \subset C$
 is a leaf, and $c$ is any constant; we illustrate this in \autoref{fig:foliation3mfd}.
 Thus the leaf space of $M$ is the product of 
 a trivalent tree with $\R$, i.e. it is a collection of 2-dimensional ``pages''
 glued together at 1-dimensional ``binders.''
 The leaf space inherits a natural Euclidean structure, so locally each page
 looks like a patch of $\R^2$.
 The pages do not carry canonical orientations, but locally choosing a
 sheet $i$ induces an orientation. This induced orientation is determined by the orientation of leaves on sheet $i$ and the ambient orientation of $M$:
 our convention is that
 the induced orientation is the \ti{opposite} of the quotient orientation.
 Note that switching the choice of sheet reverses the leaf space orientation.
 
 \subsection{The \texorpdfstring{$q$}{q}-nonabelianization map for \texorpdfstring{$N=2$}{N=2}} \label{sec:qnab-rules}

Now we are ready to define the $q$-nonabelianization map $F$.

Suppose given a framed oriented link $L$ in $M = C \times \R$, 
with standard framing.
Then $F([L])$ is given by
\begin{equation}\label{q-ab-eqn}
	F([L]) = \sum_{\tL} \alpha(\tL) [\tL],
\end{equation}
where $\tL$ runs over all links in $\tM$ built out of the following local constituents:
\begin{itemize}
	\item \ti{Direct lifts} of segments of $L$ to $\tM$: these are just the preimages of those strands under the covering map. 
	\insfigsvg{direct-lifts}{0.27}{The direct lift of a segment of $L$ to sheet $i$ of the covering $\tM$.}
	
	\item When a segment of $L$ intersects a critical leaf, $\tL$ may include a \ti{detour} along the critical leaf, as shown in \autoref{fig:detour}.
	\insfigsvg{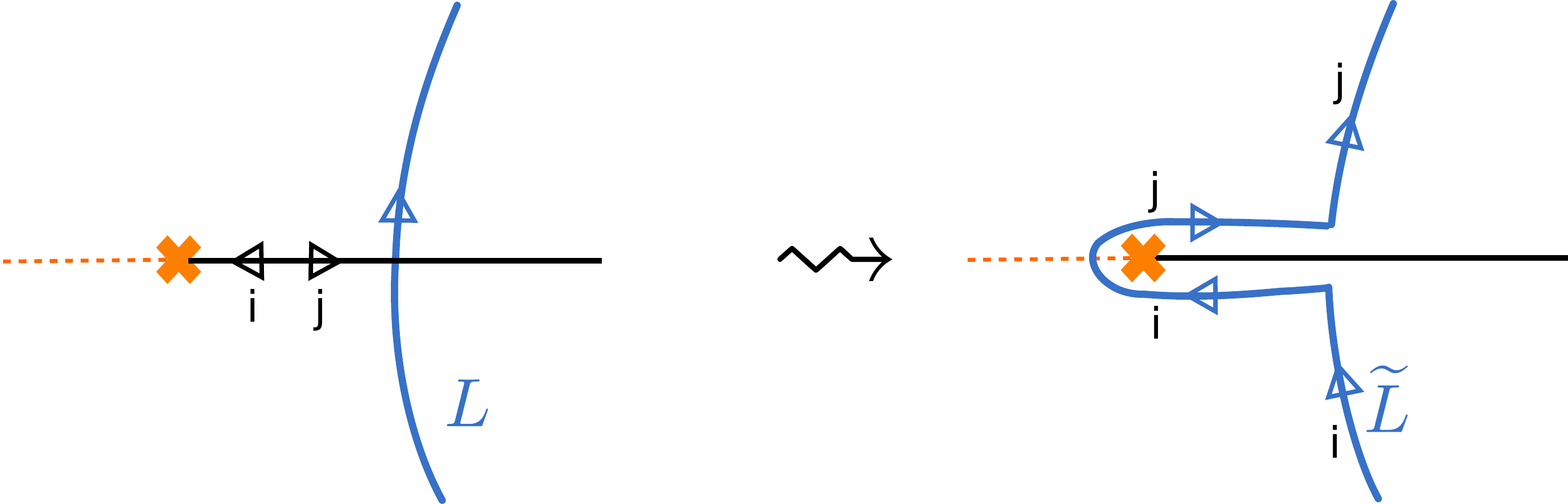}{0.32}{A detour lift of a segment of $L$ which crosses a critical leaf.}
	Note that the orientation of the detour segments in $\tL$ is constrained to match the orientations on the
	critical leaf; so e.g. in the situation of \autoref{fig:detour} we can have a detour
	from sheet $i$ to sheet $j$, but not from sheet $j$ to sheet $i$.

	\item When two segments of $L$ intersect a single $ij$-leaf in $M$, $\tL$ can have an extra 
	\ti{lifted exchange} consisting of two new segments running along the lifts of the leaf to $\tM$, as illustrated in the following figure:
	\insfigsvg{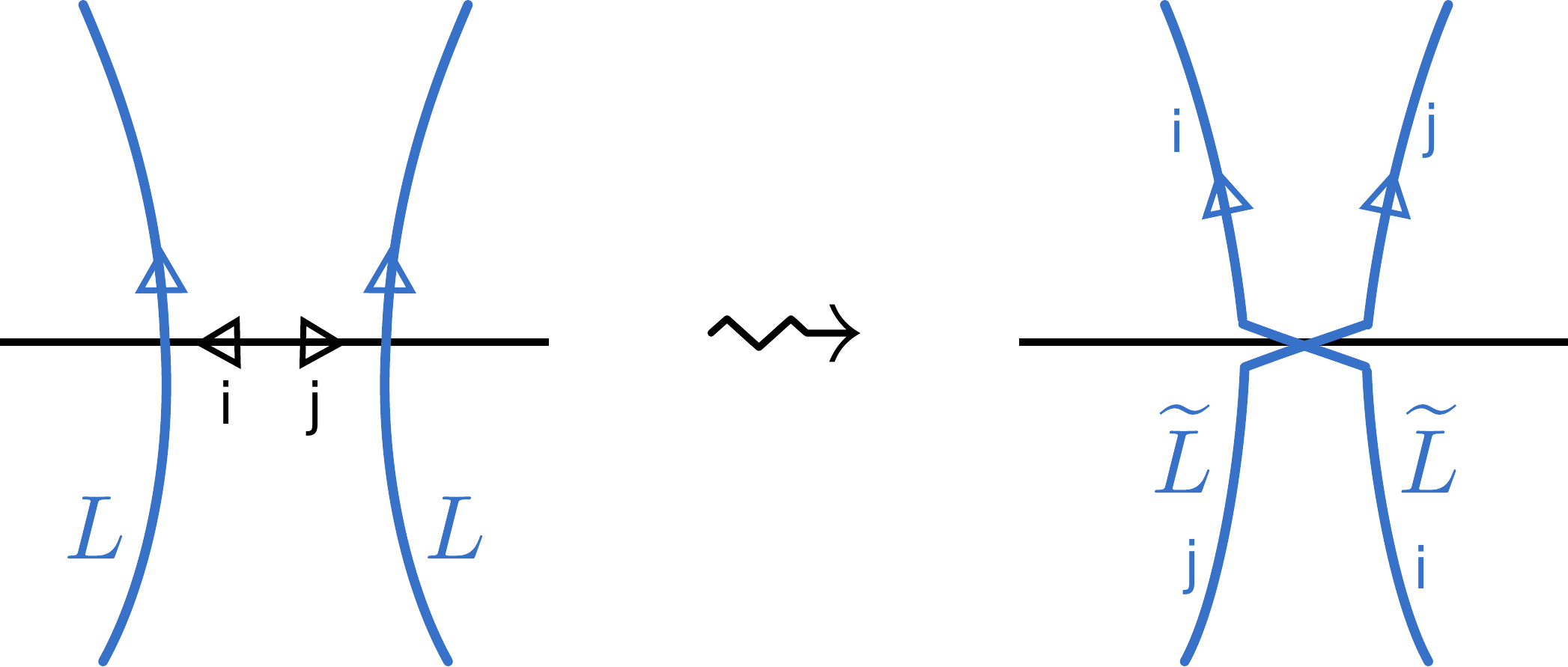}{0.31}{A lift including an exchange connecting two segments of $L$ which cross the same $ij$-leaf.}
	Again, the orientation of the lifted exchange in $\tL$ is constrained to match the
	orientations on the $ij$-leaf.
\end{itemize}

We assume (by making a small perturbation if necessary) that $L$ is sufficiently generic that detours 
and exchanges can only occur at finitely many places.
(In particular, we always make a perturbation such that $L$ is transverse
to the ``fixed height'' slices $C \times \{ x^3 = c \}$, 
since otherwise exchanges could 
occur in 1-parameter families instead of discretely.)
Once this is done, the sum over $\tL$ is a finite sum.

For each $\tL$ the corresponding weight $\alpha(\tL) \in \Z[q^{\pm 1}]$
is built as a product of elementary local factors,
as follows:
\begin{itemize}
	\item 
	At every place where the projection of $L$ onto $C$ is tangent to a leaf, we get a contribution $q^{\pm\frac{1}{2}}$ to
	$\alpha(\tL)$, with the sign determined by the figure below:
	\insfigsvg{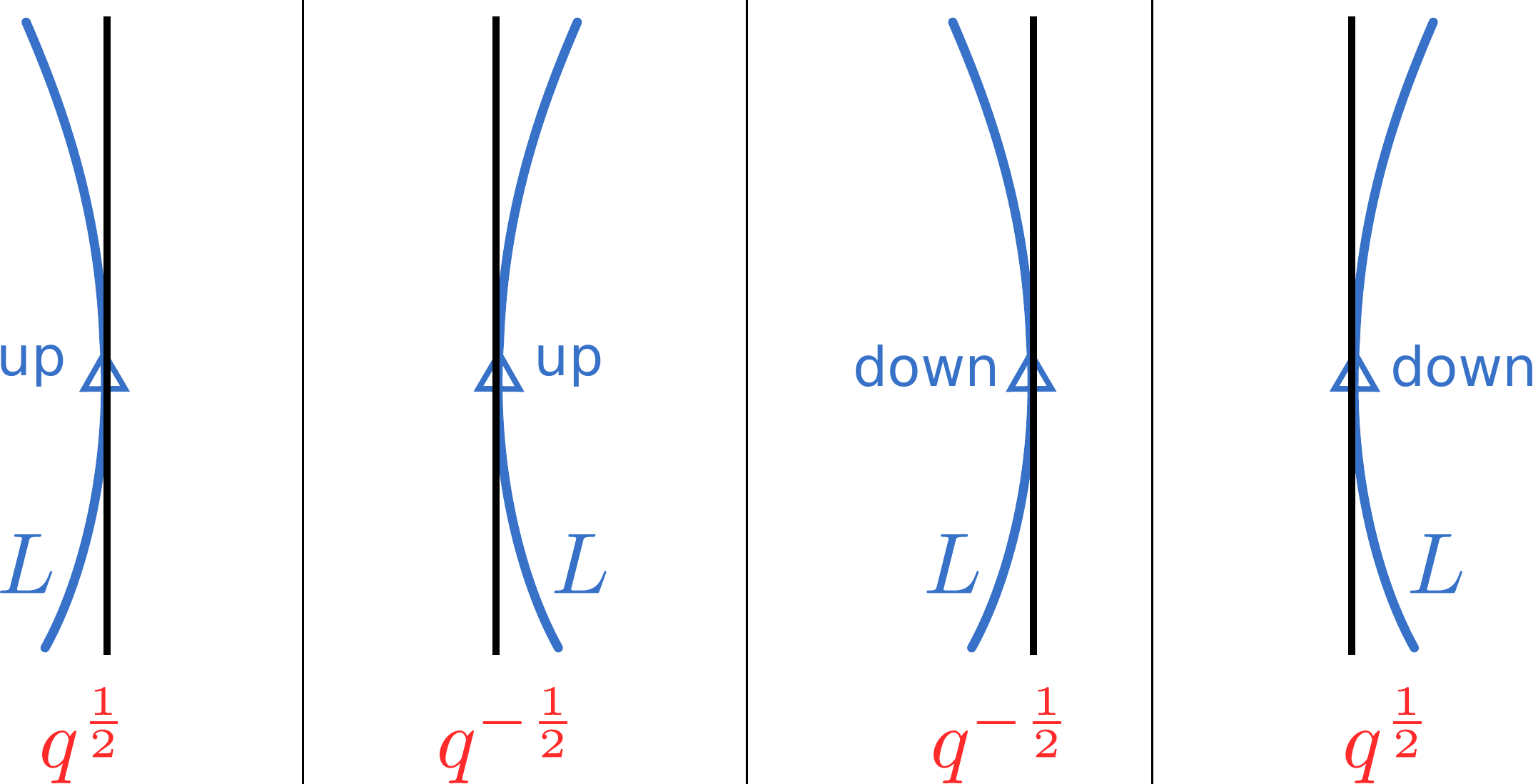}{0.31}{Framing factors contributing to the overall weight $\alpha(\tL)$. The black line denotes a leaf of the WKB foliation. 
	Here the word ``up'' or ``down'' next to a segment of $L$ indicates the behavior
	in the $x^3$-direction, which is not directly visible
	in the figure otherwise, since the figure shows the projection to $C$.}
	(Note that this is an overall factor, depending only on $L$,
	not on $\tL$.)
	
	\item Each detour in $\tL$ contributes a factor of $q^{\pm\frac{1}{2}}$ to $\alpha(\tL)$, with the sign determined by the tangent vector to $L$ at the point where a strand of $L$ meets the critical leaf. The factor is shown in \autoref{fig:detour-factor} below.
	\insfigsvg{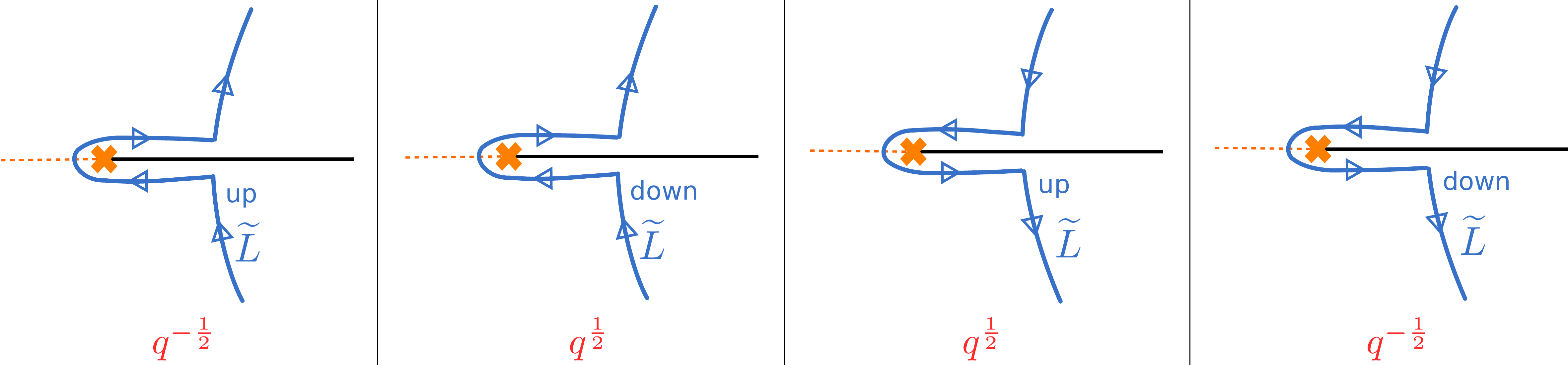}{0.3}{Detour factors contributing to the overall weight $\alpha(\tL)$. Notation is as in \autoref{fig:framing-factor-gl2} above.} 
	
	\item Each exchange in $\tL$ contributes a factor to $\alpha(\tL)$.
	This factor depends on two things: first, it depends whether the two
	legs of $L$ cross the exchange in the same direction or in opposite
	directions when viewed in the standard projection; second, it depends
	whether the crossing in the leaf space projection of $L$ is an overcrossing
	or an undercrossing. See \autoref{fig:exchange-factor} for the
	factors in the four possible cases.
	\insfigsvg{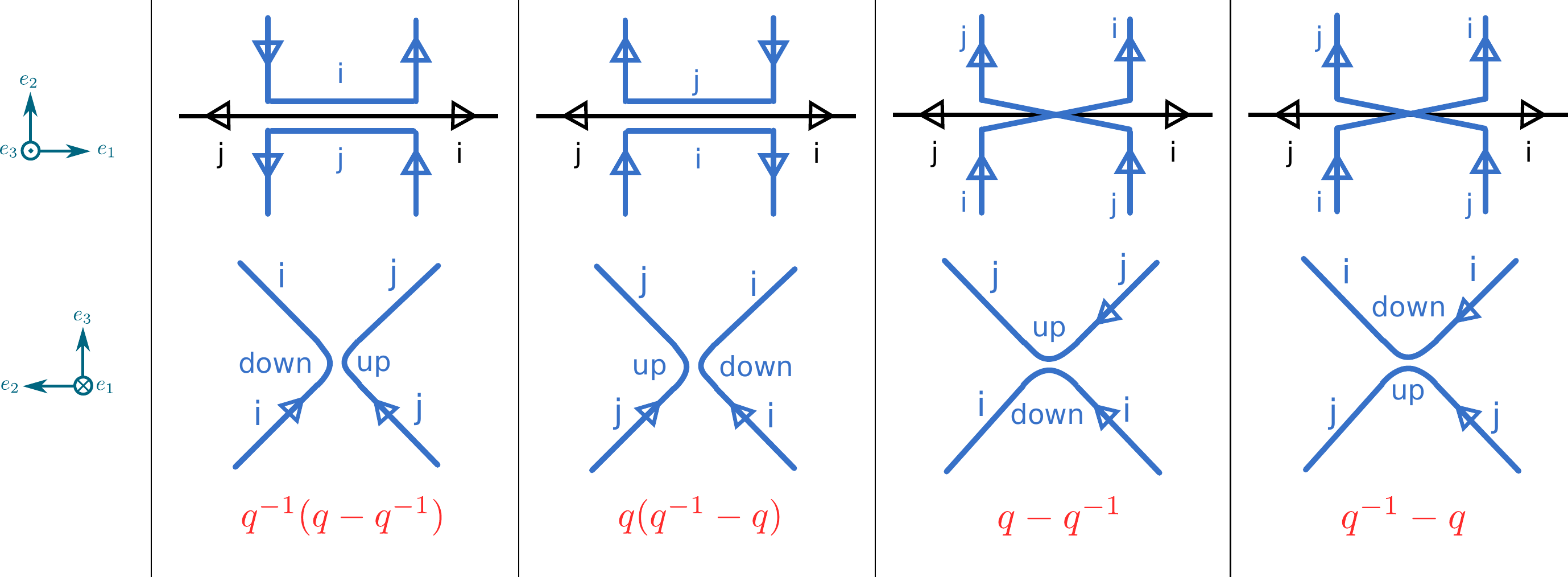}{0.45}{Exchange factors contributing to the overall
	weight $\alpha(\tL)$. The exchange factor depends on more data than we can represent
	in a single projection: we show the standard projection on top and the leaf space
	projection below. The picture represents $\tL$ rather than $L$, 
	so in the leaf space instead of a crossing we see its resolution.
	The words ``up'' and ``down'' here describe the
	behavior in the $x^1$-direction, since that is the direction not
	visible in the leaf space projection; hence ``up'' means pointing out of the paper and ``down'' means pointing into the paper.}
	
	\item Finally there is a contribution from the ``winding'' of $\tL$,
	or more precisely the winding of its projection to the leaf space 
	of the foliation. This winding is defined as follows.
	Recall that the leaf space consists of ``pages''
	each of which has a 2-dimensional Euclidean structure, glued together at
	1-dimensional ``binders.''
	After perturbing $\tL$ so that
	it meets each binder at a right angle, we can define a $\half \Z$-valued 
	winding number for the restriction of $\tL$ to each page. (Recall that 
	$\tL$ is lifted to one of the two sheets of
	$\tM \to M$, say sheet $i$, and this picks out the $i$-orientation on
	the leaf space; we use this orientation to define the winding number.)
	Summing up the winding of $\tL$ on all of the pages, 
	we get a $\Z$-valued total winding $w(\tL)$. We include a factor
	$q^{w(\tL)}$ in the weight $\alpha(\tL)$.

\end{itemize}

For practical computations, 
it is convenient to have a way of computing the
winding factors $q^{w(\tL)}$ without explicitly drawing the leaf space
projections. Here is one scheme that works. We consider all of the places
where the projection of $\tL$ to $C$ is tangent to a leaf
of the foliation.\footnote{We emphasize that we have to consider the full $\tL$,
not only the segments lifted directly from $L$; the winding does receive 
contributions from detours and exchanges.}
For each such place we assign a factor $q^{\pm \frac12}$ as indicated
in \autoref{fig:winding-contribs}.

\insfigsvg{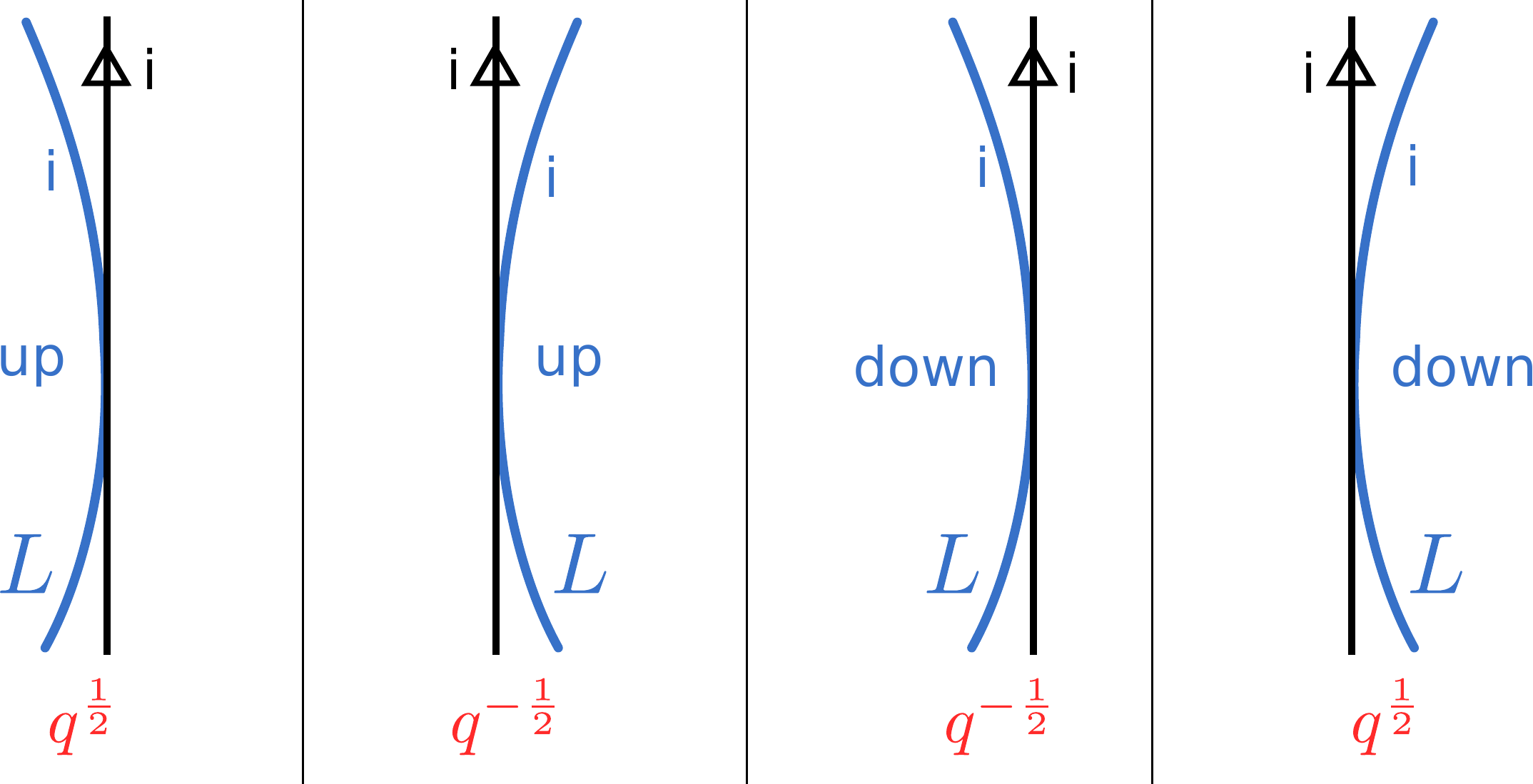}{0.32}{Factors contributing to the winding
factor $q^{w(\tL)}$. The notation is as in \autoref{fig:framing-factor-gl2}, namely ``up" and ``down" refer to the tangency of $L$ in the $x^3$-direction.
The difference from \autoref{fig:framing-factor-gl2} is that here we use the orientation of 
the $ij$-leaf and ignore that of the projection of 
$L$, while in \autoref{fig:framing-factor-gl2}
we used the orientation of the projection of $L$ and ignored that of the $ij$-leaf.}

Although the elementary local factors can 
involve half-integer powers of $q$, 
the total weight $\alpha(\tL)$ is valued in $\Z[q^{\pm 1}]$. 

\subsection{Simple unknot examples}\label{sec:unknotexample}

In this section we illustrate concretely how $q$-nonabelianization works,
in the simplest possible class of examples: we compute
$F([K])$ where $K$ is the unknot in standard framing. In this case
we have $[K] = (q + q^{-1}) [\cdot]$
in $\Sk(M,\fgl(2))$, and so since $F$ factors through
$\Sk(M,\fgl(2))$, the answer must be
\begin{equation} \label{eq:unknot-expected}
F([K]) = q + q^{-1}.
\end{equation}
The details of how this works out depend on what the WKB
foliation of $C$ looks like and how $K$ is positioned relative to that foliation. In this section we describe how it works in some simple cases. 
We give more interesting 
unknot examples in \autoref{sec:unknotscritleaves} below.

First let us consider the case $C = \C$, with a trivial double covering
$\tC \to C$, for which the WKB foliation is just given by straight lines
in the $x^1$-direction. We 
place the unknot $K$ such that its projections to the $x^1$-$x^2$ plane and the $x^2$-$x^3$ plane are as shown in \autoref{fig:unknot-4-0}.
\insfigsvg{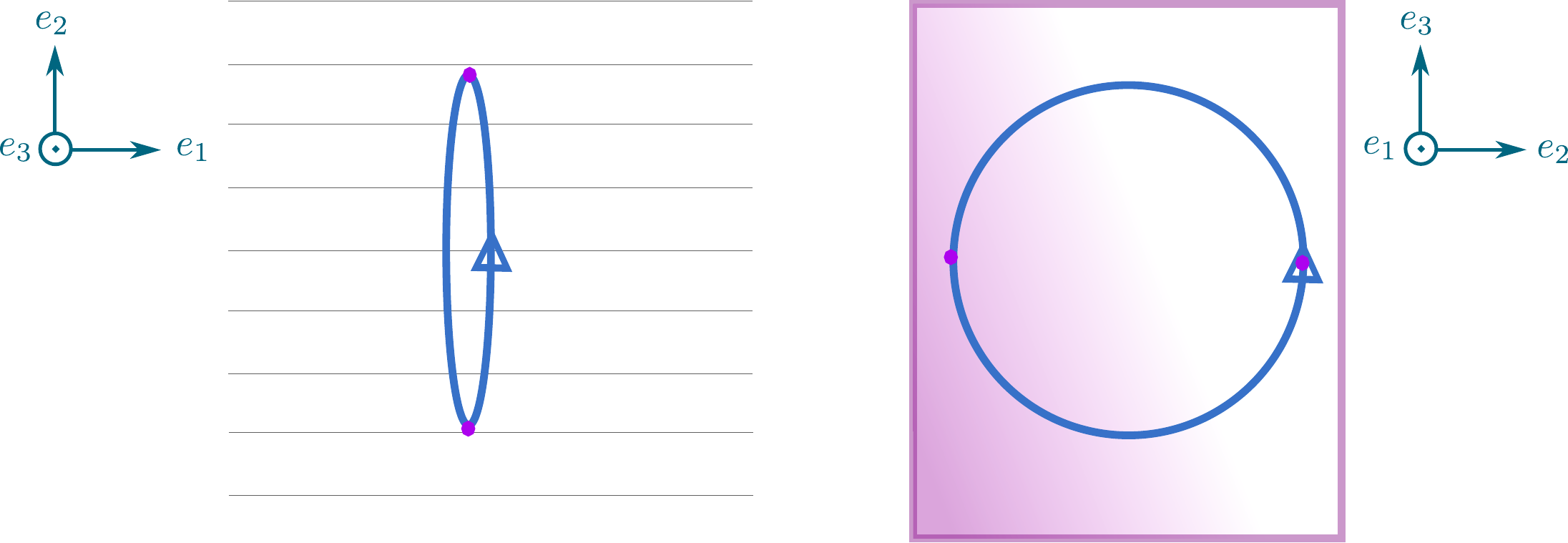}{0.55}{An unknot $K$.
Left: the projection of $K$ to $C$.
The projections of leaves of the WKB foliation are shown in light gray.
Right: the projection of $K$ to the leaf space of the foliation. We use purple dots to mark the two points on $K$ where the projection of $K$ to $C$ is tangent to a leaf. We will omit this information in the following figures.}

This is the simplest situation possible: there are no possible detours since
the covering $\tM \to M$ is unbranched, 
and there are no exchanges since each leaf meets $K$ at most once.
Thus the only contributions to $F([K])$ come from
the direct lifts $\tK_1$, $\tK_2$ to the two sheets. 
Their weights are given simply by the
winding factors, which are $q$ and $q^{-1}$ respectively, all other 
contributions being trivial. Finally, each lift $[\tK_i]$ has 
self-linking number zero and thus is equivalent
to the class $[\cdot]$ in $\Sk(\tM,\fgl(1))$.
We summarize the situation in the table:

\begin{table}[h]
	\centering
	\begin{tabular}{|c|c|c|c|c|c|c|} \hline
		lift    & framing & exchange              & detour    & winding & [lift]           & total \\
		\hline \hline
		$\tK_1$    & $1$   & $1$                & $1$         & $q$     & $1$       & $q$ \\
		$\tK_2$    & $1$   & $1$                & $1$         & $q^{-1}$     & $1$    & $q^{-1}$ \\
		\hline
	\end{tabular}
\end{table}
\noindent
Thus we indeed get the expected answer \eqref{eq:unknot-expected}.
(In fact, the need to get this answer was our original motivation for including
the winding factor in the $q$-nonabelianization map; see also \cite{Gaiotto:2011nm}
which includes a similar factor for a similar reason.)

Next we consider a slightly more interesting case: again we take 
$C = \C$ with a foliation by straight lines in the $x^1$-direction,
but now take $K$ as shown in \autoref{fig:unknot-0-0}.

\insfigsvg{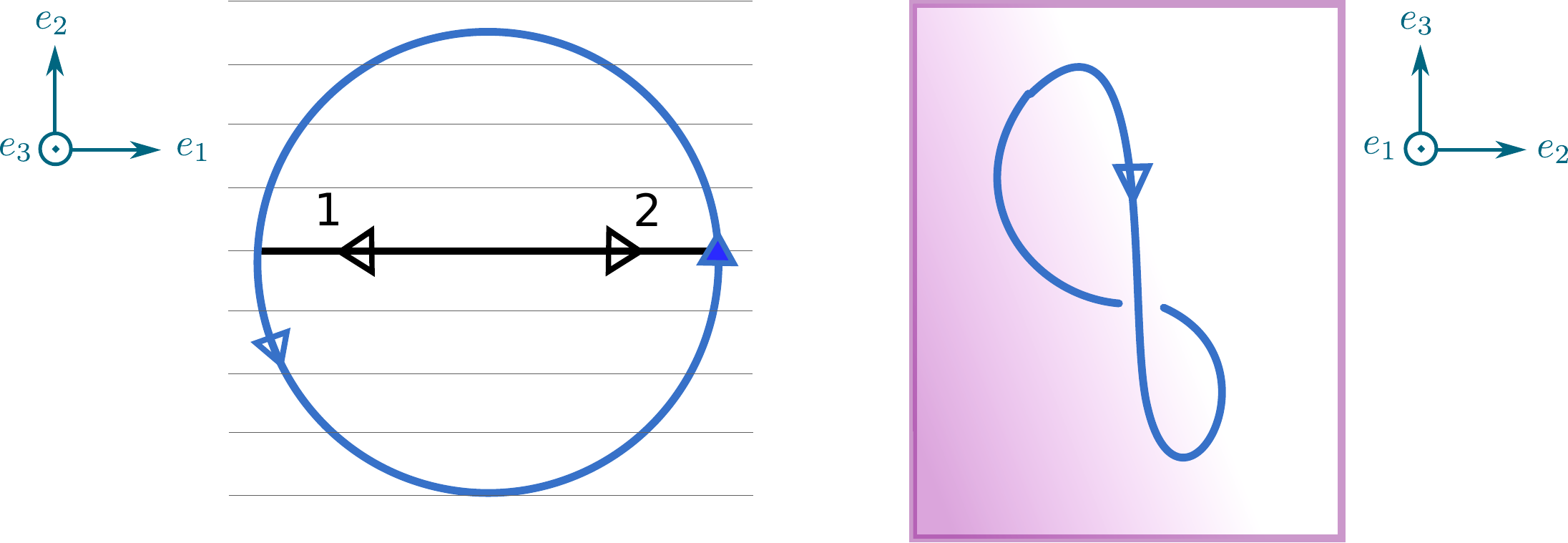}{0.45}{An unknot $K$.
Left: the projection of $K$ to $C$. The height ($x^3$) coordinate of $K$ is taken as follows: as we travel 
counterclockwise around $K$ the height 
monotonically increases, except for a small neighborhood of the filled
arrow, where the height decreases.
The projections of generic leaves of the WKB foliation are shown in light gray.
The projection of the leaf segment along which an exchange may occur is 
shown in black. Right: the projection of $K$ to the leaf space of the foliation.
The position of the crossing in this projection corresponds to the location of 
the potential exchange.}

In this case our path-lifting rules lead to three possible lifts:
\begin{itemize}
\item	We could lift the whole link $K$ to sheet $1$ or sheet $2$; this gives
	two lifts $\tK_1$ and $\tK_2$. Either of these lifts is contractible on $\tM$ and has
	blackboard framing, so $[\tK_1] = [\tK_2] = [\cdot]$ in $\Sk(\tM, \fgl(1))$.
	Each of these lifts has framing factor $q$, from the two places where the projection of $K$
	is tangent to the foliation of $C$. Each of these lifts has total winding zero, as we see 
	from the leaf space projection on the right side of \autoref{fig:unknot-0-0}; thus
	the winding factor is trivial. (Another convenient way to count the winding is to use the rules of \autoref{fig:winding-contribs}. The two tangencies contribute $q^{\frac{1}{2}}$ and $q^{-\frac{1}{2}}$ respectively, giving the total winding factor $1$.)
		Thus these lifts have $\alpha(\tK_1) = \alpha(\tK_2) = q$, so they
		each contribute $q [\cdot]$ to $F([K])$.

\item There is also a more interesting possibility shown in \autoref{fig:unknot-0-1}.
Again $[\tK_3] = [\cdot]$ in $\Sk(\tM, \fgl(1))$ and the total framing factor is $q$.
This time, however, the total winding is $-2$ instead of zero; this arises because
$\tK_3$ is divided into one loop on sheet $1$ and one on sheet $2$, and the
$1$-orientation and $2$-orientation of the leaf space are opposite, so the
windings from these two parts add instead of cancelling. Thus we get a winding
factor $q^{-2}$.
There is also an exchange
factor of $q(q^{-1} - q)$, as we read off from \autoref{fig:exchange-factor}; here
we use the fact that the leaf space crossing is an undercrossing. 
Combining all these factors, this lift contributes
$(q^{-1} - q) [\cdot]$ to $F([K])$.

\insfigsvg{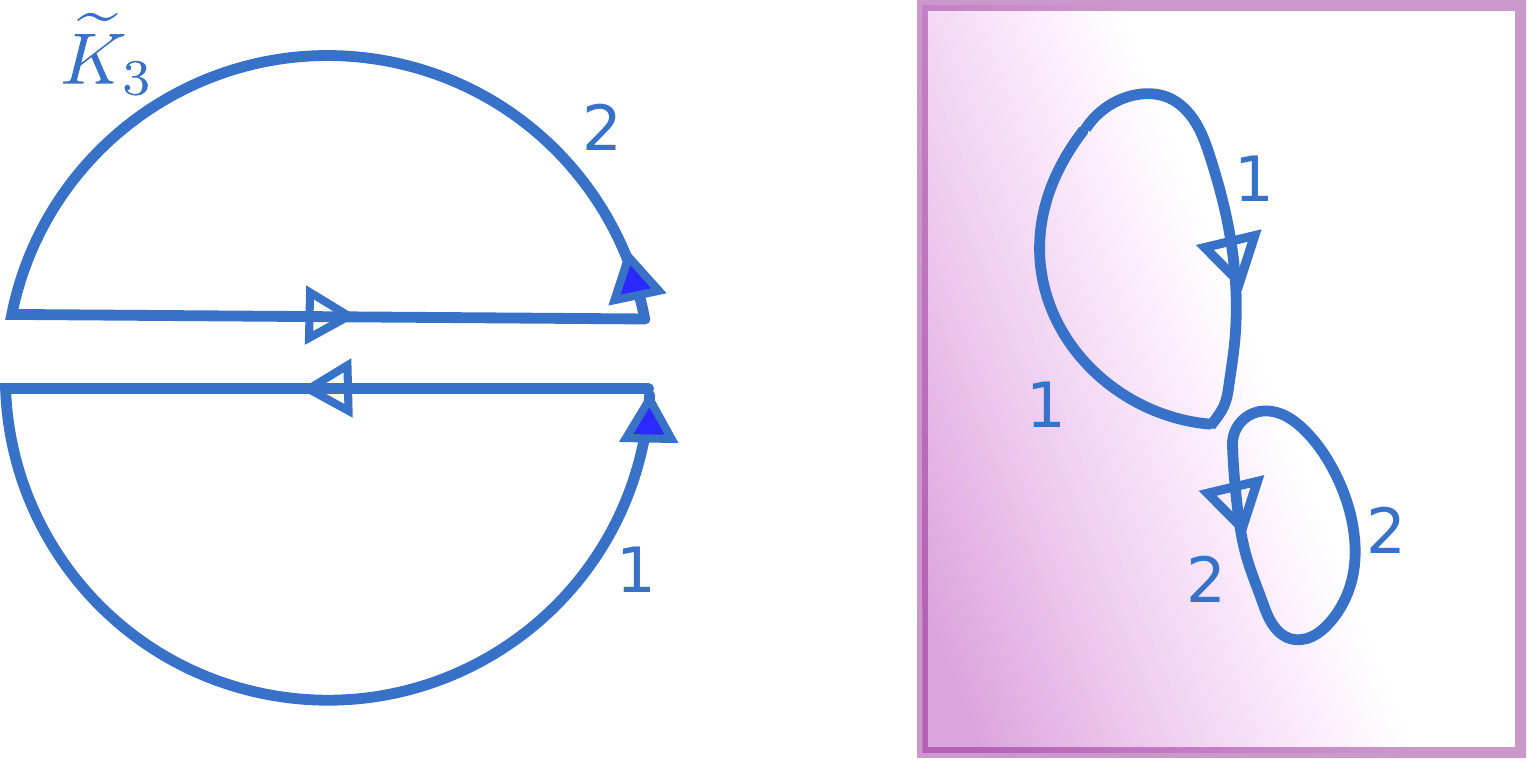}{0.4}{A lift $\tK_3$ of $K$ involving an exchange.  Left: standard projection. Right: leaf space projection. Referring to the standard projection, the top
half of $K$ is lifted to sheet $2$ of $\tM$ while the bottom half is lifted to sheet $1$. The $1$-orientation
of the leaf space is the standard orientation of the plane, 
while the $2$-orientation is the opposite.}

\end{itemize}

\begin{table}[h]
\centering
\begin{tabular}{|c|c|c|c|c|c|c|} \hline
lift    & framing & exchange              & detour    & winding & [lift]       & total \\
\hline \hline
$\tK_1$    & $q$   & $1$                & $1$         & $1$     & $1$          & $q$ \\
$\tK_2$    & $q$   & $1$                & $1$         & $1$     & $1$          & $q$ \\
$\tK_3$    & $q$   & $q(q^{-1}-q)$      & $1$         & $q^{-2}$& $1$          & $q^{-1}-q$ \\
\hline
\end{tabular}
\end{table}

Combining these three lifts we get
\begin{equation}
	F([K]) = q + q  + (q^{-1} - q)  = q^{-1} + q
\end{equation}
as expected.

As we remarked earlier, in this case our computation is similar to 
the state sum model reviewed in \autoref{sec:state-sum}, applied to the ``figure-eight
unknot'' diagram we obtained by projecting to the leaf space 
(\autoref{fig:unknot-0-0}, right.)
There is a slight difference: 
the state sum model computes 
with the blackboard framing in leaf space, which
in this case differs by one unit from our standard framing.
Thus the state sum model gives $q^{-2}(q^{-1} + q)$
instead of our result $q^{-1} + q$.
Looking into the details of the computation one sees that the relative factor $q^2$ 
comes from two different places:
our computation includes an extra $q$ in the factors associated to the
crossing, and also includes the framing factor $q$ which has no direct analogue
in the state sum model.

\section{Examples}
\subsection{Knots in \texorpdfstring{$\R^3$}{3R}} \label{sec:Jonespolynomials}

In \autoref{sec:unknotexample} we have shown how our $q$-nonabelianization map $F$ correctly produces the Jones polynomial for the 
simplest unknots in $\R^3$. 
In this section we show how it works in a few more intricate examples,
with more interesting knots placed in more interesting positions relative to the WKB foliations.
In all cases we have to get the Jones polynomial: this follows from the fact that $F$ is a well
defined map of skein modules, which we prove in \autoref{sec:isotopyproof} below.
Nevertheless it is interesting and reassuring to see how it works out explicitly
in some concrete examples.

\subsubsection{Unknots}\label{sec:unknotscritleaves}

We first look at an unknot $K$ whose projection onto $C$ is a small loop around a branch point
of the covering $\tC \to C$, as
shown in \autoref{fig:unknot-1-0}. 
This case is more interesting since we will meet detours as well as exchanges. 

\insfigsvg{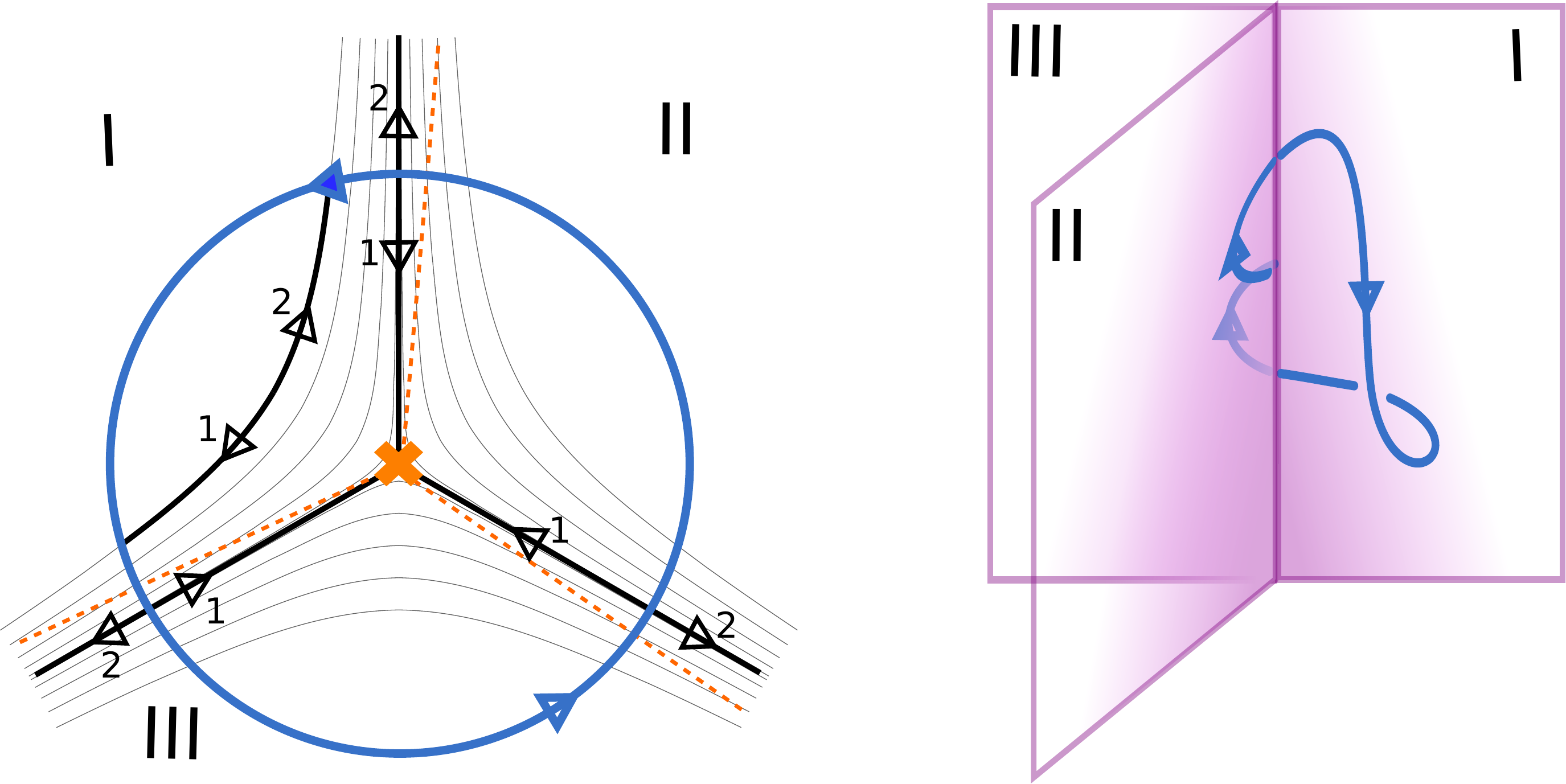}{0.25}{An unknot $K$ encircling one strand of the branch locus of the covering $\tM \to M$.
	Left: the projection of $K$ to $C$. The height coordinate of $K$ is taken as follows: as we travel 
	counterclockwise around $K$ the height monotonically increases, except for a small neighborhood of the filled
	arrow on $K$, where the height decreases.
	The projections of generic leaves of the WKB foliation are shown in light gray.
	The projections of special leaves, along which a detour or exchange may occur, are shown in black. Right: the projection of $K$ to the leaf space of the foliation. The vertical direction in the leaf space corresponds to the height direction.}

In this case a direct lift of $K$ is not allowed, since such a lift would not give 
a closed path on $\tM$: we need to include an odd number of detours 
to get back to the initial sheet. Indeed, according to 
our rules there are five lifts $\tK_n$ contributing to $F([K])$:

\begin{itemize}
	\item There are three lifts which involve a single detour each, 
	shown in \autoref{fig:unknot-1-a} below:
	\insfigsvg{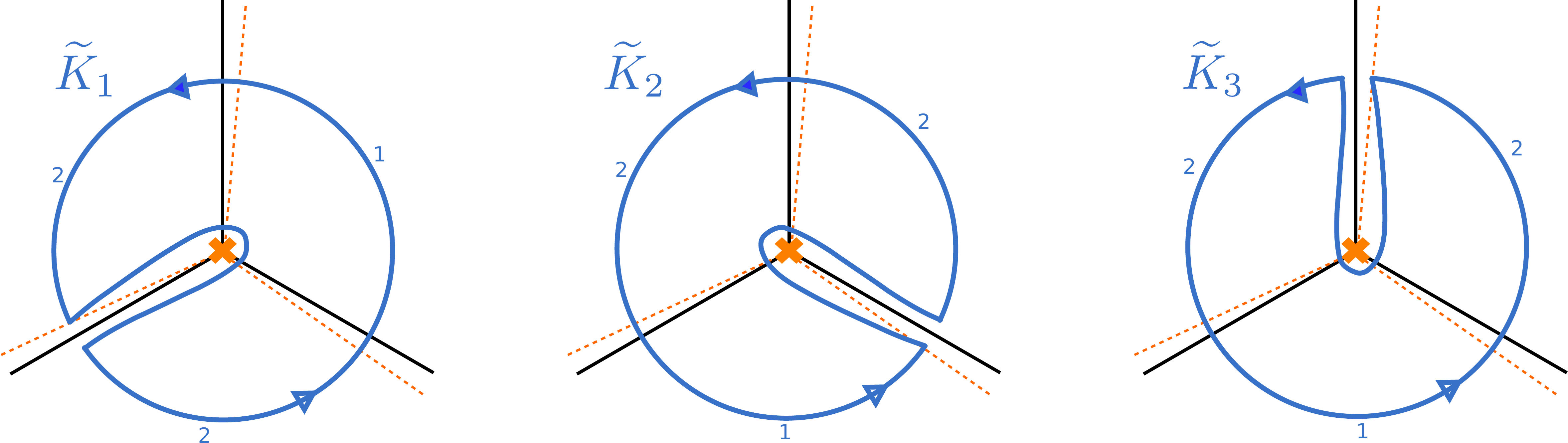}{0.21}{Three lifts $\tK_n$ of $K$.}
	\noindent Each of these lifts is contractible in $\tM$, 
	and equipped with standard framing, so in $\Sk(\tM,\fgl(1))$ we have
	\begin{equation}
	[\tK_1]=[\tK_2]=[\tK_3]=[\cdot]=1.
	\end{equation}
	Moreover, each of these lifts has a total framing factor $q^{3/2}$ and detour factor
	$q^{-1/2}$. Finally, each of these lifts has total winding $w(\tK_n) = 0$. 
	Thus each of these lifts contributes $q$ to $F([K])$.
	
	\item There is one lift $\tK_4$ involving both a detour and an exchange,
	shown in \autoref{fig:unknot-1-b}:
	\insfigsvg{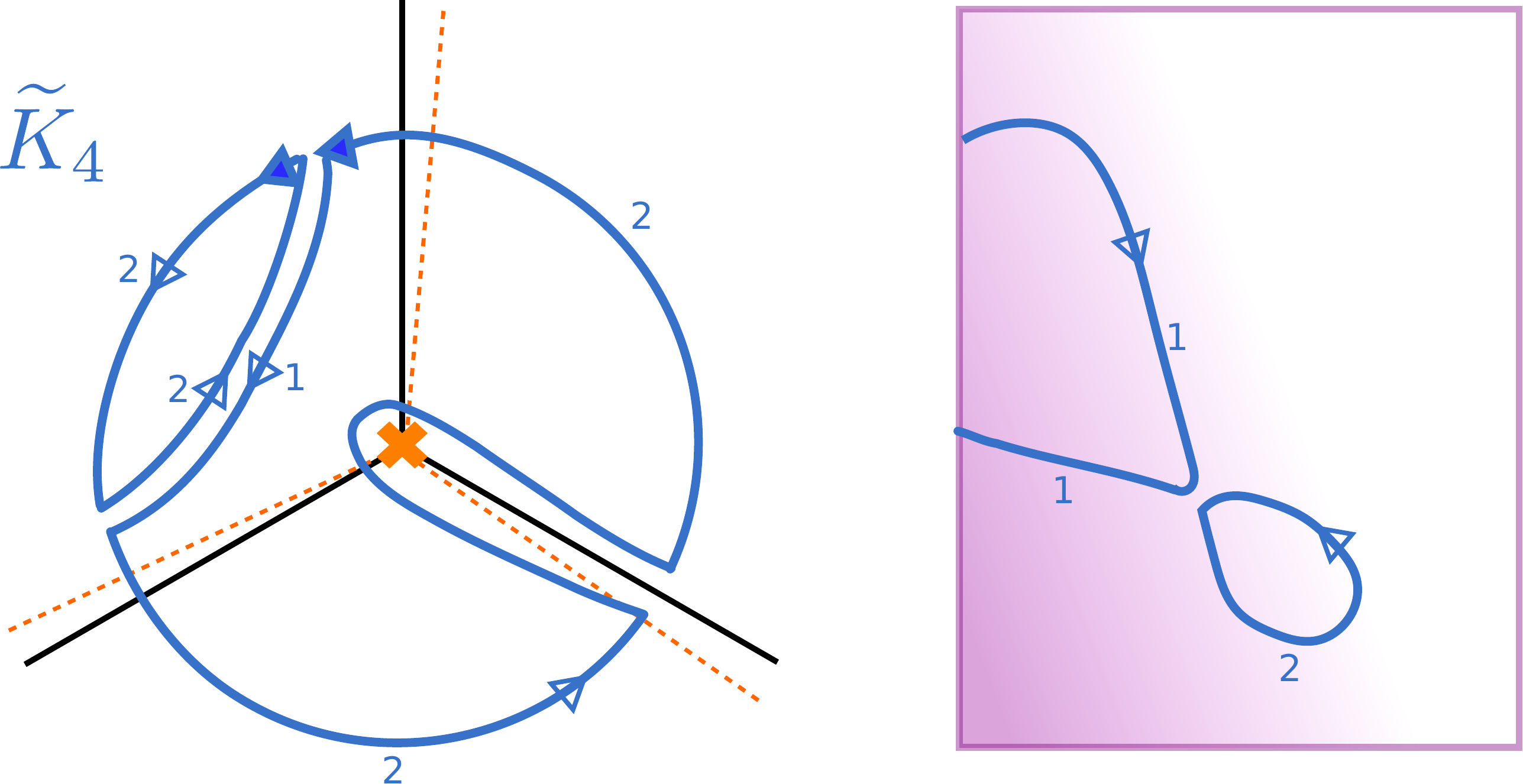}{0.24}{A lift $\tK_4$ of $K$.  Left: the standard projection. Right: the part of the leaf space projection involving the exchange.}
	
	In $\Sk(\tM,\fgl(1))$ again we have $[\tK_4]=1$. The framing factor, detour factor, and winding factor for this lift are $q^{3/2}$, $q^{-1/2}$, and $q^{-2}$ respectively. The exchange carries a factor of $q(q^{-1}-q)$. Combining all these factors, altogether this lift contributes $q^{-1}-q$.
	
	\item Finally there is one lift $\tK_5$ involving three detours,
	shown in \autoref{fig:unknot-1-c}.
	\insfigsvg{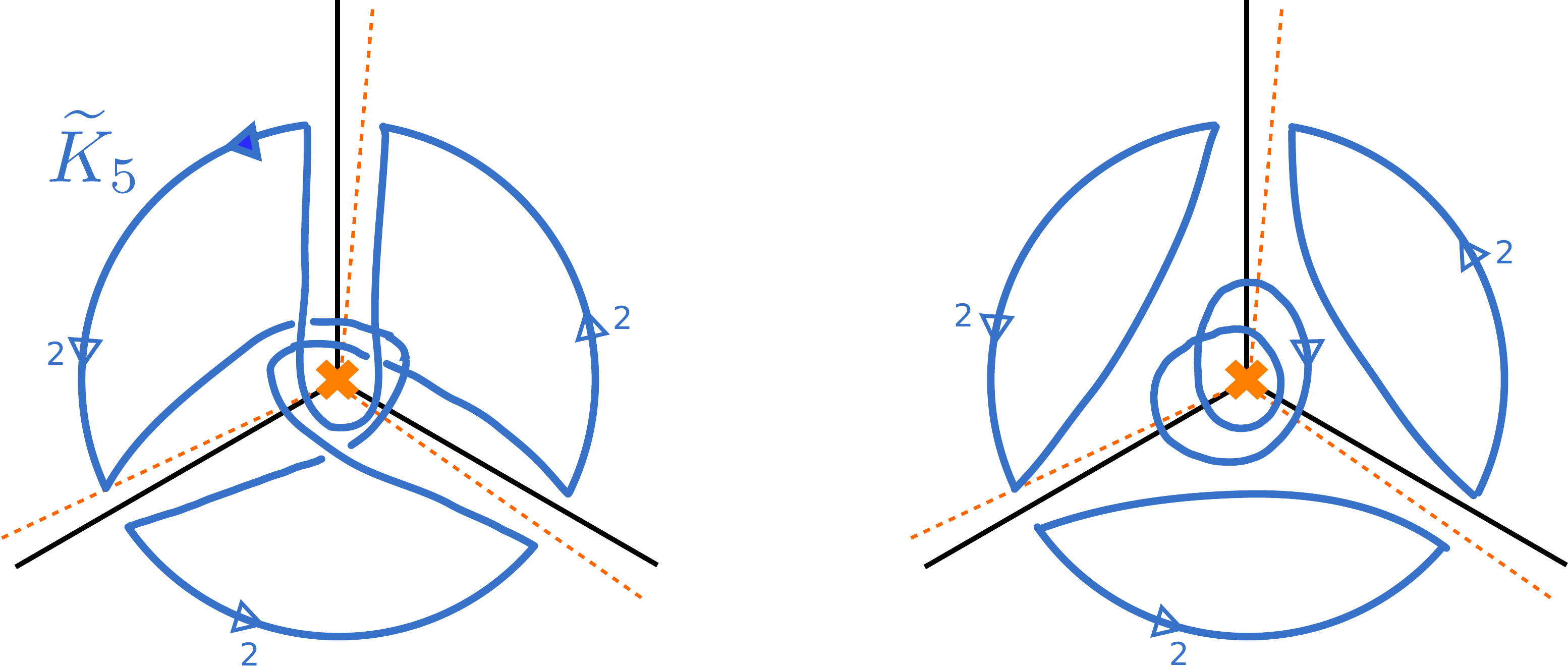}{0.24}{A lift $\tK_5$ of $K$. Left: the standard projection. Right: a link obtained
		from $\tK_5$ by applying $\fgl(1)$ skein relations.}
	
	Resolving crossings using the $\fgl(1)$ skein relations, 
	we find that $[\tK_5]$ is $q$ times the class of the link shown at the
	right of \autoref{fig:unknot-1-c}; in turn the class of that link is $-[\cdot]$ (the minus sign comes from deleting
	the loop in the middle, which winds once around the branch locus in $\tM$); so altogether
	we get $[\tK_5] = -q$.
	The framing factor is $q^{3/2}$, and the detour factor is $q^{-3/2}$. 
	Finally, the total winding of $\tK_5$ is zero, so there
	is no winding factor. Thus altogether this lift contributes $-q$.
\end{itemize}

\begin{table}[h]
	\centering
	\begin{tabular}{|c|c|c|c|c|c|c|} \hline
		lift    & framing & exchange              & detour    & winding & [lift]           & total \\
		\hline \hline
		$\tK_1$    & $q^{3/2}$   & $1$                & $q^{-1/2}$& $1$     & $1$       & $q$ \\
		$\tK_2$    & $q^{3/2}$   & $1$                & $q^{-1/2}$& $1$     & $1$    & $q$ \\
		$\tK_3$    & $q^{3/2}$   & $1$                & $q^{-1/2}$& $1$     & $1$ & $q$ \\
		$\tK_4$    & $q^{3/2}$   & $q(q^{-1}-q)$      & $q^{-1/2}$& $q^{-2}$     & $1$ & $q^{-1} - q$ \\
		$\tK_5$    & $q^{3/2}$   & $1$                & $q^{-3/2}$& $1$     & $-q$ & $-q$ \\
		\hline
	\end{tabular}
\end{table}

Putting everything together, the image of $[K]$ under $q$-nonabelianization is
\begin{equation*}
F([K])=q+q+q+(q^{-1}-q)+(-q)=q+q^{-1}.
\end{equation*}
Again this matches the expected answer.

Next we look at an unknot $K$ whose projection to $C$ is a loop around two branch points
of $\tC \to C$, shown in \autoref{fig:unknot-2-0}.

\insfigsvg{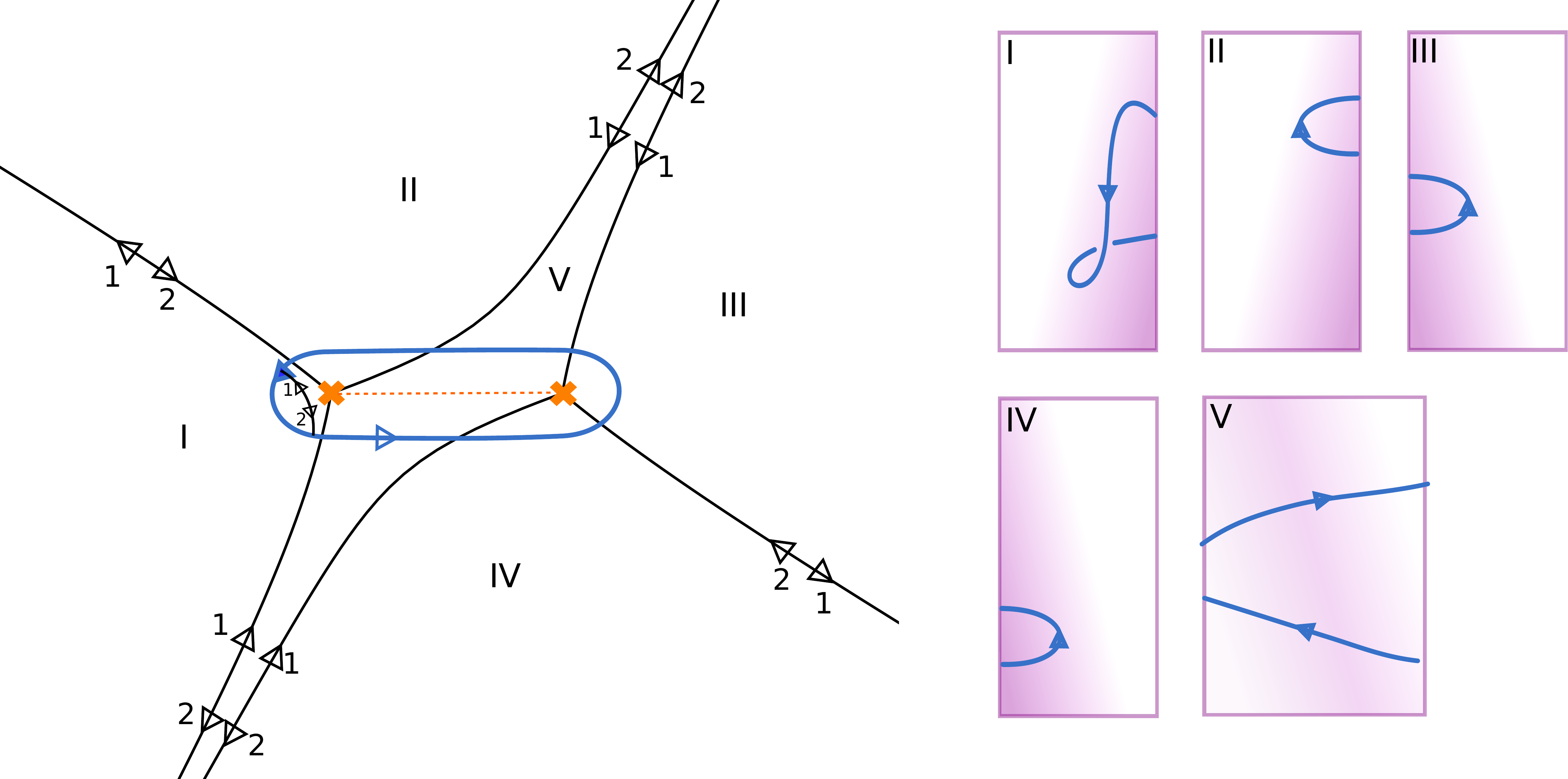}{0.2}{An unknot $K$ encircling two strands of the branch
	locus of the covering $\tM \to M$. Left: the projection of $K$ to $C$. Right:
	the projections of $K$ to five pieces of the leaf space. (For simplicity, we 
	do not show the leaf spaces glued together.)}

This is the most detailed example which we will work out by hand.
There are in total 17 lifts. Two of them, $\tK_1$ and $\tK_2$, are the direct lifts
of $K$ to the two sheets of $\tM$. 
The next four lifts $\tK_3$, $\tK_4$, $\tK_5$, $\tK_6$ each involve two detours at the same branch point:
\insfigsvg{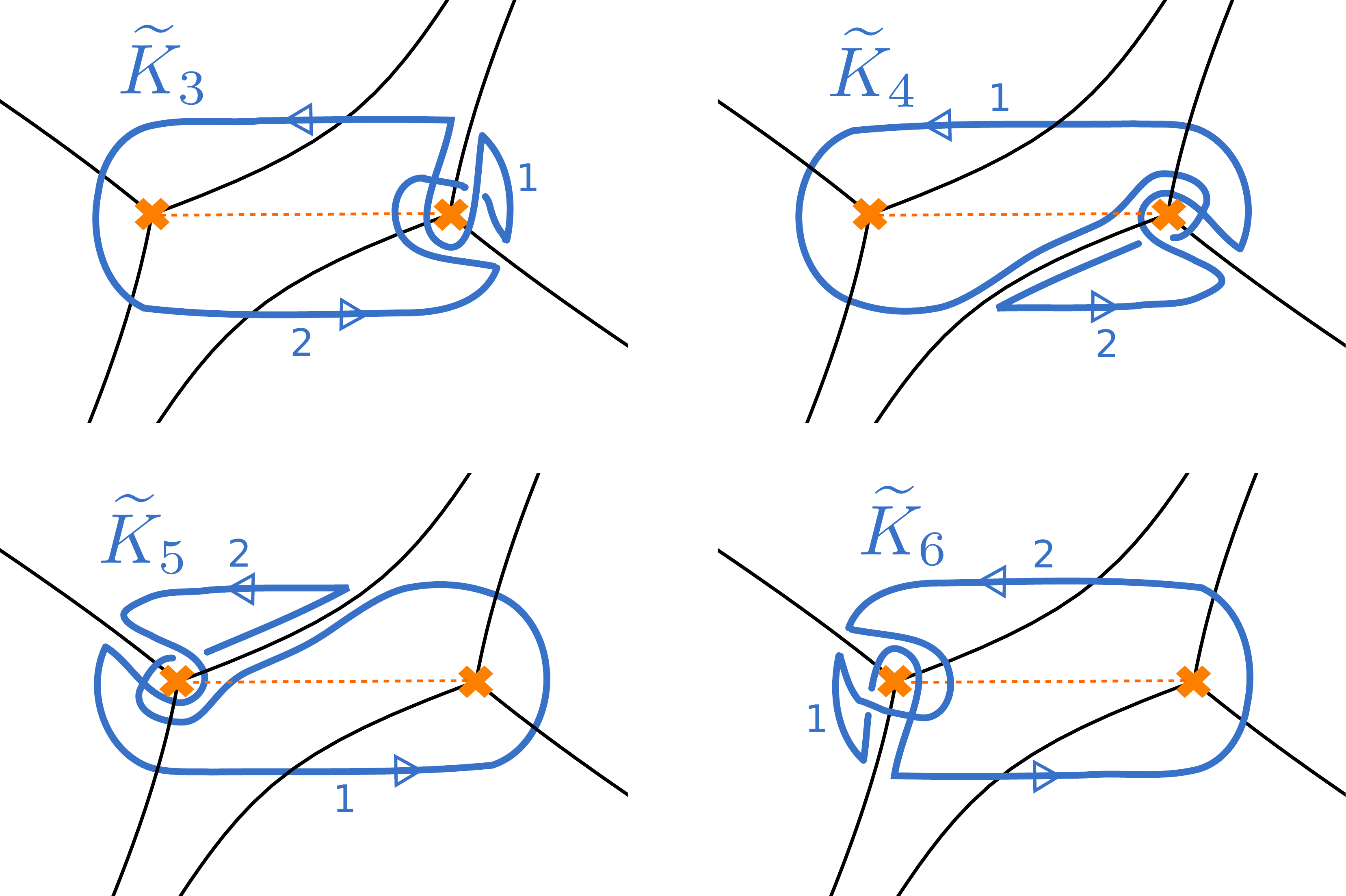}{0.22}{Four lifts of $K$ to $\tC$ which contribute
	to $F([K])$.}

The next four lifts $\tK_7$, $\tK_8$, $\tK_9$, $\tK_{10}$ each 
involve two detours at two different branch points:
\insfigsvg{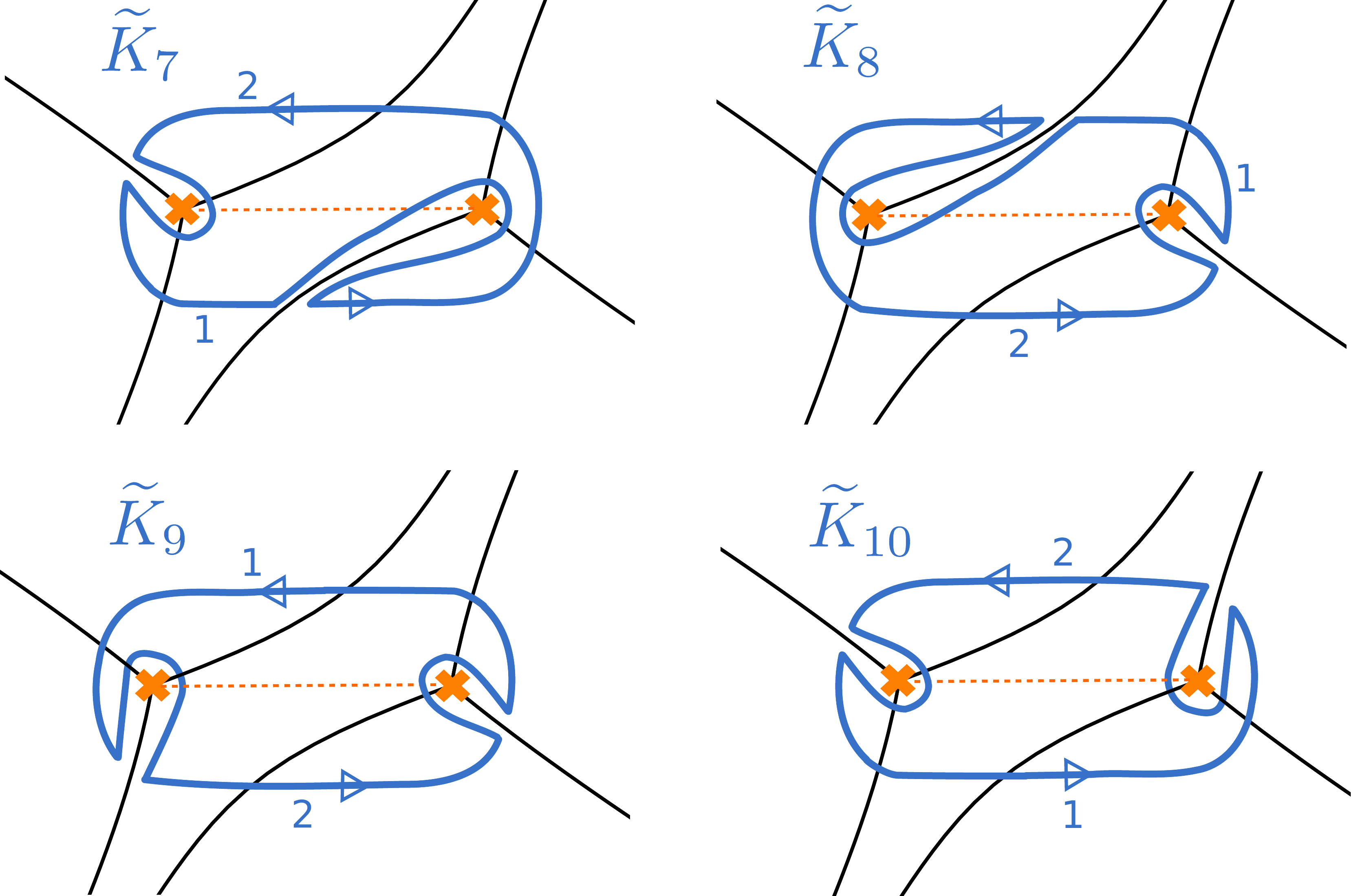}{0.24}{Four more lifts of $K$ to $\tC$ which contribute
	to $F([K])$.}

The lifts $\tK_{11}$ and $\tK_{12}$ have three detours at one branch point and one detour at the other branch point:
\insfigsvg{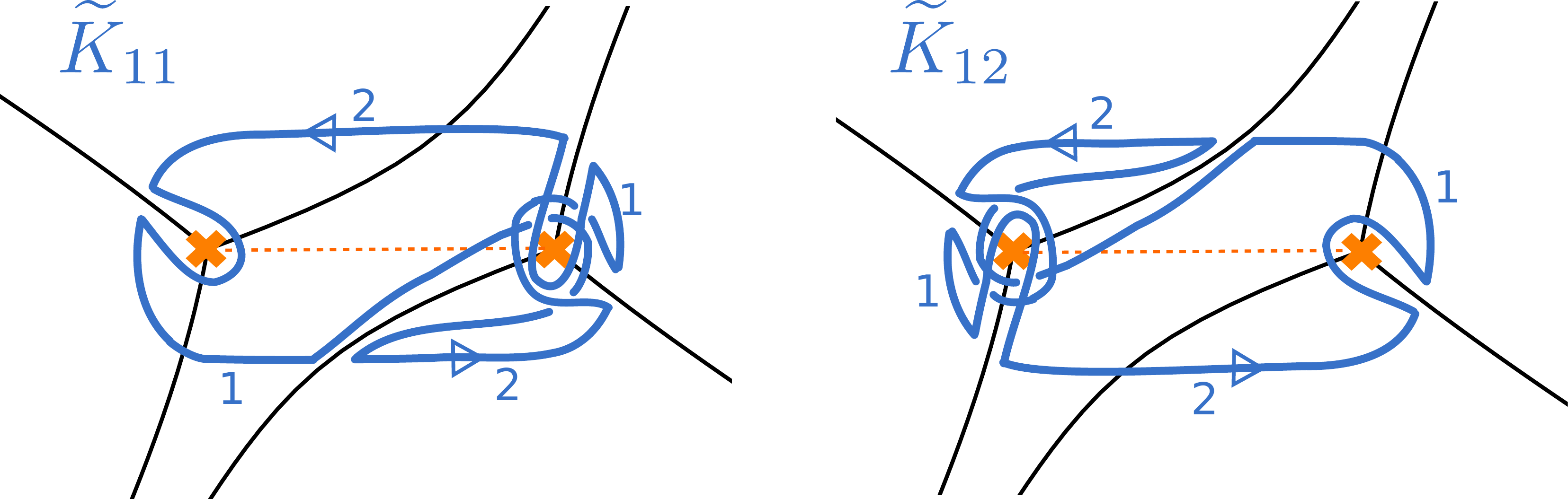}{0.24}{Two more lifts of $K$ to $\tC$ which contribute
	to $F([K])$.}

The next two lifts $\tK_{13}$ and $\tK_{14}$ have two detours at each branch point:
\insfigsvg{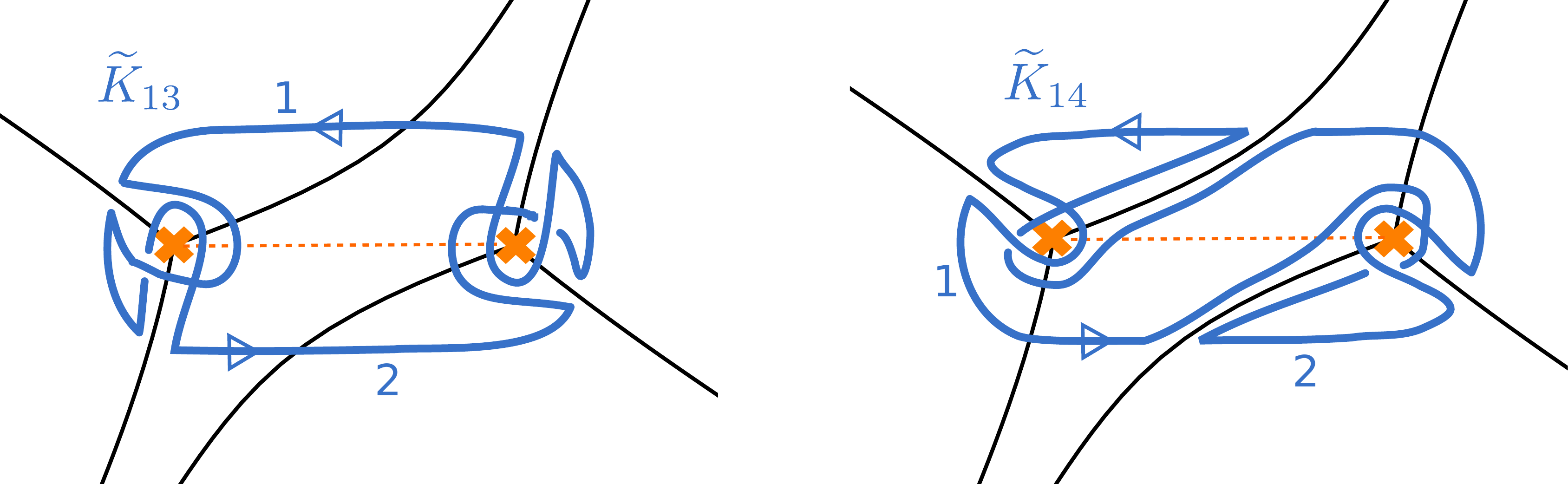}{0.24}{Two more lifts of $K$ to $\tC$ which contribute
	to $F([K])$.}

Finally there are three lifts $\tK_{15}$, $\tK_{16}$ and $\tK_{17}$ which have an exchange path:

\insfigsvg{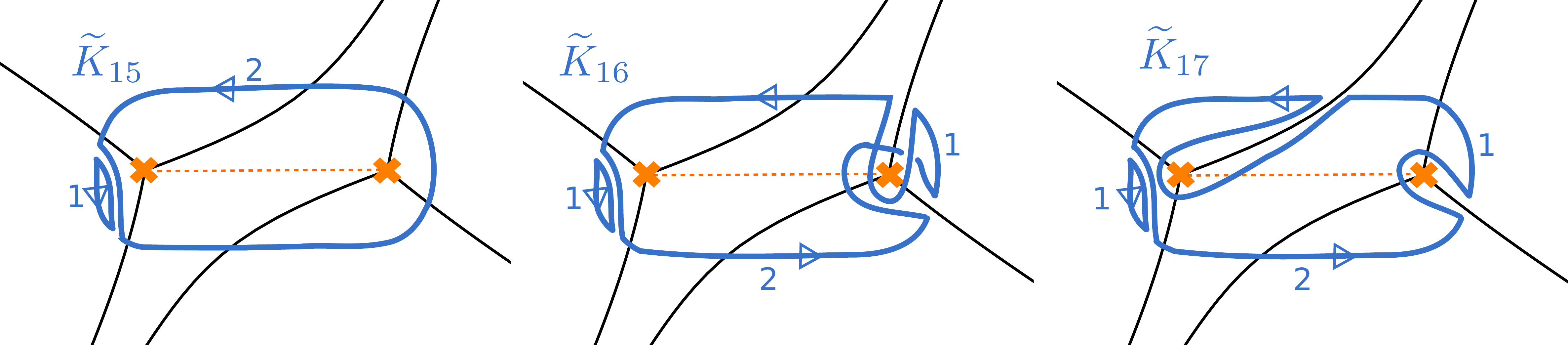}{0.24}{Three more lifts of $K$ to $\tC$ which contribute
	to $F([K])$.}

This is our first example in which there is a nontrivial homology class on $\tC$,
and thus the contributions to $F([K])$ can be more interesting than just multiples
of the unknot on $\tC$: they can involve the other quantum torus generators.
Explicitly, consider the oriented loop $p$ in \autoref{fig:two-branch-points-gamma}. 
According to the rules of \autoref{sec:quantumtorus} we define the
quantum torus generator $X_\gamma = [p] \in \Sk(\tM,\fgl(1))$.
\insfigsvg{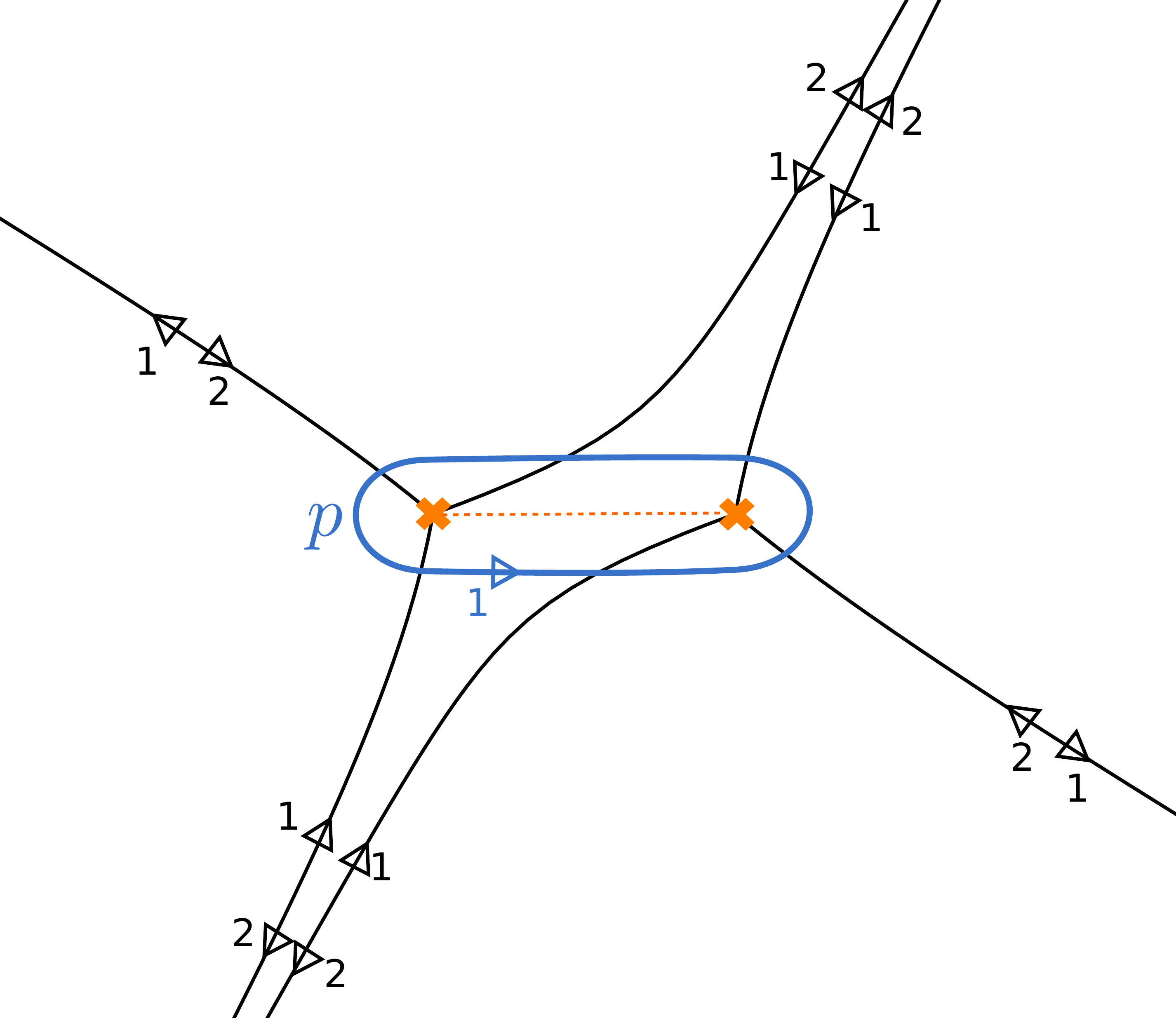}{0.16}{A cycle $p$ on $\tC$, 
	with $[p] = X_\gamma$ in $\Sk(\tM,\fgl(1))$.}
\noindent The contributions from some of the lifts will involve the variable $X_\gamma$.
We will not describe in detail the computations for all 17 lifts; see the
table below for the results.

\begin{table}[h]
	\centering
	\begin{tabular}{|c|c|c|c|c|c|c|} \hline
		lift    & framing & exchange              & detour    & winding & [lift]           & total \\
		\hline \hline
		$\tK_1$    & $q^2$   & $1$                & 1         & $1$     & $X_\gamma$       & $q^2 X_\gamma$ \\
		$\tK_2$    & $q^2$   & $1$                & 1         & $1$     & $X_{-\gamma}$    & $q^2 X_{-\gamma}$ \\
		$\tK_3$    & $q^2$   & $1$                & $q^{-1}$  & $1$     & $-q X_{-\gamma}$ & $-q^2 X_{-\gamma}$ \\
		$\tK_4$    & $q^2$   & $1$                & $q^{-1}$  & $1$     & $-q X_{\gamma}$ & $-q^2 X_{\gamma}$ \\
		$\tK_5$    & $q^2$   & $1$                & $q^{-1}$  & $1$     & $-q X_{\gamma}$ & $-q^2 X_{\gamma}$ \\
		$\tK_6$    & $q^2$   & $1$                & $q^{-1}$  & $1$     & $-q^{-1} X_{-\gamma}$ & $- X_{-\gamma}$ \\
		$\tK_7$    & $q^2$   & $1$                & $q^{-1}$  & $1$     & $1$ & $q$ \\
		$\tK_8$    & $q^2$   & $1$                & $q^{-1}$  & $1$     & $1$ & $q$ \\
		$\tK_9$    & $q^2$   & $1$                & $q^{-1}$  & $1$     & $1$ & $q$ \\
		$\tK_{10}$ & $q^2$   & $1$                & $q^{-1}$  & $1$     & $1$ & $q$ \\
		$\tK_{11}$ & $q^2$   & $1$                & $q^{-2}$  & $1$     & $-q$ & $-q$ \\
		$\tK_{12}$ & $q^2$   & $1$                & $q^{-2}$  & $1$     & $-q$ & $-q$ \\
		$\tK_{13}$ & $q^2$   & $1$                & $q^{-2}$  & $1$     & $X_{-\gamma}$ & $X_{-\gamma}$ \\
		$\tK_{14}$ & $q^2$   & $1$                & $q^{-2}$  & $1$     & $q^2 X_{\gamma}$ & $q^2 X_{\gamma}$ \\
		$\tK_{15}$ & $q^2$   & $q(q^{-1}-q)$      & $1$       & $q^{-2}$& $X_{-\gamma}$ & $q(q^{-1}-q) X_{-\gamma}$ \\
		$\tK_{16}$ & $q^2$   & $q(q^{-1}-q)$      & $q^{-1}$  & $q^{-2}$& $-q X_{-\gamma}$ & $q(q-q^{-1}) X_{-\gamma}$ \\
		$\tK_{17}$ & $q^2$   & $q(q^{-1}-q)$      & $q^{-1}$  & $q^{-2}$& $1$ & $q^{-1} - q$ \\
		\hline
	\end{tabular}
\end{table}

Summing the 17 terms together gives once again the expected answer,
\begin{align*}
F([K]) = q + q^{-1}.
\end{align*}
In particular, note that all the terms proportional to $X_\gamma$ cancel
among themselves, as do those proportional to $X_{-\gamma}$.
This had to happen, since $K$ is contained in a ball; in such cases
we always just get the polynomial \eqref{eq:homfly-relation-intro} for $K$,
just as for $M = \R^3$.
In more interesting examples where $K$ represents a nontrivial class
in $\pi_1(M)$, the $X_\gamma$ will not cancel out.

\subsubsection{Trefoils}
In \autoref{sec:state-sum} we obtained the 
Jones polynomial for a left-handed trefoil using the state-sum model. We could equally well apply our $q$-nonabelianization map to a left-handed trefoil $K_{\text{trefoil}}$ in a single domain, equipped with standard framing. The calculation goes through in a similar fashion to the state-sum model. 
Explicitly, there are $6$ lifts, whose contributions sum to
\begin{align*}
F([K_{\text{trefoil}}])&=q^{-3}+q^{-3}+(q^{-1}-q)q^{-2}+(q^{-1}-q)q^{-2}+(q^{-1}-q)q^2+(q^{-1}-q)^3q^{-2}\\
&=q^{-5}+q^{-3}+q^{-1}-q^3\\
&=q^{-3\times 2}P_{\text{HOMFLY}}(K_{\text{trefoil}}, a=q^2, z=q-q^{-1}),
\end{align*}
which matches \eqref{eq:state-sum-formula}.

A more interesting example is a trefoil in the neighborhood of a branch point as shown in \autoref{fig:trefoilbp1}. There are in total $18$ lifts, whose contributions sum up to give the expected answer,
\begin{align*}
F([K_{\text{trefoil}}])=&\ (q^{-1}-q)-q(q^{-1}-q)^2+q^{-1}(q^{-1}-q)^2-q+q^{-1}+q^{-1}+q^{-5}+(q^{-1}-q)\\
&+q^{-2}(q^{-1}-q)-q^{-2}(q^{-1}-q)+q^{-1}(q^{-1}-q)^2+q^{-2}(q^{-1}-q)^3\\
&-q^{-3}(q^{-1}-q)^2+q^{-1}+q^{-3}-q ^{-3}+q^{-2}(q^{-1}-q)-q^{-3}\\
=&\ q^{-5}+q^{-3}+q^{-1}-q^3.
\end{align*}

\insfigsvg{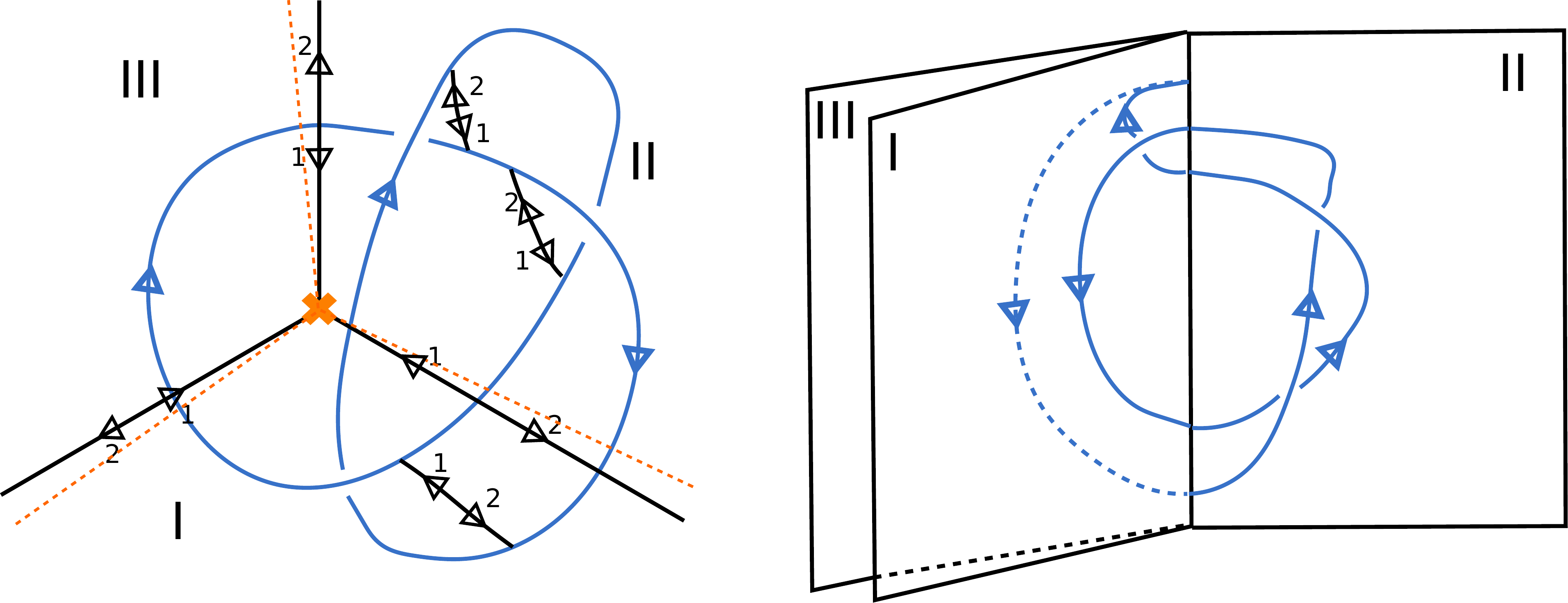}{0.3}{A left-handed trefoil in the neighborhood of a branch point in standard projection (left) and leaf space projection (right).} 

\autoref{fig:trefoilbp2} shows another example of a left-handed trefoil knot in the neighborhood of a branch point. Here there are in total $30$ lifts, whose contributions 
sum up to
give once again the expected answer $q^{-5}+q^{-3}+q^{-1}-q^3$.

\insfigsvg{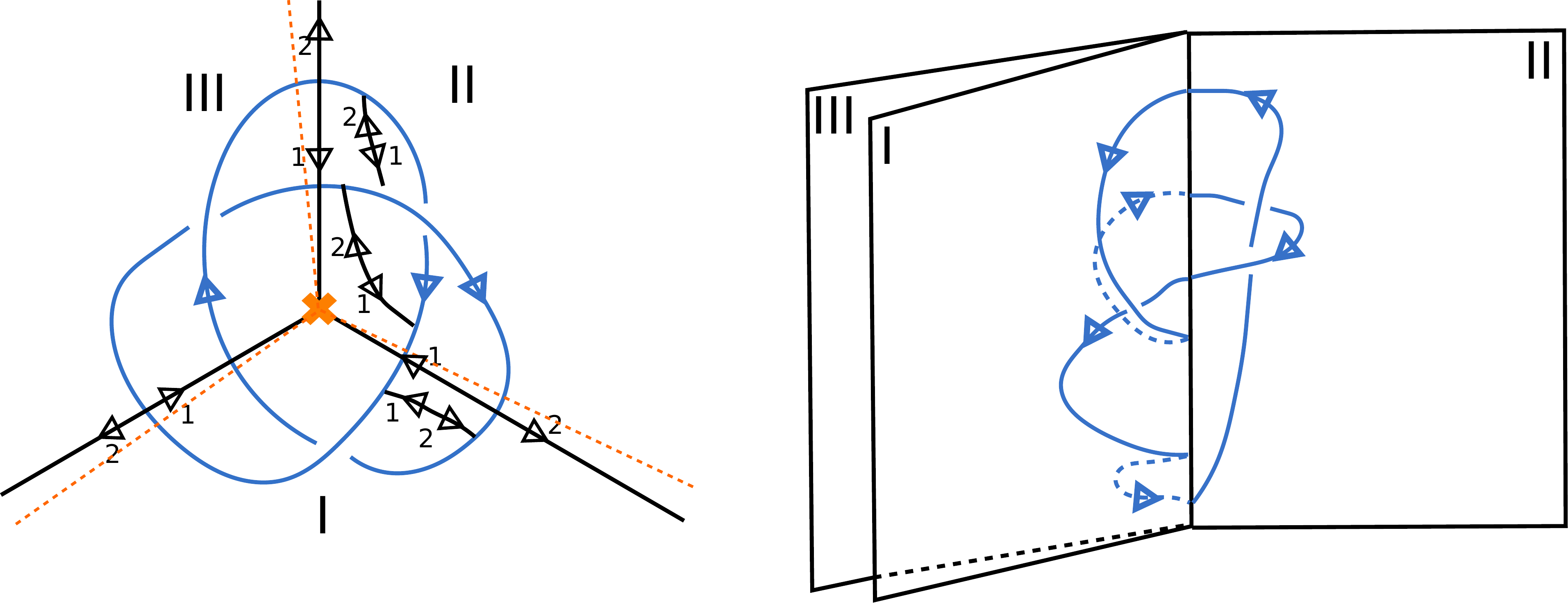}{0.25}{Another left-handed trefoil in the neighborhood of a branch point in standard projection (left) and leaf space projection (right).}

\subsubsection{Figure-eight knot}

In \autoref{fig:figure8knot} we show a figure-eight knot $K_{\text{figure-8}}$ in standard projection and leaf space projection. There are in total $47$ lifts, whose contributions sum up to 
\begin{equation}
F([K_{\text{figure-8}}]) = q^5+q^{-5},
\end{equation}
matching $P_{\text{HOMFLY}}(K_{\text{figure-8}},a=q^2,z=q-q^{-1})$ as expected.\footnote{Although 
the standard projection of $K_{\text{figure-8}}$ has crossing number $4$, its writhe is $0$.}

\insfigsvg{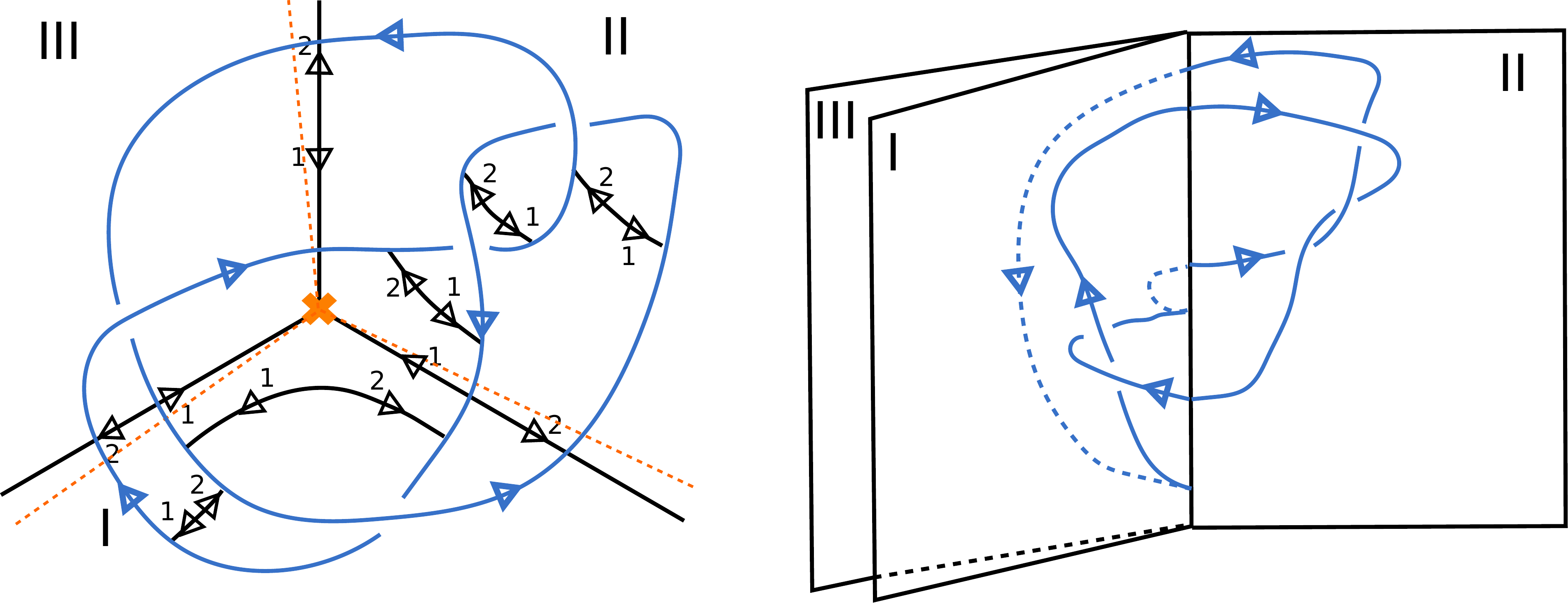}{0.25}{A figure-eight knot in the neighborhood of a branch point in standard projection (left) and leaf space projection (right).}

\subsection{A pure flavor line defect}

\insfigsvg{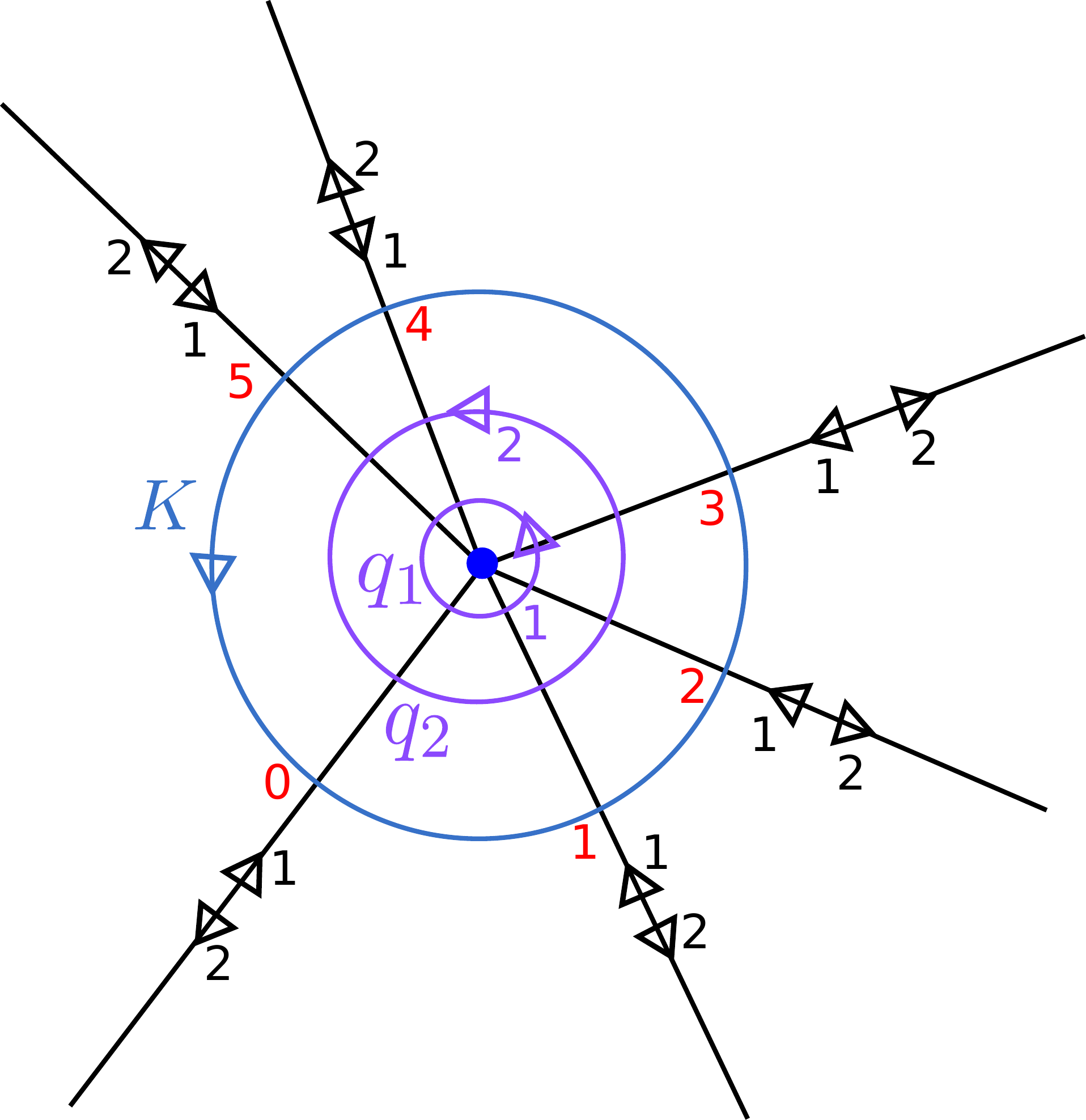}{0.25}{An unknot $K$ whose standard projection is a small loop around a puncture (denoted as a blue dot) on $C$. For simplicity, in this and following examples,  we do not draw the leaf space projection. Instead, we specify $x^3$-coordinates (in red) at points where the unknot meets a critical leaf. Between these points the $x^3$-coordinate varies as simply as possible:
	this means it varies monotonically, except for critical points which we place
	immediately before crossing a critical leaf. We also show cycles $q_1$, $q_2$ representing homology classes $\mu_1$, $\mu_2$ in $H_1(\tC,\Z)$ respectively.}

Now we begin to consider examples of links which are homotopically nontrivial in $M$. 

The simplest such example is a loop $K$ whose standard projection encircles a puncture on $C$, as illustrated in \autoref{fig:flavorline}.\footnote{The number of critical leaves going into the puncture depends on the example. 
However, that number is not important in this example, since there are
no possible detours.} $K$ corresponds to a pure flavor line defect in a theory of class $S$.

In this case the only lifts of $K$ allowed are the direct lifts $\tK_1$ and $\tK_2$ on sheet $1$ and sheet $2$ respectively. Moreover, the framing factor and winding factor for each of these are trivial, 
so we simply have $\alpha(\tK_1)=\alpha(\tK_2)=1$ and thus
\begin{equation}
F([K])=X_{\mu_1}+X_{\mu_2}.
\end{equation}

So far we have been considering class $S$ theories of type $\fgl(2)$, but in the following we will also discuss class $S$ theories of type $\fsl(2)$, in order to be able to compare directly
to previous results in the literature.
The projection from $\fgl(2)$ to $\fsl(2)$ is discussed in
\autoref{sec:sl2}; roughly it amounts to 
replacing $X_\gamma \mapsto X_{\half (\gamma - \sigma(\gamma))}$,
where $\sigma$ denotes the 
deck transformation exchanging the two sheets
of $\tC \to C$. In the following examples we first obtain the generating function
$F([K])$ in a class $S$ theory of type $\fgl(2)$, then apply this projection to get the result in the $\fsl(2)$ theory.

For a first example, we revisit the pure flavor line defect, now in a theory of class $S$ of type $\fsl(2)$. The projection identifies $\mu_1 \sim -\mu_2:=\mu$, and the generating function is
\begin{equation}
F([K])=X_{\mu}+X_{-\mu}.
\end{equation}

\subsection{\texorpdfstring{$\SU(2)$}{SU(2)} \texorpdfstring{$\cN= 2^*$}{N=2} theory}\label{sec:puncturedtorus}

Next we consider the $\SU(2)$ $\cN=2^*$ theory. This theory is obtained by giving a mass $m$ to the adjoint hypermultiplet in the $\SU(2)$ $\cN=4$ theory. Its class $S$ construction is given by compactifying the 6d (2,0) $A_1$ theory on a once-punctured torus $C$.

We choose $m$ and the coupling $\tau$ such that the WKB foliation is as shown in \autoref{fig:puncturedtorus-0}. We consider a line defect corresponding to a loop $K$ whose projection to $C$ wraps both A-cycle and B-cycle once,
as shown in \autoref{fig:puncturedtorus-0}.

\insfigsvg{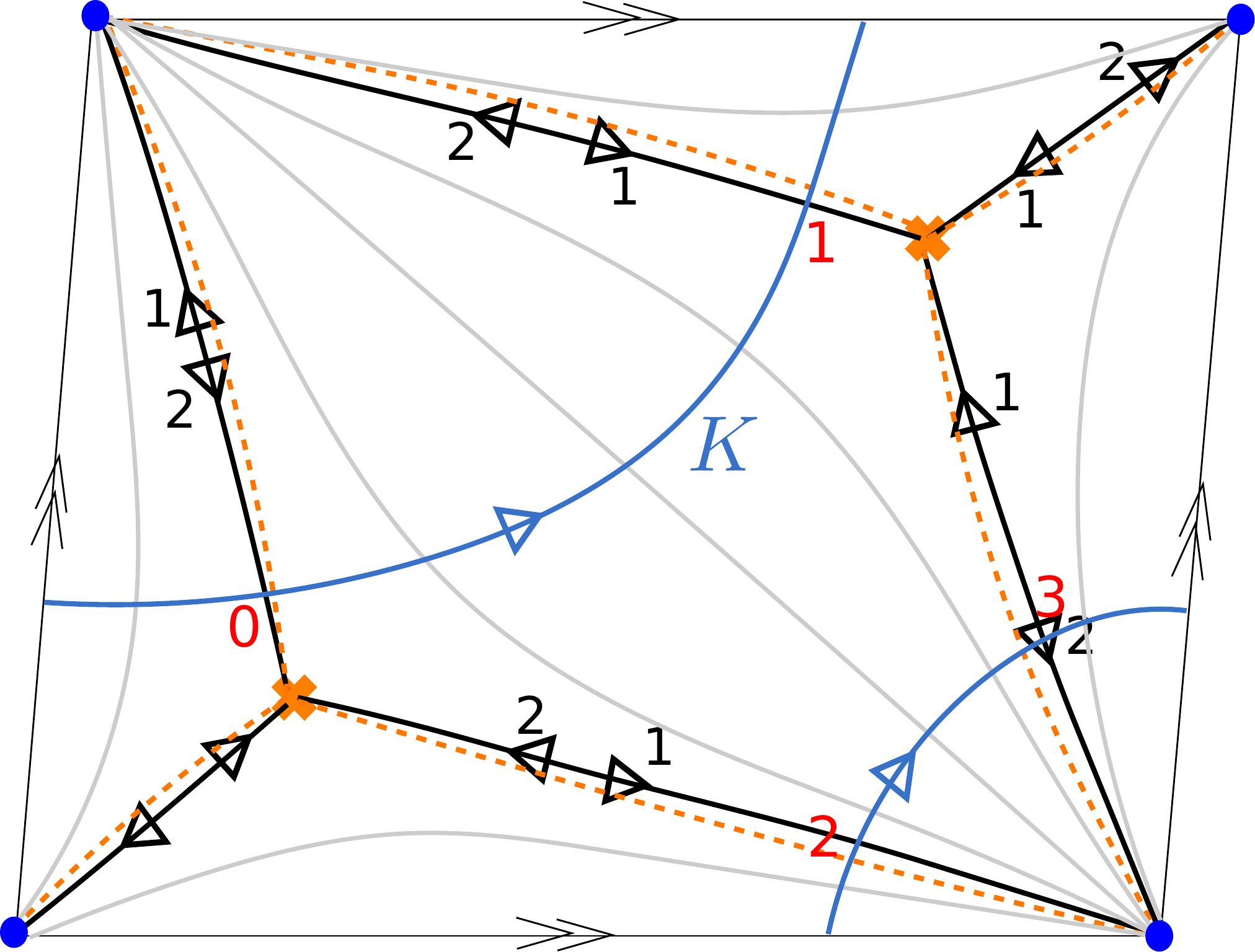}{0.25}{WKB foliation on a torus with one puncture (blue dot). We also show the standard projection of a loop $K$, whose $x^3$-profile is specified by the red numbers, following
the convention we introduced in \autoref{fig:flavorline}.} 

\insfigsvg{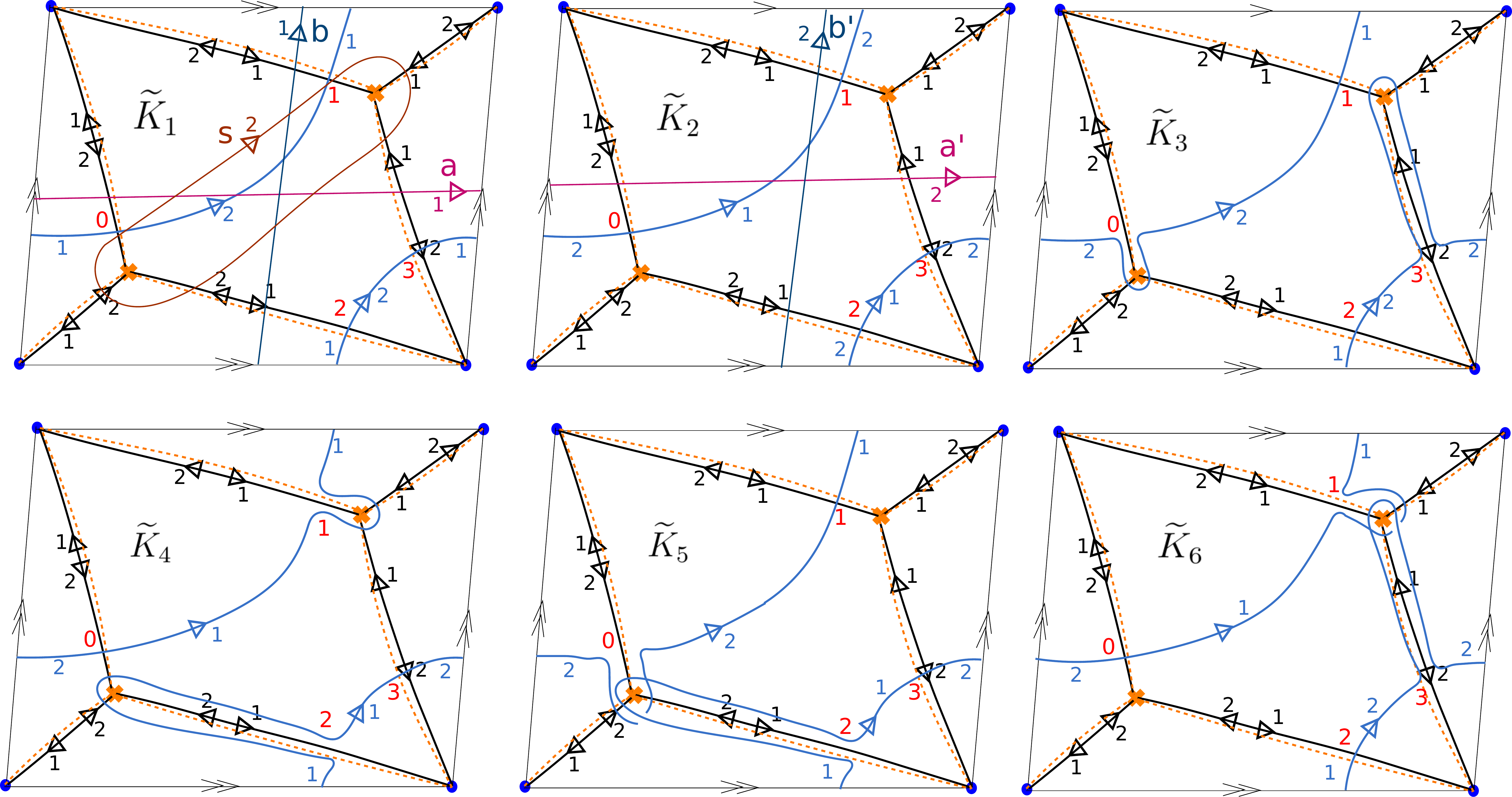}{0.20}{The six lifts $\tK_i$ of $K$. We also show a basis $\{\gamma_1,\gamma_2,\gamma_3,\gamma_4,\gamma_5\}$ for $H_1(\tC,\Z)$ with representatives $s,a,b,a',b'$ respectively.}

We will first compute the generating function in the $\fgl(2)$ case. There are in total six lifts $\tK_i$ as illustrated in \autoref{fig:puncturedtorus-1}. For each $\tK_i$, the framing factor is trivial,
while the winding factor and detour factors cancel out, so $\alpha(\tK_i)=1$. The question of finding $F([K])$ thus reduces to expressing $[\tK_i]$ in terms of $X_{\gamma_i}$. Then according to the
crossing-counting rules explained in \autoref{sec:quantumtorus}, we have
\begin{align*}
[\tK_1]&=X_{\gamma_2+\gamma_3},\quad [\tK_2]=X_{\gamma_4+\gamma_5},\\ [\tK_3]&=X_{\gamma_1+\gamma_3+\gamma_4},\quad
[\tK_4]=X_{-\gamma_1+\gamma_3+\gamma_4}, \\
[\tK_5]&=-q^{-1}X_{\gamma_3+\gamma_4},\quad
[\tK_6]=-qX_{\gamma_3+\gamma_4}
\end{align*}
(recall that the factors of $q$ come from the genuine crossings, and factors of $-1$ come from
``non-local crossings.'')
Summing these up gives
\begin{equation*}
F([K]) = X_{\gamma_2+\gamma_3}+X_{\gamma_4+\gamma_5}+X_{\gamma_1+\gamma_3+\gamma_4}+X_{-\gamma_1+\gamma_3+\gamma_4}-\big(q+q^{-1}\big)X_{\gamma_3+\gamma_4}.
\end{equation*}
This result obeys the expected positivity and monotonicity properties
discussed in \autoref{sec:positivity}.

To obtain the spectrum in the $\fsl(2)$ theory, we just perform the projection $\rho$,
which has the effect of identifying $\gamma_4 \sim -\gamma_2$ and $\gamma_5 \sim -\gamma_3$. The resulting generating function is
\begin{equation} \label{eq:gf}
F([K])=X_{\gamma_2+\gamma_3}+X_{-\gamma_2-\gamma_3}+X_{\gamma_1-\gamma_2+\gamma_3}+X_{-\gamma_1-\gamma_2+\gamma_3}-(q+q^{-1})X_{-\gamma_2+\gamma_3}.
\end{equation}
This agrees (modulo some shifts in conventions) with \cite{Gabella:2016zxu}, where the same line defect
was considered. We also remark that using the traffic rules of \cite{Gaiotto:2010be} one could compute the vacuum expectation value of this line defect, which agrees with the classical limit of our generating function \eqref{eq:gf}.

\subsection{\texorpdfstring{$\SU(2)$}{SU(2)} with \texorpdfstring{$N_f= 4$}{Nf=4} flavors}\label{sec:tetrahedron}

\insfigsvg{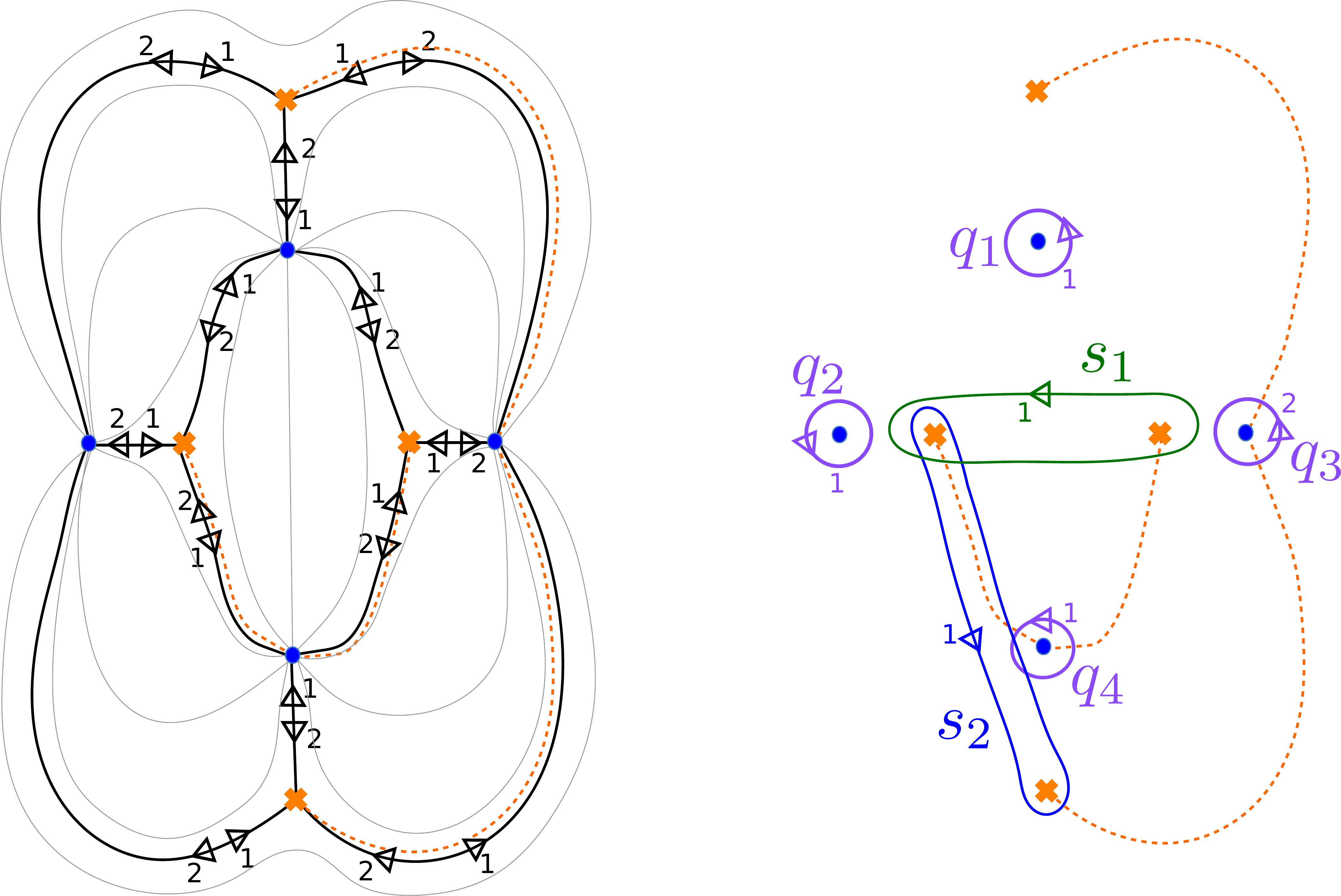}{0.2}{Left: WKB foliation structure on $\mathbb{CP}^1$ with four punctures (denoted as blue dots) with the parameters in \eqref{tetrahedronphi}.
	Right: a basis $\{\gamma_1,\gamma_2,\mu_1,\mu_2,\mu_3,\mu_4\}$ for $H_1^\odd(\tC,\Z)$ with representative cycles $s_1,s_2,q_1,q_2,q_3,q_4$ respectively.}

As our next example, we take $C$ to be $\mathbb{CP}^1$ with four punctures, corresponding to $\cN=2$ $\SU(2)$ SYM with four fundamental hypermultiplets. 
Traditional $\frac{1}{2}$-BPS line defects in this theory have been systematically studied by many people, for example \cite{Drukker:2009id, Drukker:2009tz,Alday:2009fs,Gaiotto:2010be}. Here 
we will consider both traditional and fat line defects. 
% We remark that, here as the projection map $\rho$ only affects pure flavor cycles so it is safe for us to directly compute the generating function in the $\fsl(2)$ theory.

Let $z$ be a coordinate on $\mathbb{CP}^1$. We choose a complex structure such that the four punctures are located at $z=1,\text{i},-1,-\text{i}$ respectively. Moreover, we pick Coulomb branch and mass parameters such that the double cover $\tC$ is
\begin{equation}
	\tC = \{\lambda: \lambda^2+\phi_2 =0\}\subset T^*C,
\end{equation}
where
\begin{equation}\label{tetrahedronphi}
	\phi_2=-\frac{z^4+2z^2-1}{2(z^4-1)^2} \, \de z^2.
\end{equation}
The WKB foliation structure on $C$ is shown in \autoref{fig:tetrahedron}. 

\insfigsvg{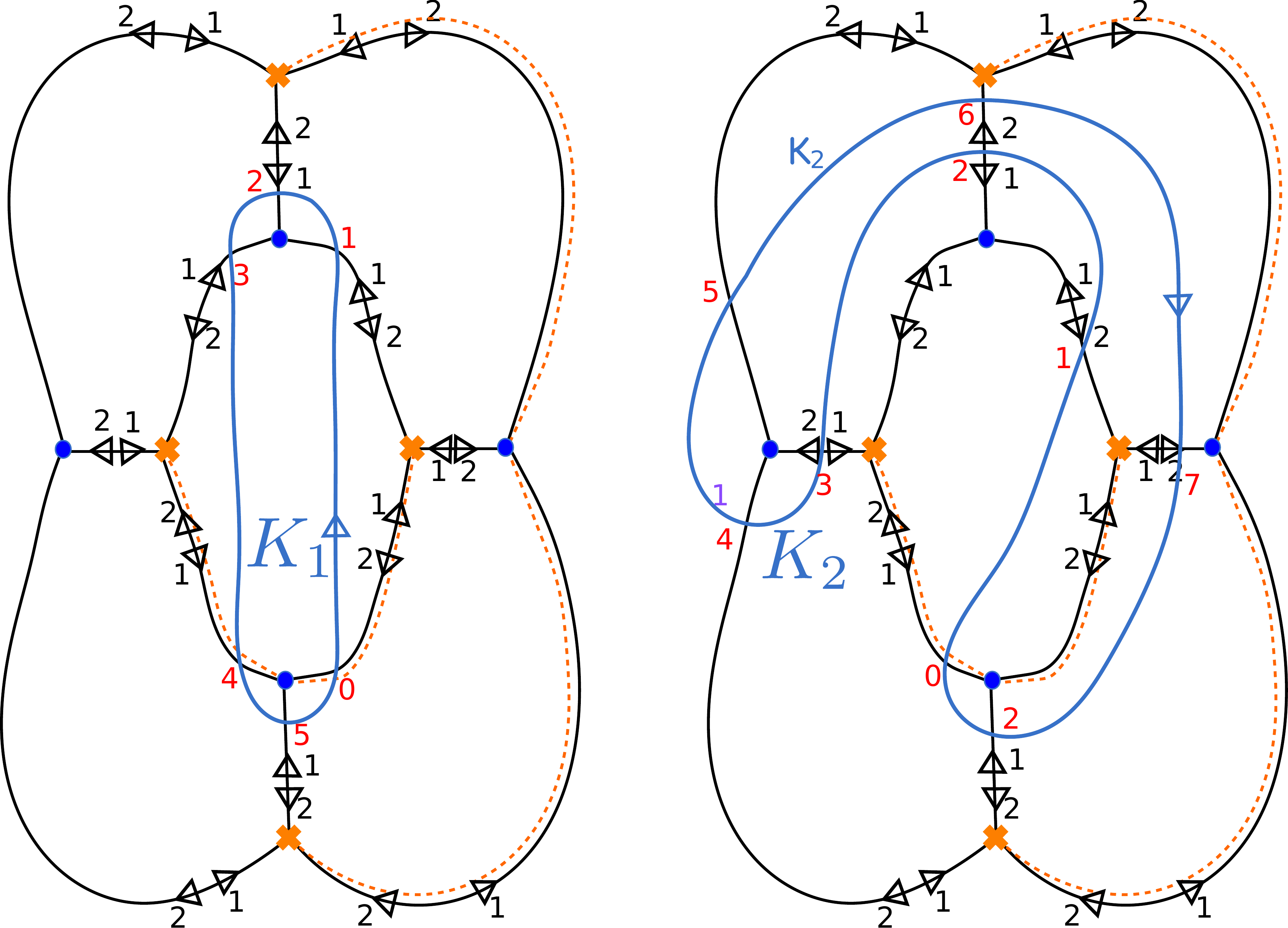}{0.2}{
	Standard projection of two loops $K_1$ and $K_2$ in $M=C\times \R$. Height profiles for these loops are specified in the same way as in \autoref{fig:flavorline}.}

As a warmup we consider a loop $K_1$ as shown on the left of \autoref{fig:tetrahedronLD1LD2}. We choose a basis $\{\gamma_1,\gamma_2,\mu_1,\mu_2,\mu_3,\mu_4\}$ for $H_1^\odd(\tC,\Z)$, where $\langle \gamma_1, \gamma_2 \rangle =-1$ and $\{\mu_i\}$ span the flavor charge lattice. Applying the rules
from \autoref{sec:qab-N=2} we obtain 11 lifts, whose total contribution is
\begin{align*}
	F([K_1]) =&\  X_{-\gamma_2-\mu_2+\mu_3}+X_{-\gamma_2-\mu_1-\mu_4}+X_{\gamma_1+\mu_1-\mu_4}+X_{-\gamma_1-\mu_1+\mu_4}+X_{\gamma_1-\gamma_2+\mu_1-\mu_4}\\
	&+X_{\gamma_1-\gamma_2-\mu_2+\mu_3-2\mu_4}+X_{\gamma_1-2\gamma_2-\mu_2+\mu_3-2\mu_4},
\end{align*}
This also agrees with the classical nonabelianization result in \cite{Gaiotto:2010be}.

As a more interesting example we consider the loop $K_2$ shown on the right of \autoref{fig:tetrahedronLD1LD2}. $K_2$ has in total 48 lifts whose contributions sum up to
\begin{align*}
	F([K_2]) = &\ X_{\gamma_1+\mu_1-\mu_3}+X_{\gamma_2+\mu_1-\mu_3}+X_{\gamma_1+\gamma_2+\mu_1-\mu_3}+X_{-\gamma_2-\mu_1+\mu_3}+X_{\gamma_1+\mu_1+\mu_3}\\
	&+X_{\gamma_1-\gamma_2+\mu_1+\mu_3}+X_{2\gamma_1-3\gamma_2-\mu_2+2\mu_3-3\mu_4}-(q+q^{-1})X_{2\gamma_1-2\gamma_2-\mu_2+2\mu_3-3\mu_4}\\
	&+X_{2\gamma_1-\gamma_2-\mu_2+2\mu_3-3\mu_4}+X_{\gamma_1-2\gamma_2-\mu_1+\mu_3-2\mu_4}+X_{\gamma_1-\gamma_2-\mu_1+\mu_3-2\mu_4}\\
	&+X_{\gamma_1+\mu_1+\mu_3-2\mu_4}-(q+q^{-1})X_{2\gamma_1+\mu_1+\mu_3-2\mu_4}-(q+q^{-1})X_{2\gamma_1-2\gamma_2+\mu_1+\mu_3-2\mu_4}\\
	&+X_{\gamma_1-\gamma_2+\mu_1+\mu_3-2\mu_4}+(2+q^2+q^{-2})X_{2\gamma_1-\gamma_2+\mu_1+\mu_3-2\mu_4}+X_{\gamma_1-\mu_2-\mu_4}\\
	&+X_{\gamma_1-\gamma_2-\mu_2-\mu_4}+X_{\gamma_1+\mu_2-\mu_4}+X_{\gamma_1-\gamma_2+\mu_2-\mu_4}+X_{\gamma_1+2\mu_1+\mu_2-\mu_4}\\
	&-(q+q^{-1})X_{2\gamma_1+2\mu_1+\mu_2-\mu_4}+X_{2\gamma_1-\gamma_2+2\mu_1+\mu_2-\mu_4}+X_{\gamma_1+\gamma_2+2\mu_1+\mu_2-\mu_4}\\
	&+X_{2\gamma_1+\gamma_2+2\mu_1+\mu_2-\mu_4}+X_{\gamma_1-2\gamma_2-\mu_2+2\mu_3-\mu_4}+X_{\gamma_1-\gamma_2-\mu_2+2\mu_3-\mu_4}.
\end{align*} 

Once again the result has the expected properties:
the coefficients of $X_\gamma$ form characters of $\SU(2)_P$ representations,
and framed BPS states that form even- (odd-) dimensional $\SU(2)_P$ representations contribute to the protected spin character with a minus (plus) sign. 

The most interesting
charge sector is the charge $2\gamma_1-\gamma_2+\mu_1+\mu_3-2\mu_4$,
where the framed BPS states form a direct sum of one-dimensional and three-dimensional $\SU(2)_P$ representations. We show the four lifts realizing these framed BPS states in \autoref{fig:tetrahedronLD2_4lifts}.

\insfigsvg{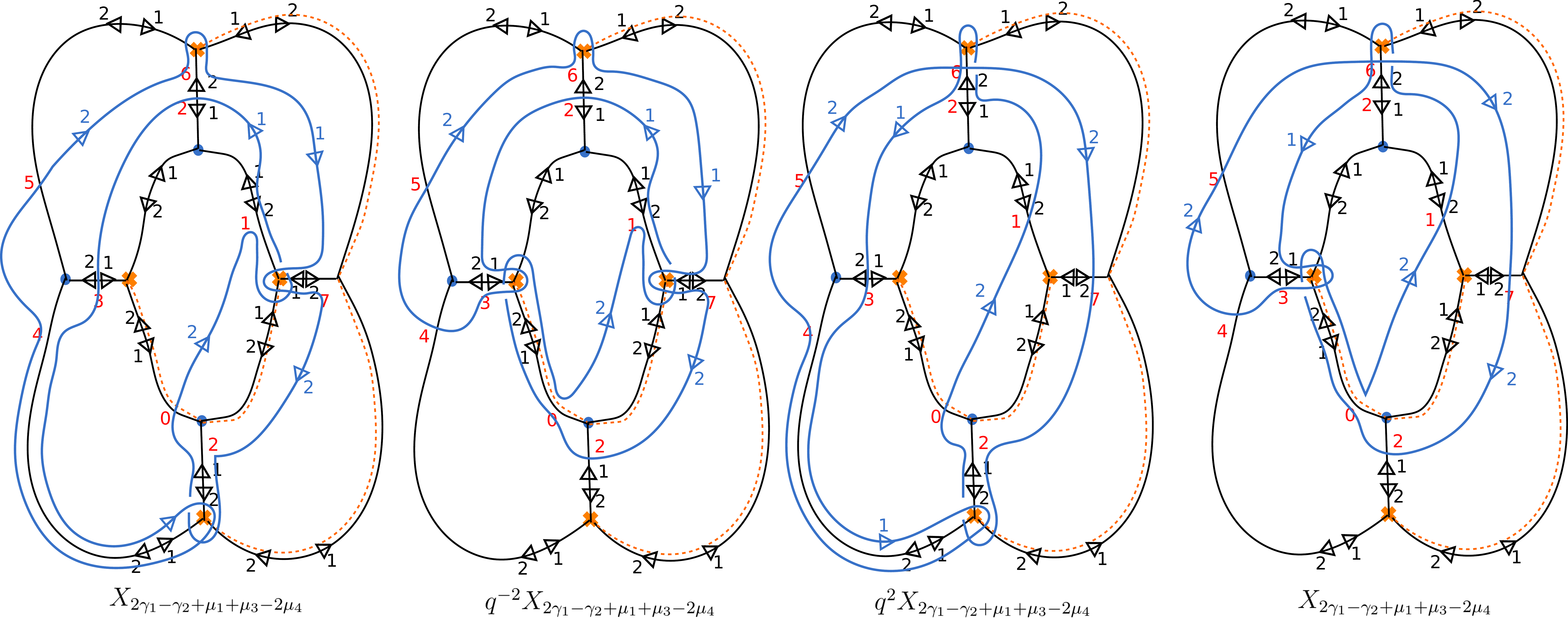}{0.2}{Lifts corresponding to framed BPS states in the charge sector $2\gamma_1-\gamma_2+\mu_1+\mu_3-2\mu_4$ and their contributions to $F([K_2])$.}

As our final example, we consider the link $K_3$ shown in \autoref{fig:tetrahedronLD3}.
$K_3$ has in total $52$ lifts. Summing up their contributions, $F([K_3])$ is 
given by a long expression, which can be conveniently written
in terms of $F([K_2])$ as follows:\footnote{This relation could be
obtained directly from the relations in the $\fsl(2)$ (Kauffman bracket) skein algebra.}
\begin{equation}
	F([K_3])=-F([K_2])+q\big(X_{\mu_2}+X_{-\mu_2}\big)\big(X_{\mu_4}+X_{-\mu_4}\big).
\end{equation}
\insfigsvg{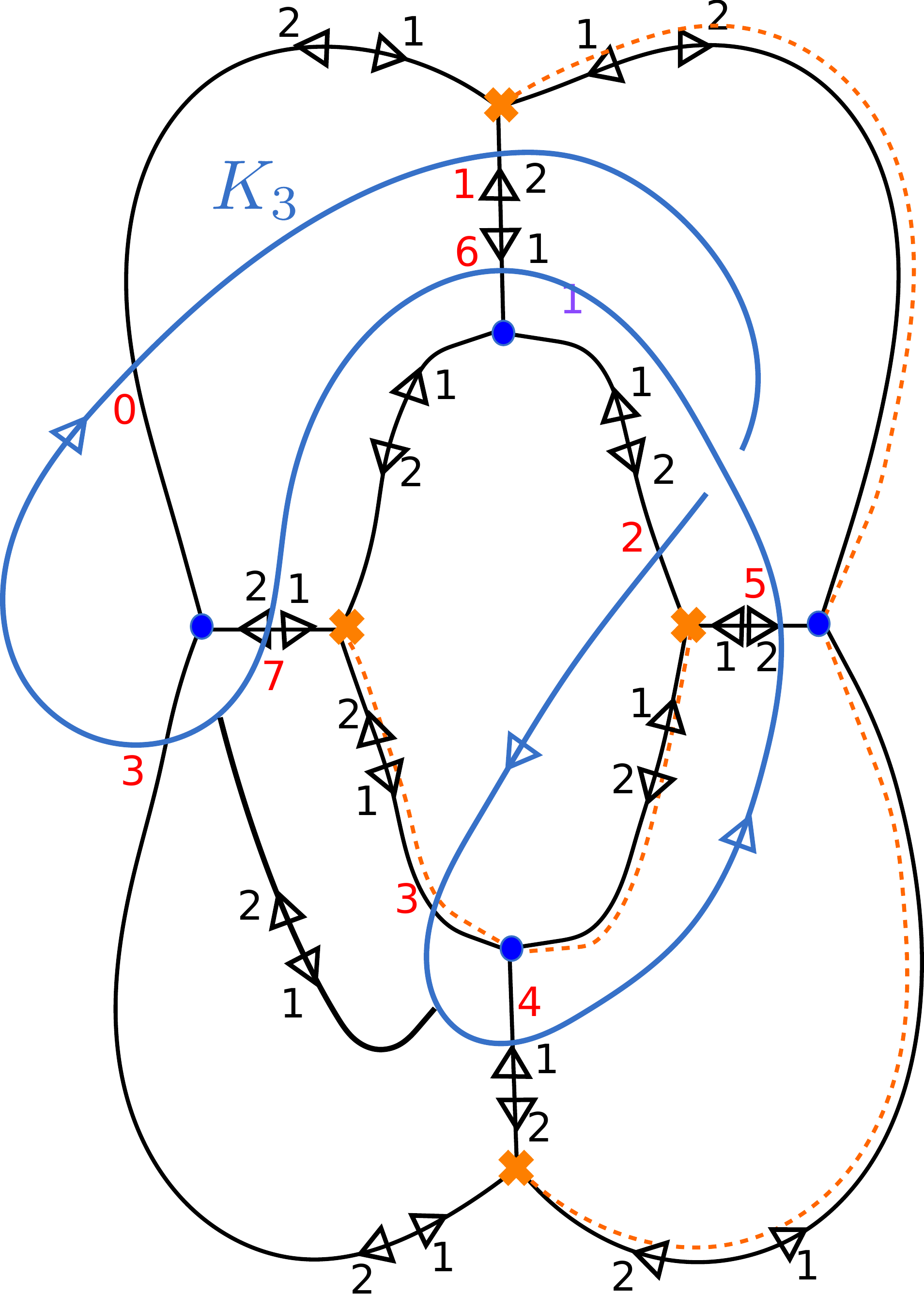}{0.22}{Standard projection of an unknot $K_3$ in $M=C\times \R$. Its height profile is specified in the same convention as in \autoref{fig:flavorline}.} 

In particular, the positivity is violated, and even if we ignore this, the coefficients of $X_\gamma$ in $F([K_3])$ do not in general form $\SU(2)_P$ characters (we have a $q$ term but no corresponding $1/q$). This is as expected since $K_3$ corresponds to a fat line defect, as the standard projection of $K_3$ contains a crossing.

\subsection{\texorpdfstring{$(A_1,A_N)$}{(A1,AN)} Argyres-Douglas theories}

In this section we briefly consider some theories whose class $S$ construction involves irregular singularities: the $(A_1, A_N)$ Argyres-Douglas theories \cite{Cecotti:2010fi}.
These theories are obtained by taking $C = \C P^1$ with an irregular singularity at $z = \infty$, which prescribes that 
the coverings $\tC$ that 
we consider have $\lambda \sim z^{\frac{N}{2} + 1} \de z$ as $z \to \infty$.\footnote{The rules we discuss in this section also apply to the $(A_1,D_N)$ Argyres-Douglas theories, which have an irregular singularity at $z = \infty$
and also a regular singularity at $z = 0$. The spin content of framed BPS states for line defects in simple $(A_1,D_N)$ theories has also been studied in \cite{Neitzke:2017cxz}.}

The local behavior of the WKB foliation in the neighborhood of an irregular singularity $P$ is very different from that near a regular singularity. 
A generic leaf asymptotes tangentially to one of $M$\footnote{For $(A_1, A_N)$ Argyres-Douglas theories, $M=N+3$; for $(A_1, D_N)$ Argyres-Douglas theories, $M=N$.} rays near $P$. If we draw an infinitesimal circle $S^1_P$ around $P$ bounding an infinitesimal disk $D_P$, these $M$ rays determine $M$ marked points $P_i$ on $S^1_P$, evenly distributed. In the examples that we will consider here $C = \mathbb{CP}^1\setminus D_P$. 

\insfigsvg{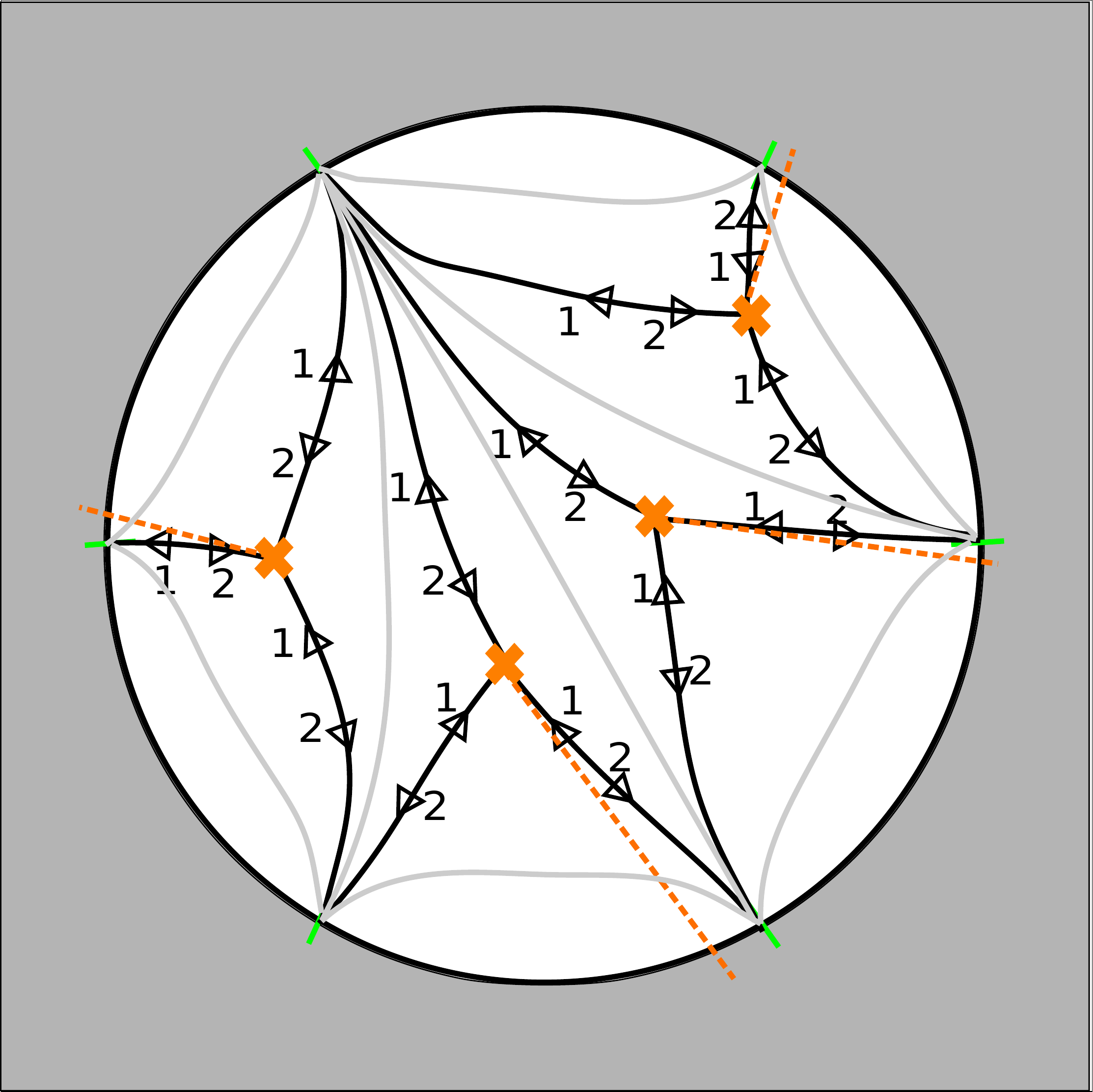}{0.25}{The WKB foliation at some point in the Coulomb branch of the $(A_1,A_3)$ Argyres-Douglas theory. Here the irregular singularity is placed at $z=\infty$. The green markers on the boundary circle correspond to $P_i$, which are determined by the six rays going into the irregular singularity.}

In the following we will use the $(A_1,A_3)$ Argyres-Douglas theory as an example. We choose a point in the parameter space of this theory such that the covering $\tC$ is given by
\begin{equation}
\tC=\Bigg\{\lambda: \lambda^2-\Bigg(z^4-\Big(\frac{3}{2}+2\text{i}\Big)z^2+\big(1-2\text{i}\big)z+\frac{3}{2}\text{i}-\frac{3}{16}\Bigg)\de z^2=0\Bigg\}
\end{equation}
where $z$ is a coordinate on $C$ such that $D_P$ is centered at $z=\infty$. \autoref{fig:WKB-A3} shows the WKB foliation at this point in the moduli space.

In the presence of an irregular singularity, line defects do not correspond to ordinary links in $C\times \R$ anymore; we also need to include links which can have some endpoints on $S^1_P \times \R$. Here we focus on flat line defects, i.e. we only consider $L'$ whose projection to $C$ does not contain any crossings. The projection of such an $L'$ to $C$ is an oriented version of what was called a {\it lamination} in \cite{Gaiotto:2010be} following
\cite{Fock-Goncharov}. A lamination is a collection of paths on $C=\mathbb{CP}^1 \setminus D_P$, which can be either closed
or open with ends on the marked points $P_i$. Each path carries an 
integer weight, subject to the constraint
that paths that carry negative weights must be open and end at two adjacent marked points $P_i$, $P_{i+1}$, and the sum of the weights of paths ending 
at each $P_i$ must be zero.
For example, \autoref{fig:WKB-lamination-A3} illustrates a lamination corresponding to a line defect in the $(A_1, A_3)$ Argyres-Douglas theory.

\insfigsvg{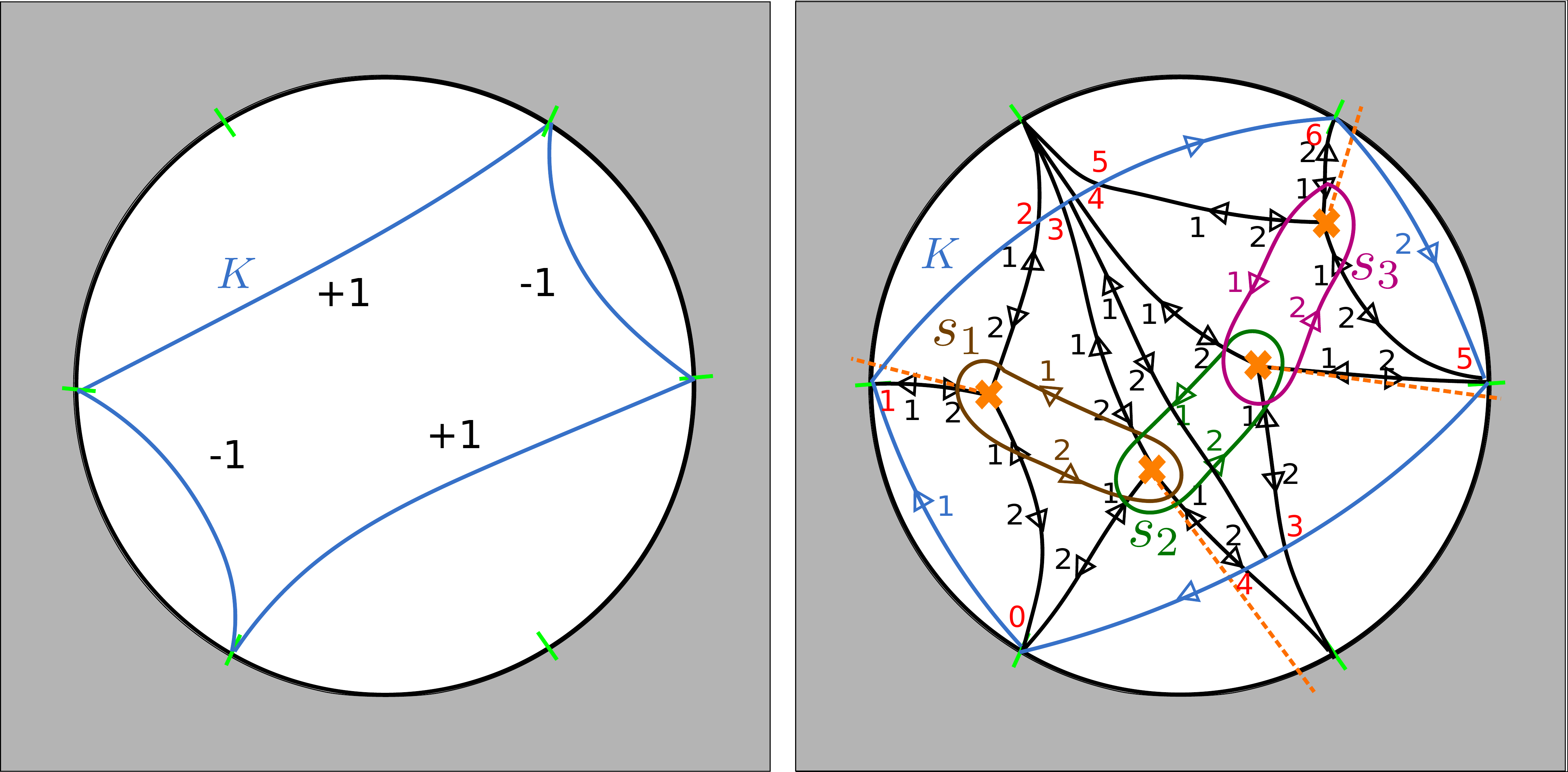}{0.3}{Left: A lamination which corresponds to a line defect in the $(A_1, A_3)$ Argyres-Douglas theory. Here the irregular singularity is placed at infinity. The green markers represent the six marked points $P_i$. Right: Standard projection of a collection $K$ of arcs corresponding to the lamination. The height profile is specified in the same convention as \autoref{fig:flavorline}. The open paths carrying $-1$ weight have a fixed lift, as indicated in blue. We also show a choice of basis $\big\{[\gamma_1],[\gamma_2],[\gamma_3]\big\}$ for $H_1(\tC,\Z)$ with representative cycles $s_1$ (brown), $s_2$ (green) and $s_3$ (purple).}

Now we state the extra ingredients in our $q$-nonabelianization rules to accommodate the presence of irregular singularities.
For simplicity we just consider the case of laminations where all weights
on paths are $\pm 1$.
\begin{itemize}
	\item Each open path carrying the weight $-1$ has to be lifted to 
	a specific sheet of $\tM$: its orientation has to match with the 
	orientation of the leaves near the boundary circle. 
	In \autoref{fig:WKB-lamination-A3} we show the lifts for the two open paths with $-1$ weight in that lamination.
	\item We enumerate all possible lifts $\tL'\subset \tM$ subject to
	these lifting constraints. Each $\tL'$ is a link in $\tM$, meeting $S^1_P\times \R$ at a finite number of points. We then apply our $q$-nonabelianization rules as usual, except that for computing $\alpha(\tL')$, we do not include winding factors or framing weights for the lifts
	of paths carrying weight $-1$.
\end{itemize}

Applying these rules to the line defect in \autoref{fig:WKB-lamination-A3}, we obtain the result:
\begin{equation}
F([K])=X_{\gamma_1}+X_{\gamma_1+\gamma_2}+X_{-\gamma_3}-(q+q^{-1})X_{\gamma_1-\gamma_3}+X_{-\gamma_2-\gamma_3}+X_{\gamma_1-\gamma_2-\gamma_3}+X_{\gamma_1+\gamma_2-\gamma_3}.
\end{equation}
This line defect was also studied in \cite{Gaiotto:2010be}; in particular the classical version of its generating function was given in (10.15) of that paper. Our $q$-nonabelianization computation refines the framed BPS state index from $2$ to $-(q+q^{-1})$ in the charge sector $\gamma_1-\gamma_3$. 

Let us take a closer look at the term $-(q+q^{-1})X_{\gamma_1-\gamma_3}$, which turns out to be the sum of contributions from three lifts $\tK_1$, $\tK_2$ and $\tK_3$ shown in \autoref{fig:A3lifts}.  
\insfigsvg{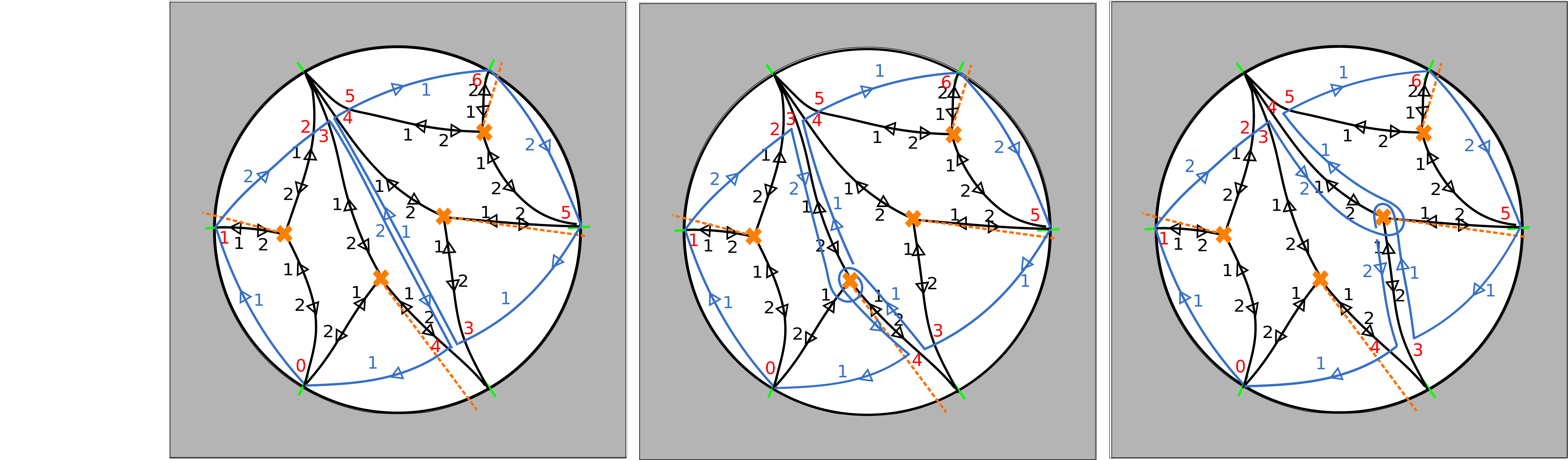}{0.21}{The standard projection of three lifts $\tK_1$ (left), $\tK_2$ (middle) and $\tK_3$ (right) whose contributions sum up to $-(q+q^{-1})X_{\gamma_1-\gamma_3}$. Here we have denoted the sheet numbers of lifted strands in blue.}

For all three lifts $\tK_i$, there are four places where the standard projection of a strand with $+1$ weight is tangent to a WKB leaf. The associated local framing weight factors cancel out, so in the end each $\tK_i$ has a trivial framing factor. The winding numbers of $\tK_1$, $\tK_2$ and $\tK_3$ are $-1$, $0$ and $-1$ respectively. Additionally $\tK_1$ has an exchange factor of $q(q^{-1}-q)$, $\tK_3$ has a detour factor of $q$, and $\tK_2$ has trivial exchange and detour factors. In summary the total weights associated with the lifts are:
\begin{equation}
\alpha(\tK_1)=q^{-1}-q,\quad
\alpha(\tK_2)=1, \quad
\alpha(\tK_3)=1.
\end{equation}
Using the sign rules introduced in \S\ref{sec:quantumtorus}, we get
\begin{equation}
[\tK_1]=X_{\gamma_1-\gamma_3},\quad
[\tK_2]=- q^{-1} X_{\gamma_1-\gamma_3},\quad
[\tK_3]=-q^{-1}X_{\gamma_1-\gamma_3}.
\end{equation}
Combining these gives
\begin{equation}
\sum\limits_{i=1}^3 \alpha(\tK_i)[\tK_i]=-(q+q^{-1})X_{\gamma_1-\gamma_3}.
\end{equation}

\section{A covariant version of \texorpdfstring{$q$}{q}-nonabelianization} \label{sec:covariant}

In \autoref{sec:qnab-rules} 
above, we described the $q$-nonabelianization map $F$ in a way that
used the special structure $M = C \times \R$. 
For example, we always chose links with standard framing,
and our explicit formulas for the weight
factors $\alpha(\tL)$ involved the projection $M \to C$.
In this section we reformulate
$F$ in a more covariant way.
This reformulation will be useful for the future 
generalization to other 3-manifolds $M$.
It will also prove to be convenient for the proof that $F$ is a map of
skein modules in \autoref{sec:isotopyproof} below.

Given a framed link $L$ in $M$, we again write
\begin{equation}
	F([L]) = \sum_{\tL} \alpha(\tL) [\tL]
\end{equation}
where $\tL$ runs over links in $\tM$ built from the same kinds of 
pieces as in \autoref{sec:qnab-rules} as above. Now, however, we need
to explain what framings we use.

\begin{itemize}
\item Direct lifts of portions of $L$ to $\tM$: these 
are equipped with a framing which just lifts the framing of
$L$, using the projection to identify the tangent spaces to $M$ and $\tM$.

\item Detours: each 
detour is equipped with a distinguished framing, as follows.
Let $p$ denote the point where the link $L$ crosses the 
critical leaf. At $p$ we have two 
distinguished tangent directions: the tangent $t_w$ 
to the critical leaf (oriented toward the branch point)
and the tangent $t_L$ to $L$.
Define $f_p := t_w \times t_L$.
Without loss of generality, we may assume that the framing of $L$ at $p$
is given by $f_p$; then the framing of $\tL$ as we approach $p$
will also be given by $f_p$.
(More generally if the framing of $L$ is given by some $f \neq f_p$,
we simply insert a rotation of the framing of
$\tL$ from along some arc $a$ from $f$ to $f_p$ as we approach $p$
along $\tL$,
and insert the opposite rotation from $f_p$ back to $f$ immediately
after we leave $p$ along $\tL$. The homotopy class of the 
resulting framing of $\tL$ is then independent
of the choice of arc $a$, since a change of $a$ cancels out.)
Then, the framing of the lifted detour must start out from
the framing $f_p$ at the beginning of the lifted detour and end again
at $f_p$ at the end of the lifted detour. To get this interpolating 
framing we just choose any trivialization of $TM$ in a neighborhood of 
the detour and then use that trivialization to extend $f_p$.

\item Lifted exchanges:
Let $p$ denote either of the two points 
where the exchange attaches to the link $L$.
At $p$ we have two distinguished tangent directions: the tangent $t_e$ 
to the exchange (oriented away from $p$ towards the exchange)
and the tangent $t_L$ to $L$. Define $f_p := t_e \times t_L$;
without loss of generality, we may assume that the framing of $L$ at $p$
is given by $f_p$ (if not we add interpolating arcs as above.)
The framing of the exchange needs to interpolate between the two
framings $f_p$ at the two ends.
Fortunately the normal bundle to the exchange has a canonical connection,
coming from the fact that the exchange is a leaf of the foliation of $M$.
We can use this connection to identify all the fibers with a single $2$-dimensional 
vector space $E$.
Then a framing of the exchange is a 
path in the circle $E / \R^+$, with its ends on the two points $[f_p]$.
There are two distinguished such paths, one going the ``short way''
around the circle, the other going the ``long way'' around.
We allow either of these framings; whichever one we choose,
we use it on both sheets of the lift.
(Thus
when we consider a diagram involving $n$ exchanges,
the sum over $\tL$ includes $2^n$ different terms
associated to this diagram, differing only in their framing.) 
\end{itemize}

The weight factor $\alpha(\tL)$ is given as follows:
\begin{itemize}
	\item 
	At every place where the framing 
	of $\tL$ is tangent to a leaf of the foliation of $M$, 
	we include a contribution $q^{\pm\frac{1}{2}}$ to
	$\alpha(\tL)$, with the sign determined as follows.
	We consider the normal planes to $\tL$ near the point
	where the tangency occurs. These planes are naturally oriented
	(since $\tL$ and $\tM$ are), and each contains two vectors,
	one given by the framing, the other by the foliation. 
	At the point of tangency these two vectors coincide;
	as we move along $\tL$ in the positive direction,
	the framing vector rotates across the foliation vector,
	either in the positive or negative direction.
	The factor is $q^{\half}$ in the positive case,
	$q^{-\half}$ in the negative case. 
	
	To illustrate how this works, we revisit the first case in \autoref{fig:framing-factor-gl2}. 
	In that case the link $\tL$ was taken to carry standard framing. In \autoref{fig:framing-factor-covariant} we show the framing vector and foliation vector in the normal plane to $\tL$ near the tangency point. The framing vector rotates across the foliation vector in the positive direction, giving the framing factor $q^{\frac{1}{2}}$.
	
	\insfigsvg{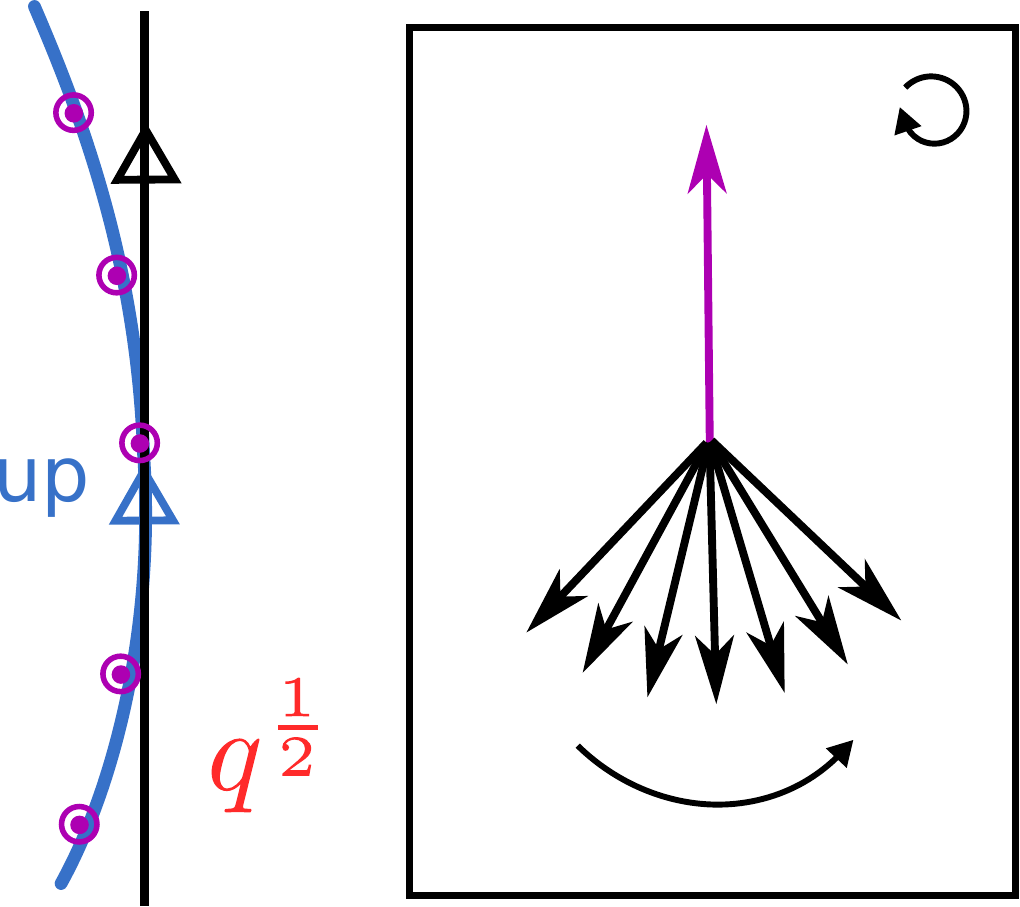}{0.35}{An example of the framing factor contribution to $\alpha(\tL)$ for $\tL$ in standard framing. The framing vector is shown in purple while the foliation vector is shown in black.}

	\item For each lifted exchange, as we have explained, there are two 
	possible framings: one going the ``short way'' and one going the ``long way.''
	When the framing goes the ``long way'' we include an extra
	factor $-1$ in $\alpha(\tL)$.

	\item Finally, the winding contribution
	$q^{w(\tL)}$ to $\alpha(\tL)$ is defined just as in \autoref{sec:qnab-rules}
	(this definition was already covariant.)

\end{itemize}

This completes our description of the covariant rules.
There are two special cases worth discussing:

\begin{itemize}
\item One can check directly that if we apply these covariant rules in the case
where $L$ has standard framing, we recover the rules of \autoref{sec:qnab-rules}
above.
(When we frame a lifted exchange, 
since we apply the same framing to the lifts on both sheets, 
the terms using the ``short way''
and ``long way'' differ by a total of $2$ units of framing.
Thus altogether the two choices of framing give lifts differing by a factor 
$-q^2$. This is the origin of
the exchange factors $\pm(q - q^{-1})$ in the rules
of \autoref{sec:qnab-rules}.)

\item Suppose $M = \R^3$ with the foliation in the $x^3$-direction.
Then we can consider equipping $L$ with ``leaf space blackboard framing,'' i.e.
choose the framing vector to point in the $x^3$-direction.
This framing is well defined provided that $L$ is nowhere parallel to the 
$x^3$-direction. 
There is a slight technicality: our covariant
rules cannot be applied directly because the framing is not generic enough.
We perturb by rotating the framing everywhere very slightly in (say) the
clockwise direction.
After so doing, the covariant rules reduce
exactly to the $N=2$ state sum rules of \autoref{sec:state-sum-details}.
In particular:
\begin{itemize}
\item The framing factor in the covariant rules is trivial
(there are no points where the framing is tangent to the foliation.)
\item The process of evaluating $[\tL] \in \Sk(\tM, \fgl(1))$ 
reduces to computing the self-linking number, 
which is given by a product over 
crossings, using the weights appearing in the first column of \autoref{fig:state-sum-formulas}.
\item The sum over framings for an exchange
produces the $q - q^{-1}$ or $q^{-1} - q$ appearing in the 
last column of \autoref{fig:state-sum-formulas}. 
For an example of how this works, see \autoref{fig:exchange-factor-covariant} below. 

\insfigsvg{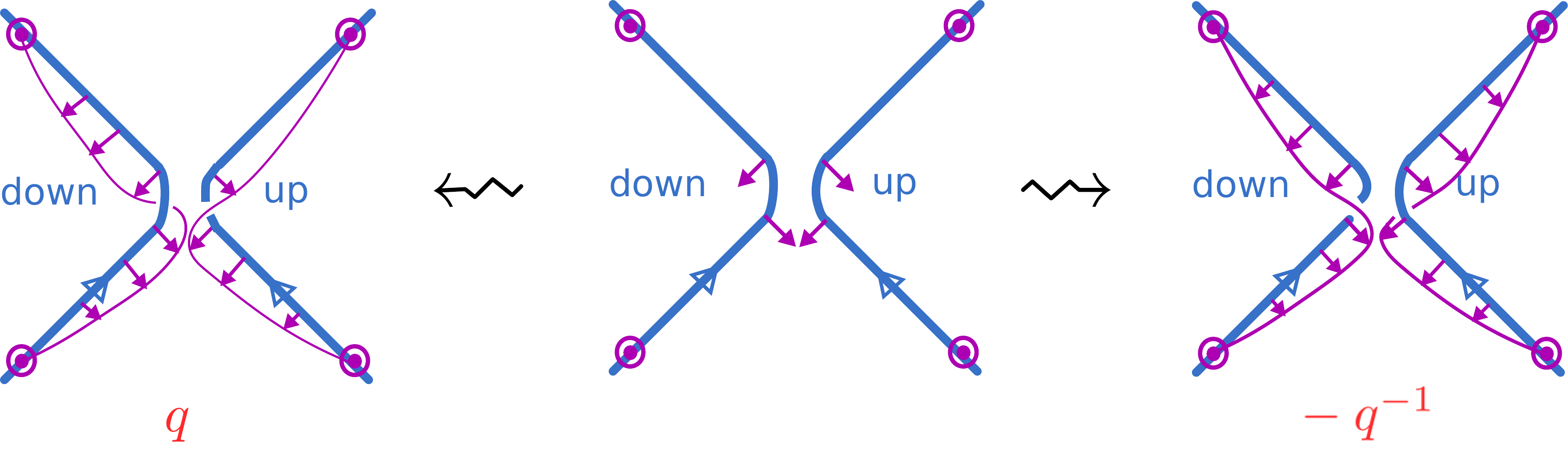}{0.35}{The sum over two framings for a lifted exchange,
with the original link in blackboard framing. Middle: the blackboard
framing far from the exchange, and the framings $f_p$ near the exchange.
Left: framing interpolated the ``short way.'' This framing differs from the blackboard framing by a factor of $q$. Right: framing interpolated the ``long way.'' This framing differs from the blackboard framing by a factor of $q^{-1}$. Summing over the two framings
gives the factor $q - q^{-1}$.}

\item The winding as defined in the covariant rules reduces to the
winding of the link projection as used in the state sum rules.
\end{itemize}
\end{itemize}

\section{Isotopy invariance and skein relations}\label{sec:isotopyproof}

In this section we give a sketch proof that $F$ is well defined, i.e. that
 $F(L)$ really only depends on the class $[L] \in \Sk(M,\fgl(N))$.

The basic strategy is as follows. First we need to prove that $F(L)$
depends only on the framed isotopy class of $L$. 
For this purpose we use the covariant formulation of $F$,
which we described in \autoref{sec:covariant}. We check first
that $F$ is covariant under changes of framing. Next we turn 
to the question of isotopy. From the covariant rules it follows immediately 
that $F(L)$ is invariant under any isotopy which does not create or destroy
exchanges and which leaves $L$ untouched in some small neighborhood of the
critical leaves. What remains is:
\begin{itemize}
	\item To deal with processes in which exchanges are created or destroyed; since exchanges correspond to crossings in the leaf space projection, this boils down to checking invariance under Reidemeister moves for that projection. We check this
	in \autoref{sec:isotopy-noncritical} below.
	\item To deal with processes which change $L$ in a neighborhood of a critical leaf;
	here again a small isotopy does not change $F(L)$, but
	there are several Reidemeister-like moves which have to be considered, which
	create or destroy detours, or move detours across one another. We check 
	invariance under these 
	in \autoref{sec:isotopy-critical} below.
\end{itemize}
In each case we are free to choose any convenient framing; in practice the way
we implement this is to choose a profile for the standard projection of $L$,
and then use standard framing. Then for the actual computations we can use the
concrete rules of \autoref{sec:qnab-rules}.

After this is done, we check that $F$ preserves the skein relations.
This check is simplified by the fact that we have already verified
isotopy invariance, so we are free to put the link $L$ in a simple position
relative to the WKB foliation.

\subsection{Changes of framing}

Suppose $L$ and $L'$ are two links in $M$ which differ only by a homotopy of the framing.
Then we have $[L] = [L']$ in $\Sk(M,\fgl(2))$, and thus we must have $F(L) = F(L')$.
This is relatively straightforward to check: as we vary the framing, the framing contributions
$q^{\pm \frac12}$ to $F(L)$ appear and disappear in cancelling pairs, while all other 
contributions vary continuously, so $F(L)$ is constant.

\subsection{Isotopy invariance away from critical leaves} \label{sec:isotopy-noncritical}

With an eye towards future generalization to $N>2$, we use the notion of $ij$-leaf space; in the case of $N=2$ we simply have $\{i,j\}=\{1,2\}$. 

In the following we denote the open strand configurations before and after each isotopy as $l_A$ and $l_B$ respectively. Correspondingly we denote its lifts before and after the isotopy as $\widetilde{l_A^\kappa}$ and $\widetilde{l_{B}^\kappa}$, where $\kappa$ runs from $1$ to the number of lifts.

\subsubsection{The first Reidemeister move}

In the following we consider the first Reidemeister move as shown in \autoref{fig:RM1-2}.

\insfigsvg{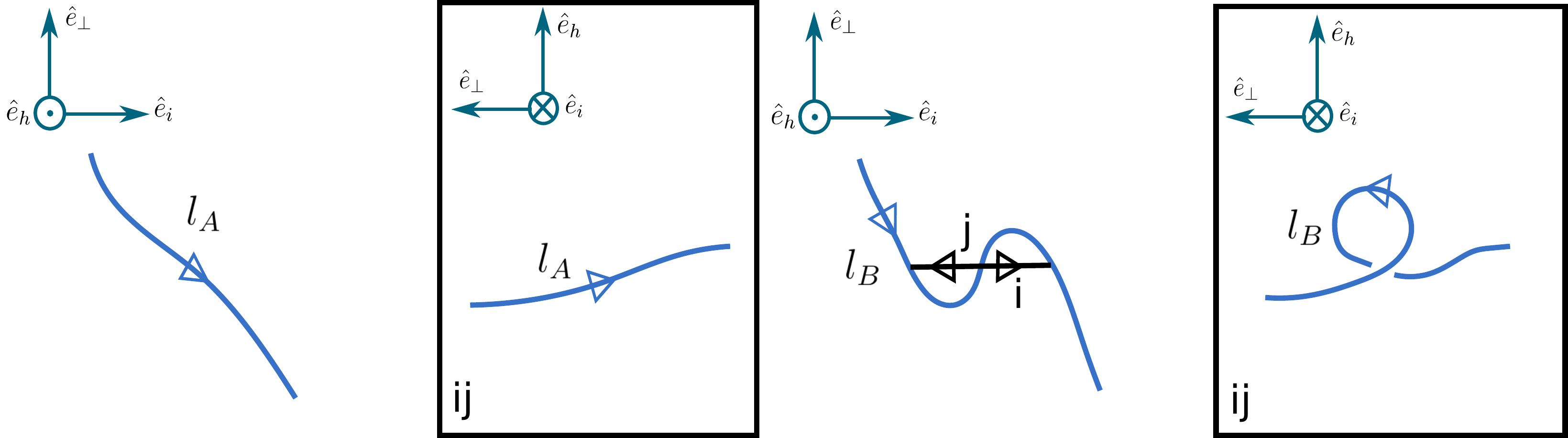}{0.45}{The first Reidemeister move. Here $\hat{e}_h$ is the unit vector along the height direction and $\hat{e}_i$ is the unit vector along the $i$-orientation of $ij$-leaves.}

\insfigsvg{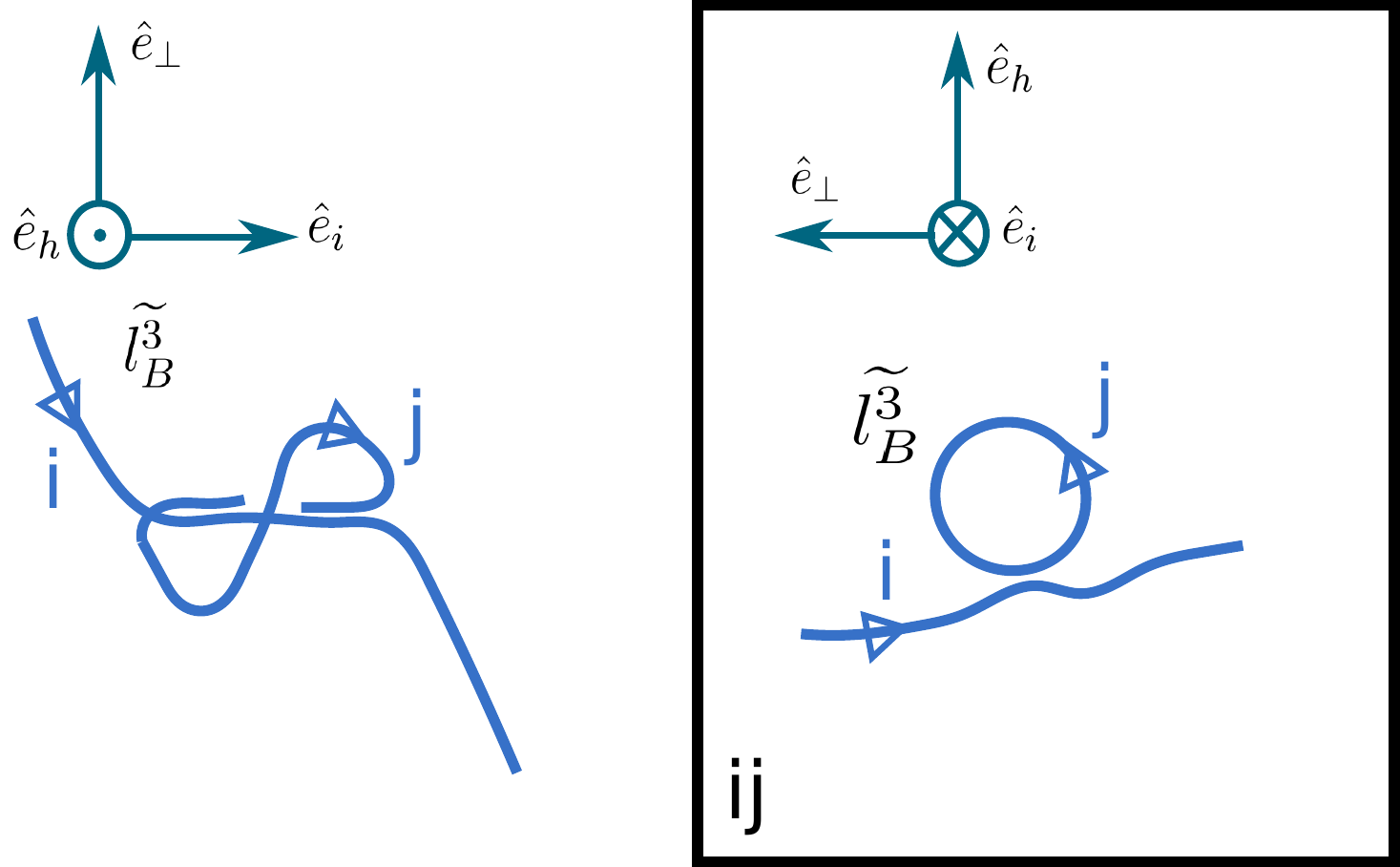}{0.5}{Lift of $l_B$ that contains an exchange.}

$l_A$ has two direct lifts to sheet $i$ and sheet $j$, denoted as $\widet{l_A^1}$ and $\widet{l_A^2}$ respectively. $l_B$ has two direct lifts $\widet{l_B^1}$ and $\widet{l_B^2}$, plus one lift $\widet{l_B^3}$ containing an exchange as shown in \autoref{fig:RM1-2-lifts}. Lifts of $l_A$ carry no framing factor while lifts of $l_B$ have a framing factor $q$. Denoting the winding of $\widet{l_A^1}$ as $w$, we have 
\begin{align*}
F([l_B])&=q^{(w+1)+1}[\widet{l_A^1}]+q^{-(w+1)+1}[\widet{l_A^2}]+(q^{-1}-q)q^{(w-1)+1}q[\widet{l_A^1}]\\
&=q^{w}[\widet{l_A^1}]+q^{-w}[\widet{l_A^2}]=F([l_A]).
\end{align*}
Here we have used $[\tl_B^1]=[\tl_A^1]$, $[\tl_B^2]=[\tl_A^2]$, $[\tl_B^3]=q[\tl_A^1]$ in $\Sk(\tM, \fgl(1))$.

\subsubsection{The second Reidemeister move}\label{2ndRM}

\insfigsvg{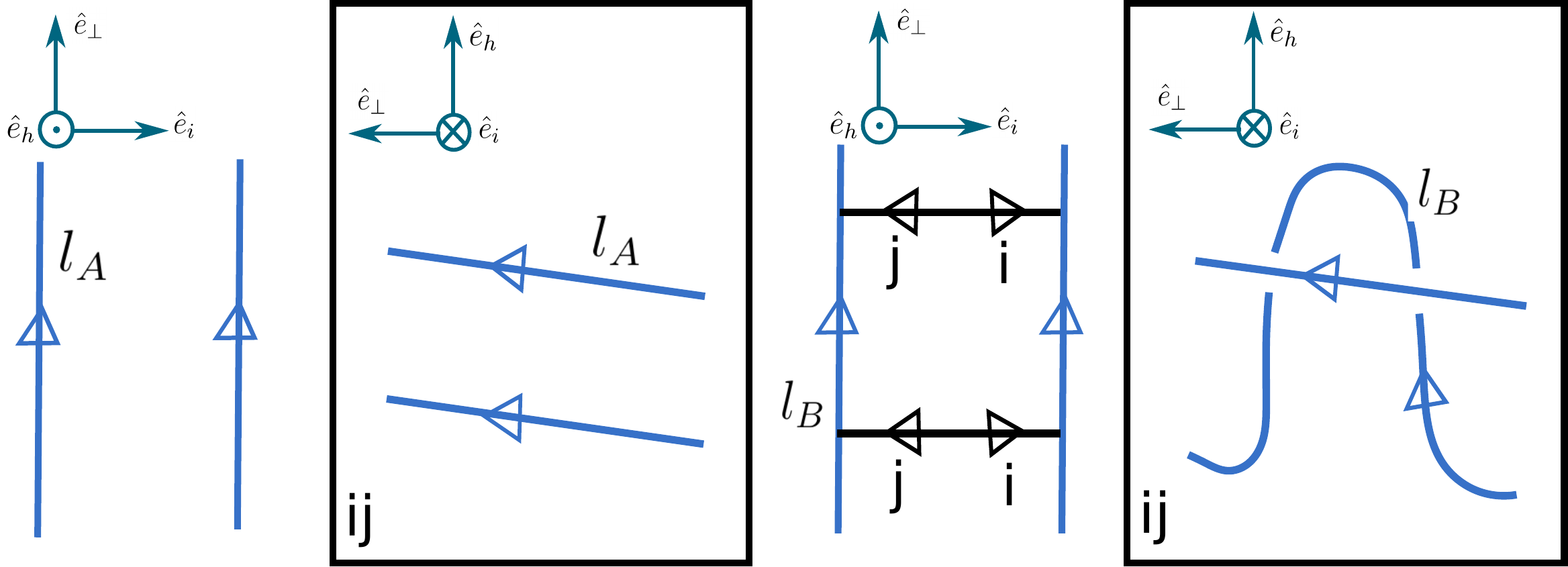}{0.5}{The second Reidemeister move.}

In \autoref{fig:RM2-1-0} we illustrate the second Reidemeister move. Here both $l_A$ and $l_B$ have four direct lifts. It is easy to see that contributions from these direct lifts match with each other. $l_B$ has two extra lifts $\widet{l_B^5}$ and $\widet{l_B^6}$ each containing an exchange,  as shown in \autoref{fig:RM2-1-0-lifts}. So we only need to prove that the contributions from these two lifts cancel with each other. This works out simply because $[\tl_B^5]=[\tl_B^6]$ in $\Sk(\tM, \fgl(1))$, and their weights differ by a minus sign due to the sign difference in exchange factors.

\insfigsvg{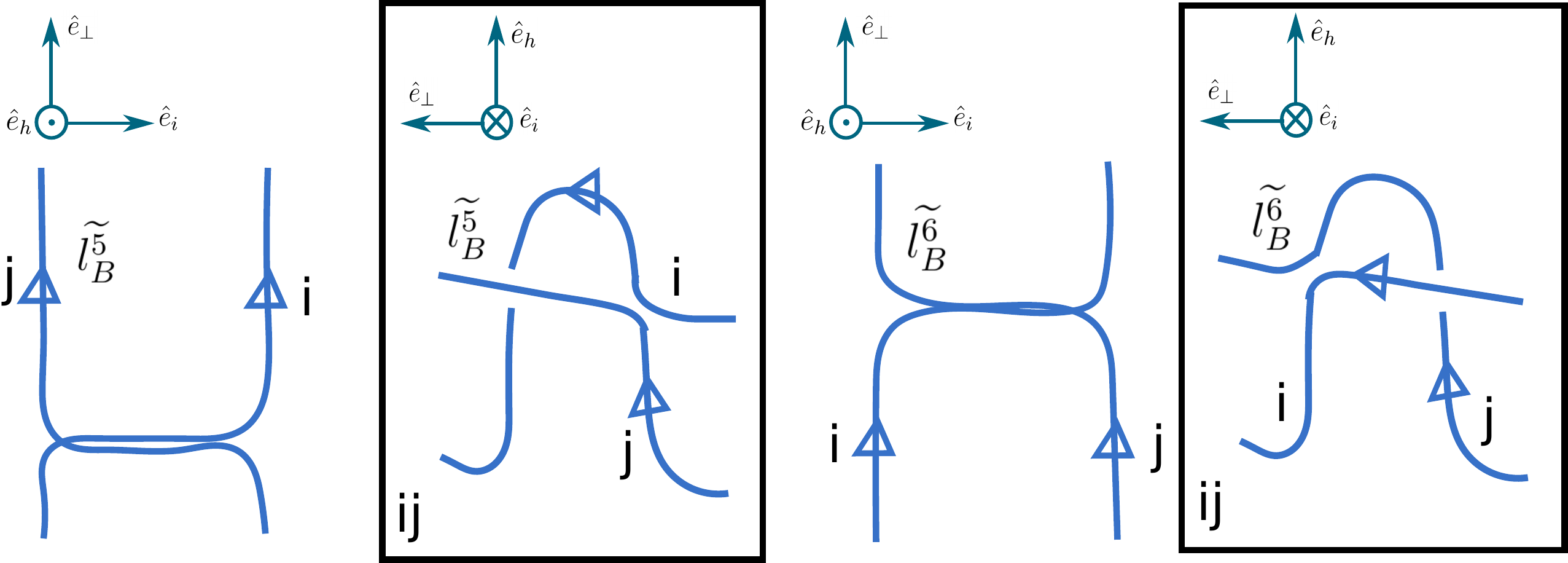}{0.5}{Two lifts of $l_B$ whose contributions cancel each other.}

\subsubsection{The third Reidemeister move}\label{3rdRM}

In this section we consider the third Reidemeister move as shown in \autoref{fig:RM3-1}. 

\insfigsvg{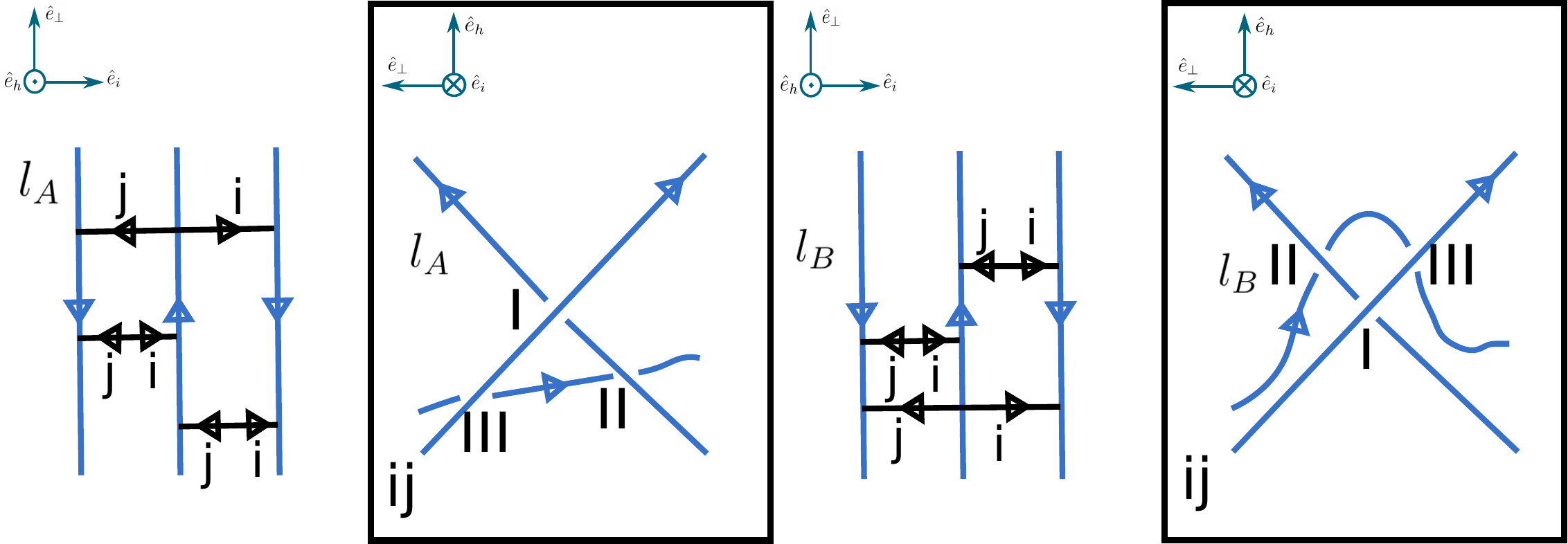}{0.55}{The third Reidemeister move.}

Here $l_A$ and $l_B$ both have fifteen lifts, eight of which are direct lifts. It's easy to match contributions from twelve lifts of $l_A$ and $l_B$ on a term-by-term basis, so we only need to show contributions from the left-over three lifts match with each other. These three lifts are illustrated in \autoref{fig:RM3-1-lifts}, where $\widet{l_{A,B}^{1,2}}$ are lifts containing an exchange at leaf space crossing III and $\widet{l^3_A},\widet{l^3_B}$ are lifts containing two exchanges at leaf space crossings I and II.

\insfigsvg{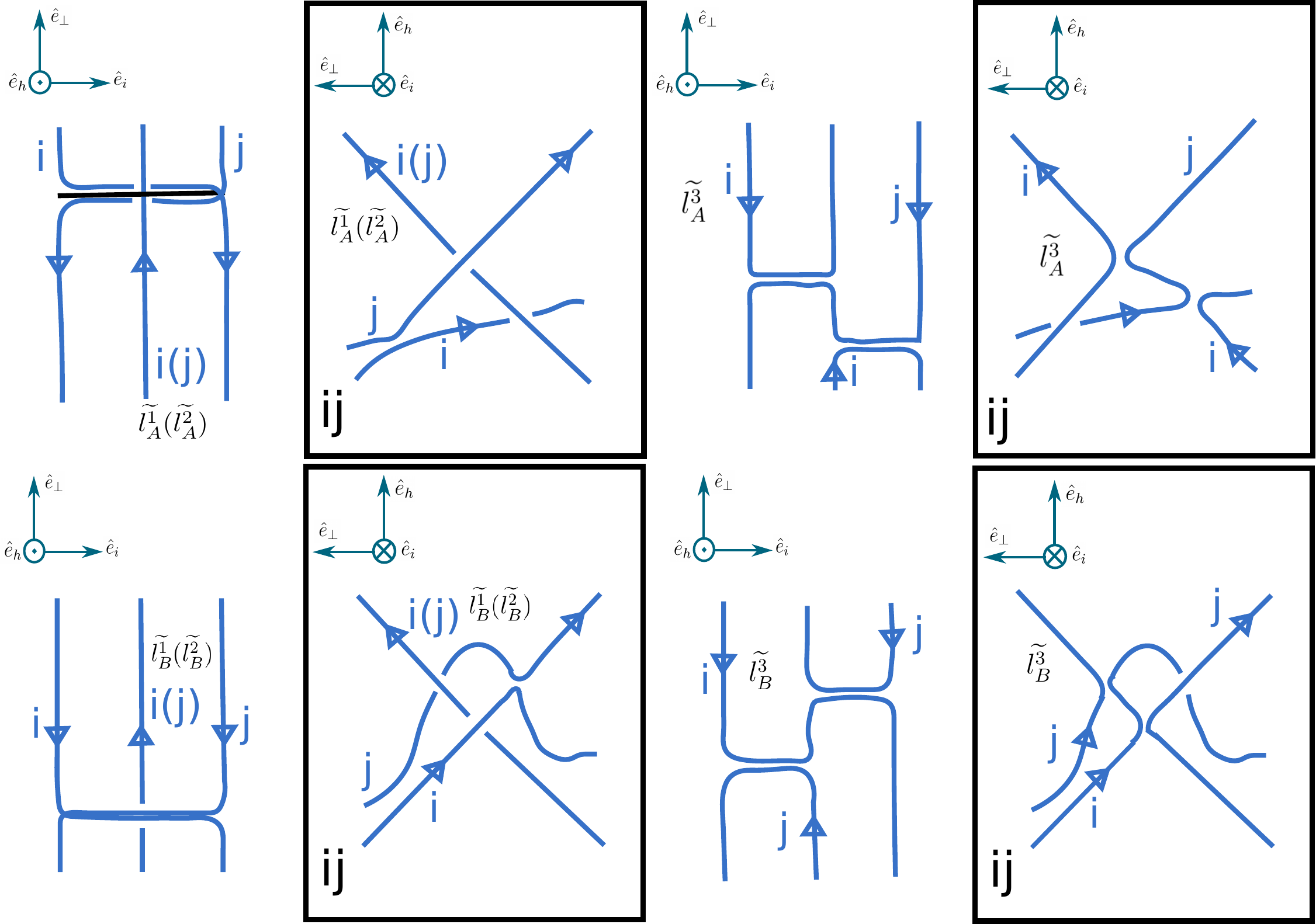}{0.55}{Three lifts of $l_A$ and $l_B$.}

First, using the relations in $\Sk(\tM, \fgl(1))$ we observe that
\begin{equation*}
[\tl_A^1]=q^{-1}[\tl_A^3], \quad [\tl_A^2]=q[\tl_B^3], \quad
[\tl_B^1]=q[\tl_A^3], \quad
[\tl_B^2]=q^{-1}[\tl_B^3].
\end{equation*}

We denote the winding of $\tl_A^3$ and $\tl_B^3$ as $w_1$ and $w_2$. Then the weight factors associated to these six lifts are given as follows:
\begin{align*}
\alpha([\tl_A^1])&=(q^{-1}-q)q^{w_1}, \quad \alpha([\tl_A^2])=(q^{-1}-q)q^{w_2},\quad
\alpha([\tl_A^3])=q(q^{-1}-q)q^{-1}(q-q^{-1})q^{w_1},\\
\alpha([\tl_B^1])&=(q^{-1}-q)q^{w_1}, \quad \alpha([\tl_B^2])=(q^{-1}-q)q^{w_2},
\quad \alpha([\tl_B^3])=q(q^{-1}-q)q^{-1}(q-q^{-1})q^{w_2}.
\end{align*} 
Therefore contributions from these three lifts do match with each other:
\begin{equation*}
\sum\limits_{i=1,2,3}\alpha([\tl_A^i])[\tl_A^i]=q(q^{-1}-q)q^{w_1}[\tl_A^3]+q(q^{-1}-q)q^{w_2}[\tl_B^3]=\sum\limits_{i=1,2,3}\alpha([\tl_B^i])[\tl_B^i].
\end{equation*}

\subsection{Isotopy invariance near critical leaves} \label{sec:isotopy-critical}

In this section we consider isotopies of open strands which involve critical leaves. 
All such isotopies can
be perturbed into combinations of four basic moves, which are
the analogues in this context of the Reidemeister moves.

Each basic move happens in a neighborhood of a branch point with three critical leaves emanating from it. In such a neighborhood, the leaf space of $M$ topologically looks like three pages glued together at a ``binder" corresponding to the branch locus. For convenience, we label these three pages as I, II, III, and illustrate the leaf space projection within each page.

\subsubsection{Moving an exchange across a critical leaf}

\insfigsvg{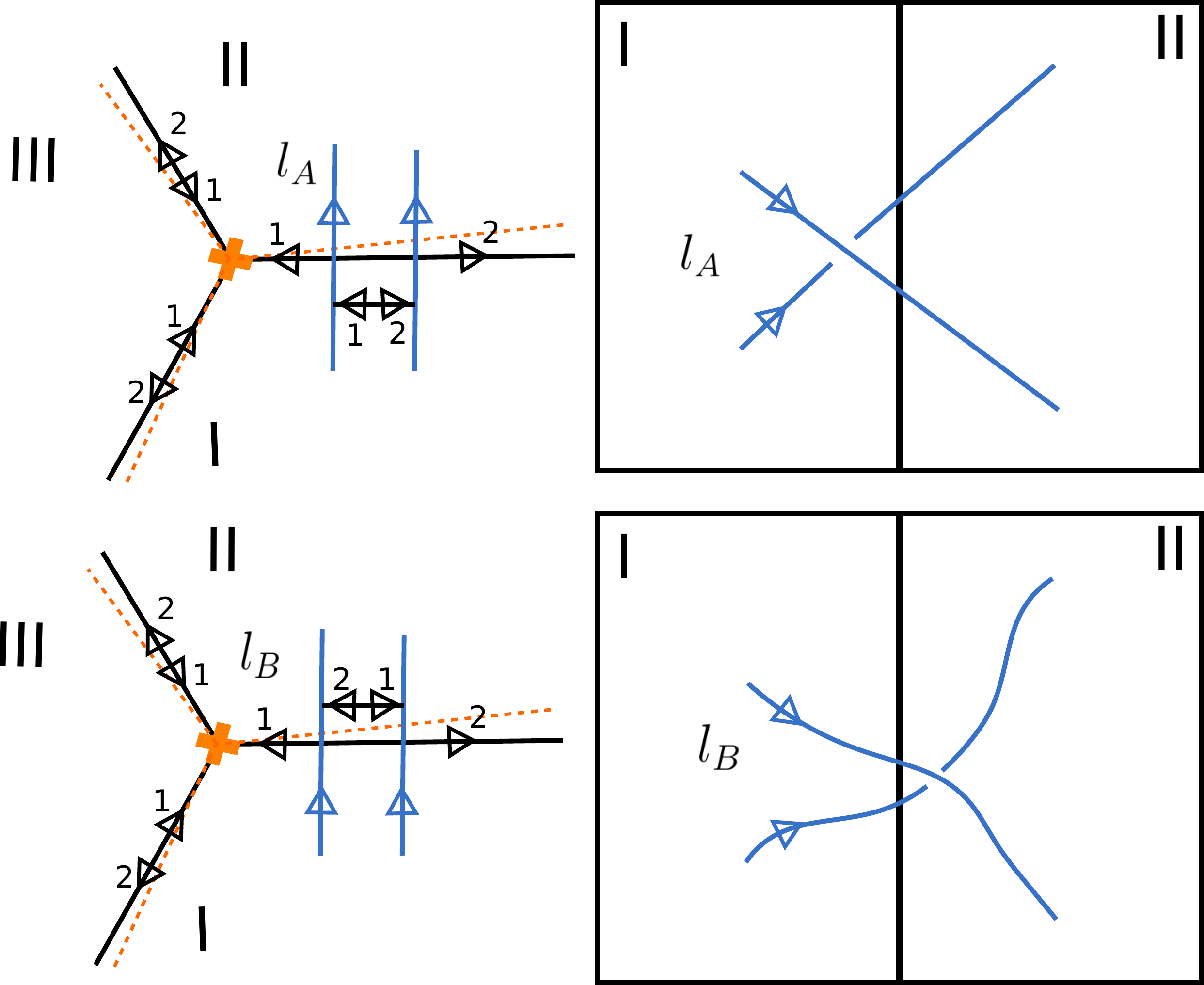}{0.3}{The first basic move in the neighborhood of a 
critical leaf.}

We first consider an isotopy which moves a leaf space crossing from one page to another, as shown in \autoref{fig:crit-1}. There are in total eleven lifts on both sides. The comparison is reduced to matching contributions from the three lifts of $l_A$ and $l_B$ shown in \autoref{fig:crit-1-lifts}.\footnote{Here and below we omit the leaf space projection of the individual lifts.}

\insfigsvg{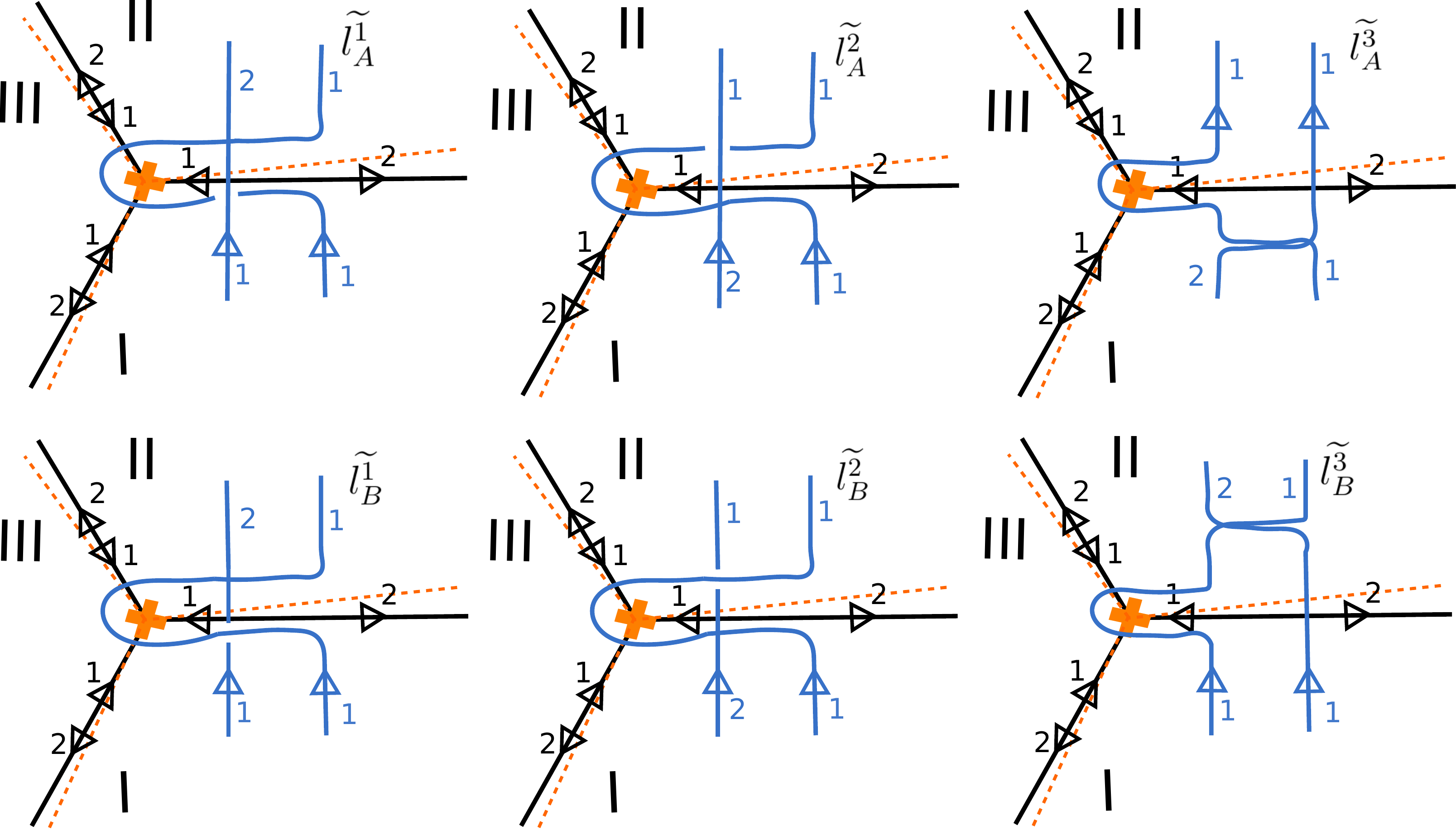}{0.3}{Three lifts of $l_A$ and $l_B$ making the same total contribution.}

We denote the weights of $[\tl_A^1]$ and $[\tl_A^2]$ as $\alpha_1$ and $\alpha_2$. The weights of all the other lifts are then given by:
\begin{equation*}\
\alpha([\tl_B^1])=\alpha_1, \quad
\alpha([\tl_B^2])=\alpha_2, \quad
\alpha([\tl_A^3])=(q-q^{-1})\alpha_2,\quad 
\alpha([\tl_B^3])=(q-q^{-1})\alpha_1.
\end{equation*}
We also have the following relations in $\Sk(\tM, \fgl(1))$:
\begin{equation*}
[\tl_B^3]=q^{-1}[\tl_A^1]=q[\tl_B^1],\quad
[\tl_A^3]=q[\tl_A^2]=q^{-1}[\tl_B^2].
\end{equation*}
Combining these we see that the contributions from these three lifts match on both sides.

\subsubsection{Height exchange for detours}

In \autoref{fig:crit-3} we show another basic move in the neighborhood of a critical leaf. Here $l_A$ has nine lifts while $l_B$ has ten lifts in total. Eight lifts of $l_A$ and eight lifts of $l_B$ match on a term-by-term basis. The remaining two lifts of $l_B$ make the same total contribution as the remaining lift of $l_A$. These terms are illustrated in \autoref{fig:crit-3-lifts}.

\insfigsvg{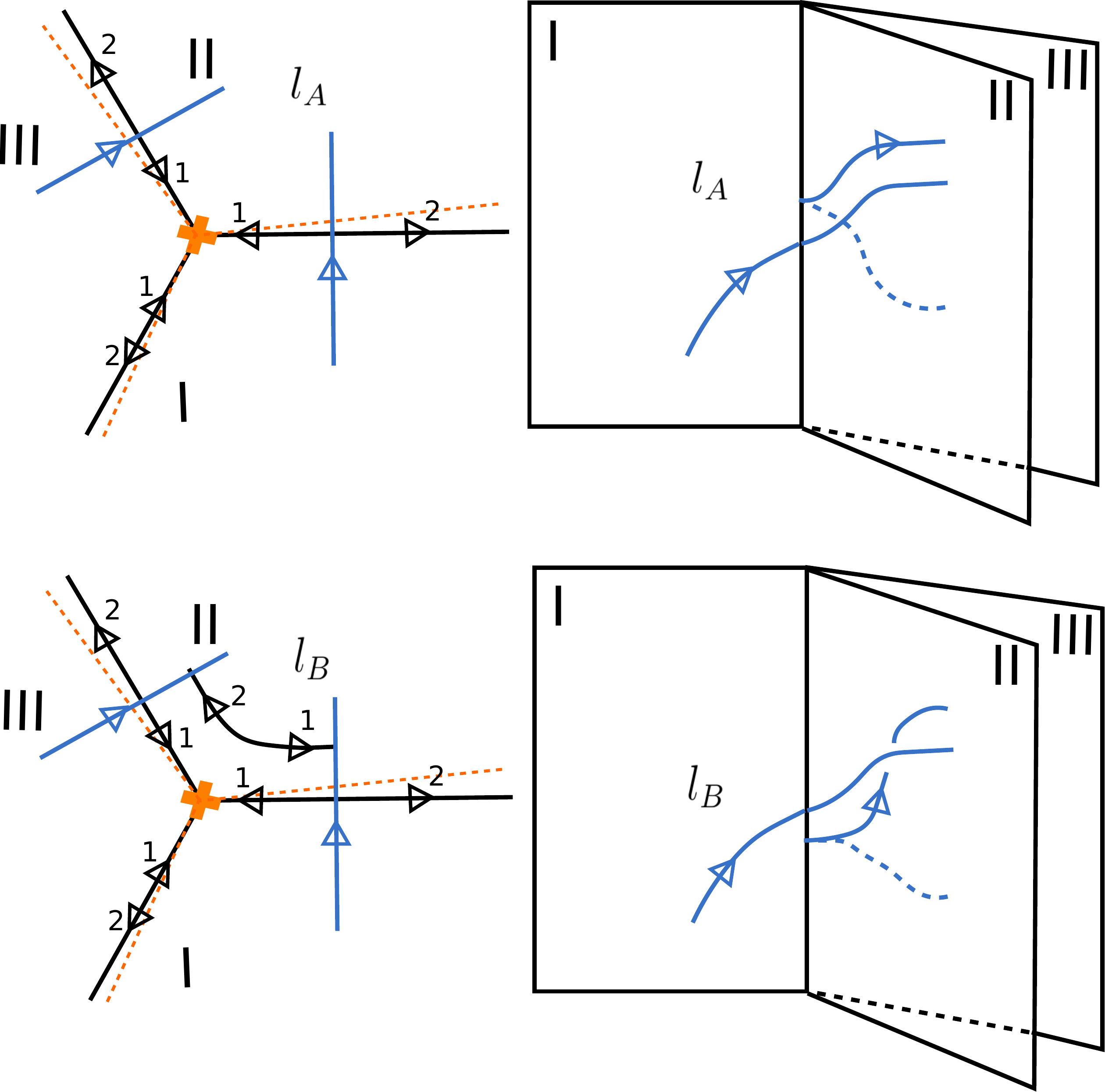}{0.33}{The height exchange for detours.}

\insfigsvg{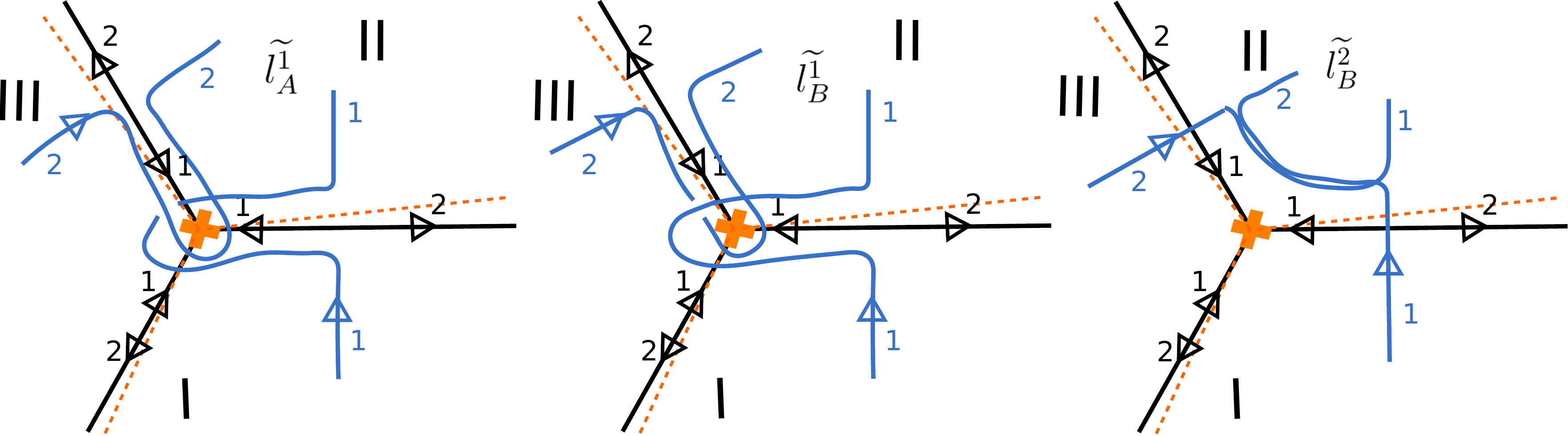}{0.36}{Two lifts of $l_B$ which contribute the same as one lift of $l_A$.}

Taking into account all the local factors, the weights associated to these lifts are related to each other as follows:
\begin{equation*}
\alpha([\tl_B^2])=(q-q^{-1})\alpha([\tl_A^1])=(q-q^{-1})\alpha([\tl_B^1]).
\end{equation*}
We obtain the desired matching using the following relation in 
$\Sk(\tM, \fgl(1))$:
\begin{equation*}
[\tl_A^1]=q[\tl_B^2], \quad [\tl_B^1]=q^{-1}[\tl_B^2],
\end{equation*}

\subsubsection{Moving a strand across a critical leaf}

In this section we consider a basic isotopy where a strand is moved across a critical leaf. This is illustrated in \autoref{fig:crit-4}. Here both $l_A$ and $l_B$ have two direct lifts with matching contributions. $l_B$ has two extra lifts involving detours, shown in \autoref{fig:crit-4-lifts}, whose contributions cancel each other. 

\insfigsvg{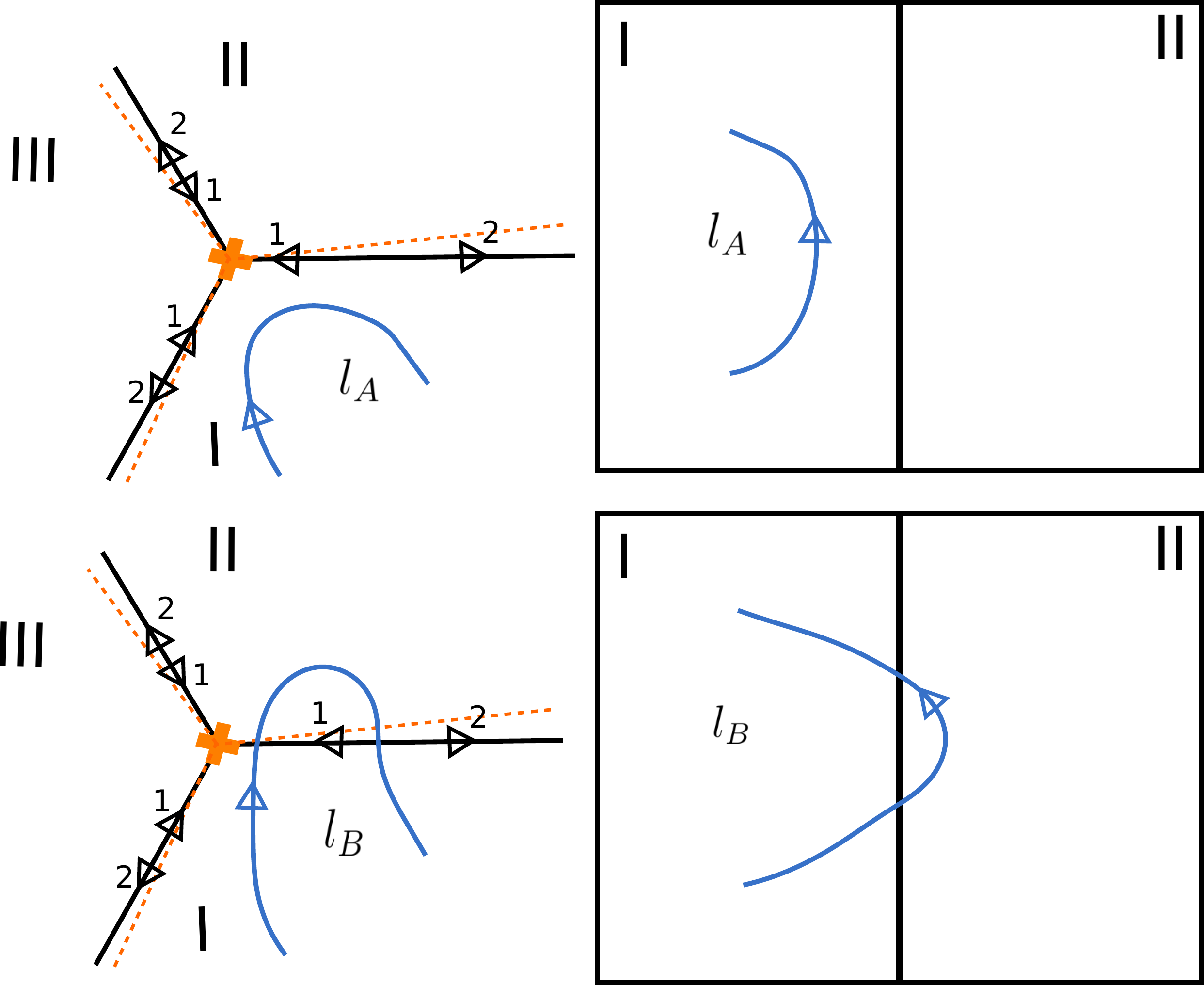}{0.3}{Moving a strand across a critical leaf.}

\insfigsvg{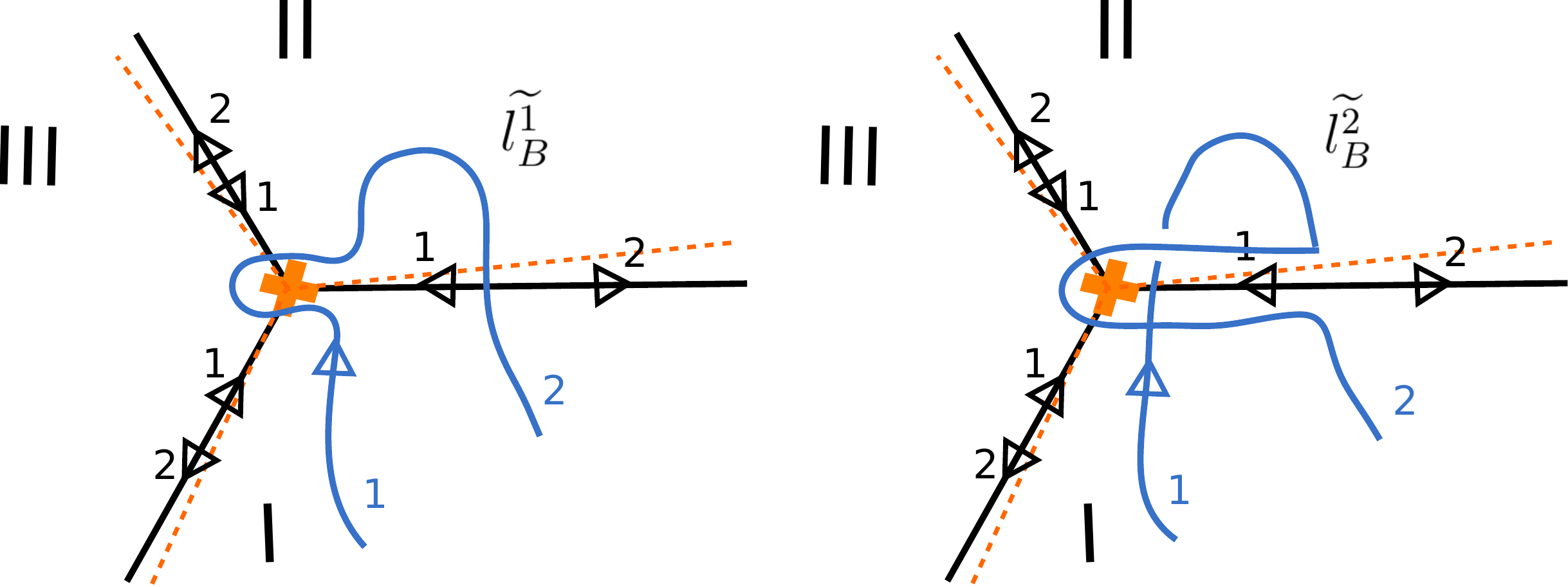}{0.35}{Two lifts of $l_B$ whose contributions cancel each other.}

It is easy to see that $\tl_B^1$ and $\tl_B^2$ have the same winding and framing weight. However, their detour weights are inverse to each other, giving $\alpha([\tl_B^1])=q^{-1}\alpha([\tl_B^2])$. Moreover, in $\Sk(\tM,\fgl(1))$ we have $[\widet{l_B^1}]=-q[\widet{l_B^2}]$, as shown in \autoref{fig:crit-4-0}. Combining these facts gives us the desired cancellation.

\insfigsvg{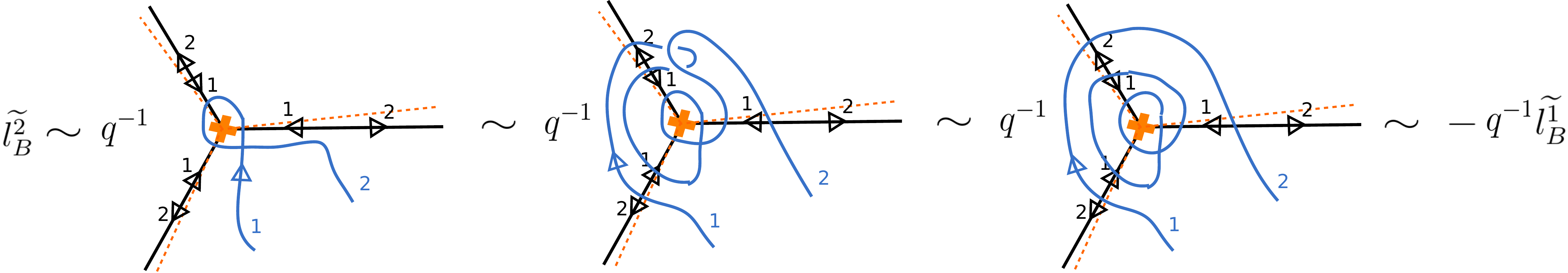}{0.3}{Moves illustrating the relation $[\widet{l_B^2}]=-q^{-1}[\widet{l_B^1}]$.}

\subsubsection{Moving a strand across a branch point}

In this section we consider the basic isotopy where a strand is moved across a branch point. This is illustrated in \autoref{fig:crit-5}.

$l_A$ has three lifts which match three lifts of $l_B$ on a term-by-term basis. These matching lifts are shown in \autoref{fig:crit-5-lifts}. We take the pair $\widet{l_A^1}$ and $\widet{l_B^1}$ as an example. They clearly have the same winding. Moreover $\widet{l_B^1}$ has an extra framing factor $q^{-1/2}$ and an extra detour factor $q^{1/2}$. Therefore $\alpha([\tl_A^1])=\alpha([\tl_B^1])$ and they make the same contribution to $F([l_{A,B}])$.

\insfigsvg{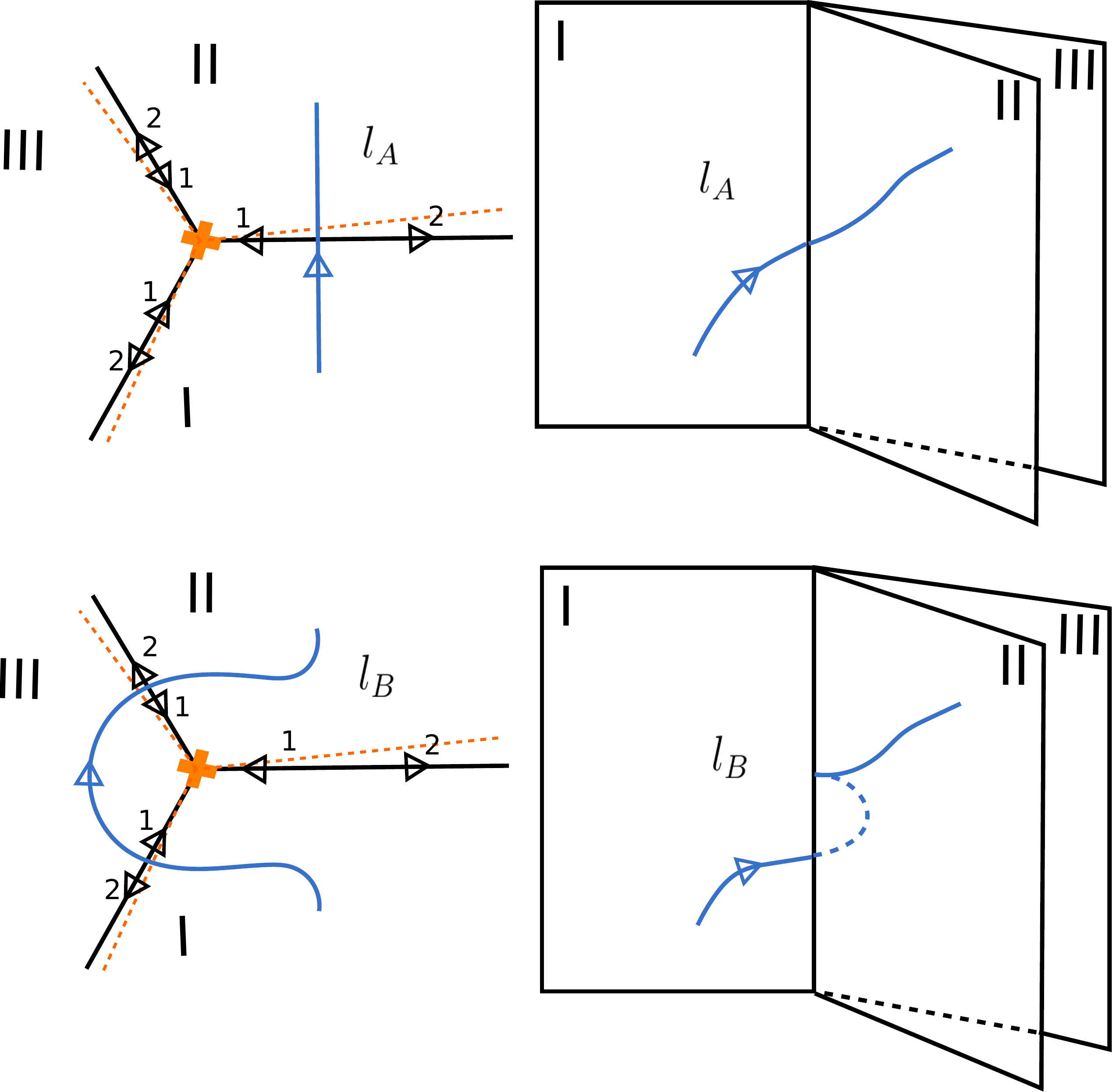}{0.3}{The fifth basic move in a neighborhood of a critical leaf. Here an open strand is moved across a branch point.}
\insfigsvg{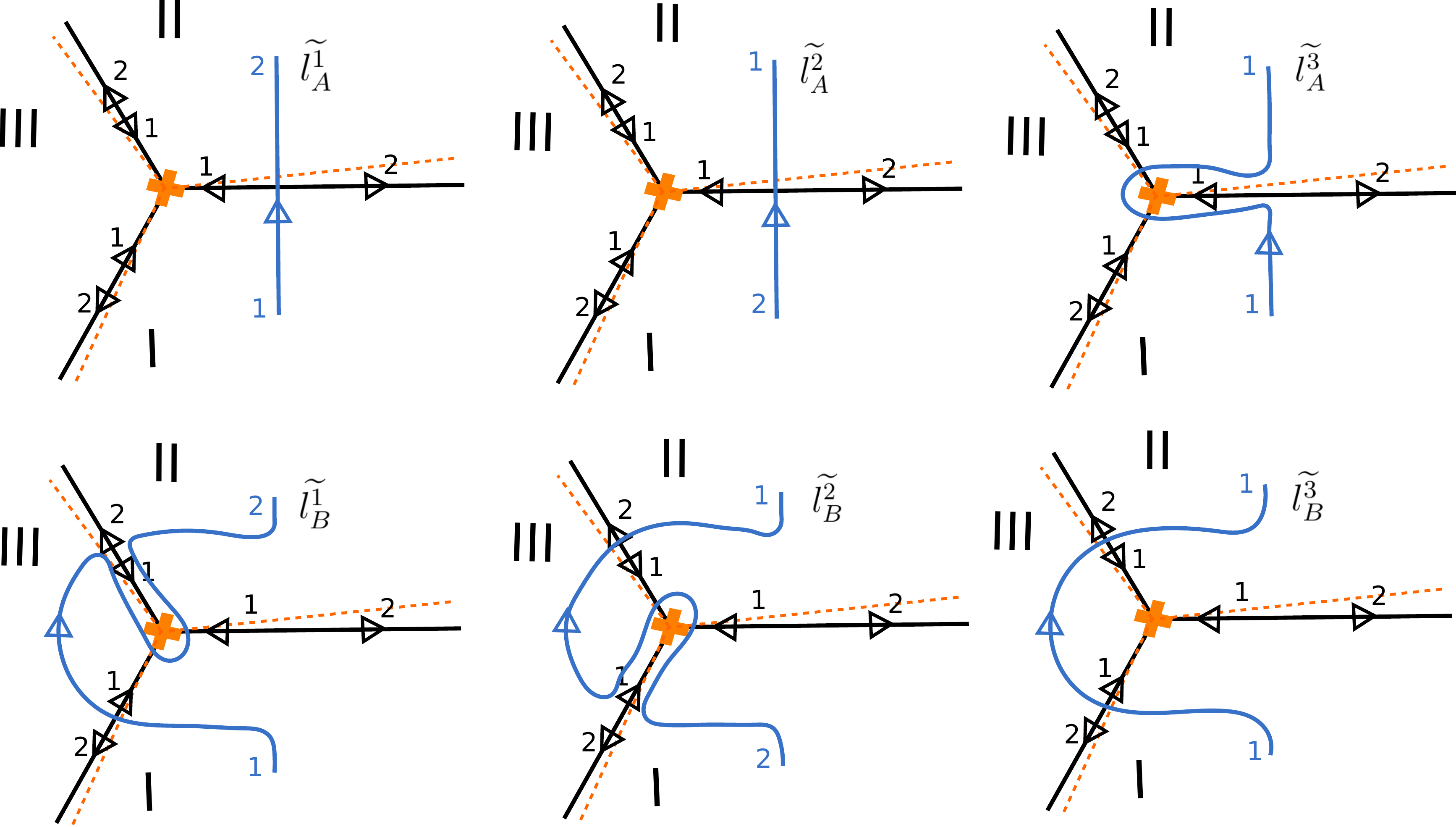}{0.27}{Three lifts of $l_A$ and $l_B$ with matching contributions.}

$l_B$ also has two more lifts whose contributions are supposed to cancel with each other. These two lifts are illustrated in \autoref{fig:crit-5-lifts-1}.

\insfigsvg{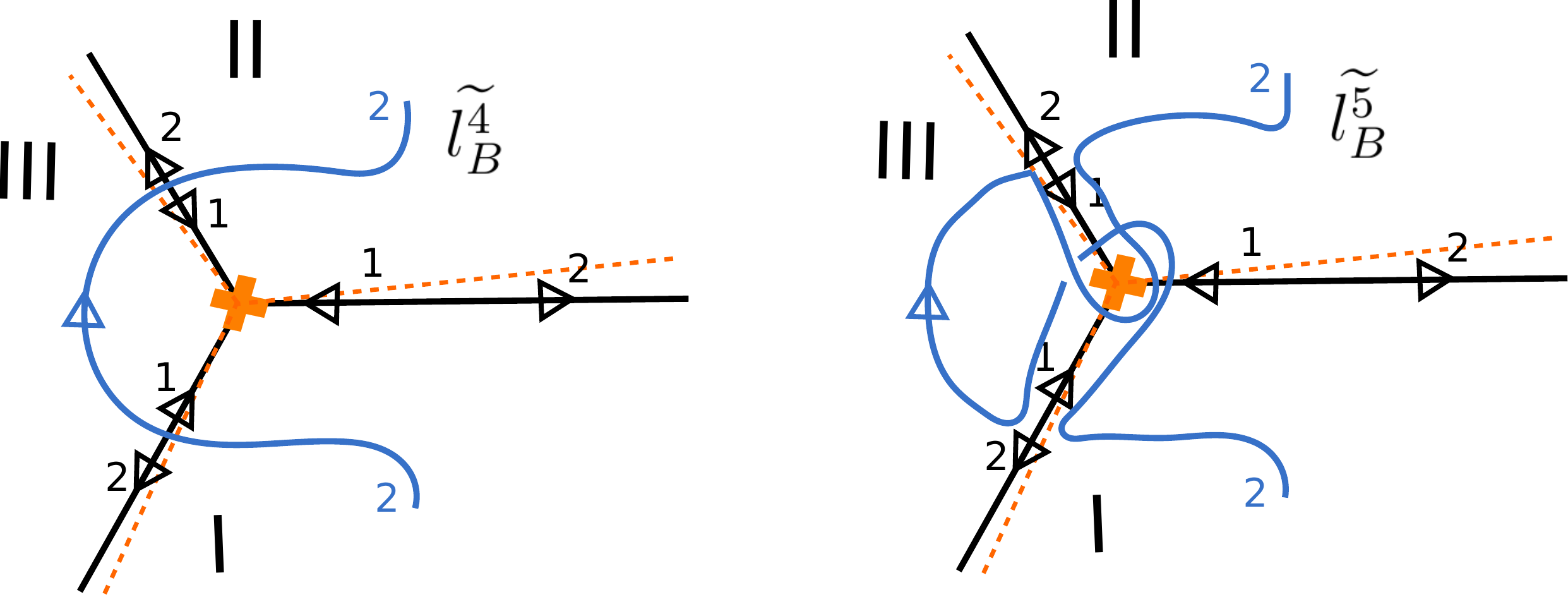}{0.35}{Two lifts of $l_B$ whose contributions cancel with each other.}

All local weights associated to $\widet{l_B^4}$ and $\widet{l_B^5}$ are the same except that $\widet{l_B^5}$ has an extra detour weight factor of $q$. Furthermore in $\Sk(\tM, \fgl(1))$ we have $[\widet{l_B^5}]=-q^{-1}[\widet{l_B^4}]$, as shown in \autoref{fig:crit-5-lifts-2}. Therefore the 
contributions from $\widet{l_B^4}$ and $\widet{l_B^5}$ cancel with each other.

\insfigsvg{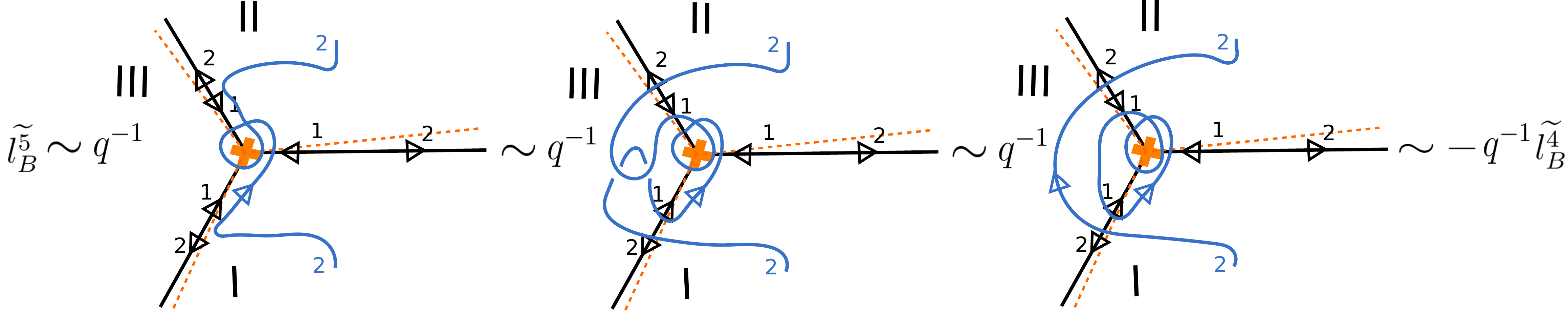}{0.3}{Moves illustrating the relation $[\widet{l_B^5}]=-q^{-1}[\widet{l_B^4}]$.}

\subsection{The skein relations}\label{sec:q-hom}

In this section we prove that the skein relations are preserved by the $q$-nonabelianization map. 

We first look at the skein relation (I) in $\Sk(M,\fgl(2))$. Without loss of generality, we pick a specific direction for the leaves, and monotonic height profiles for the open strands\footnote{We can safely do so as we have already proved the $q$-nonabelianization map is isotopy invariant.}, as shown in \autoref{fig:gl2skeinI12proj}.

\insfigsvg{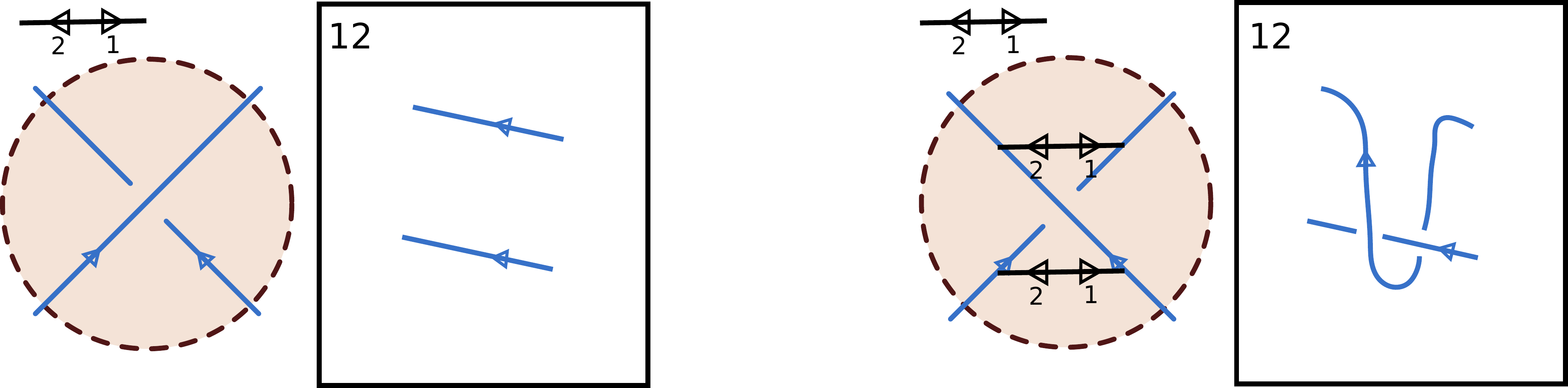}{0.25}{The two terms on the LHS of skein relation (I) in $\Sk(M,\fgl(2))$. On the right, we show the loci where exchanges can occur.}

\insfigsvg{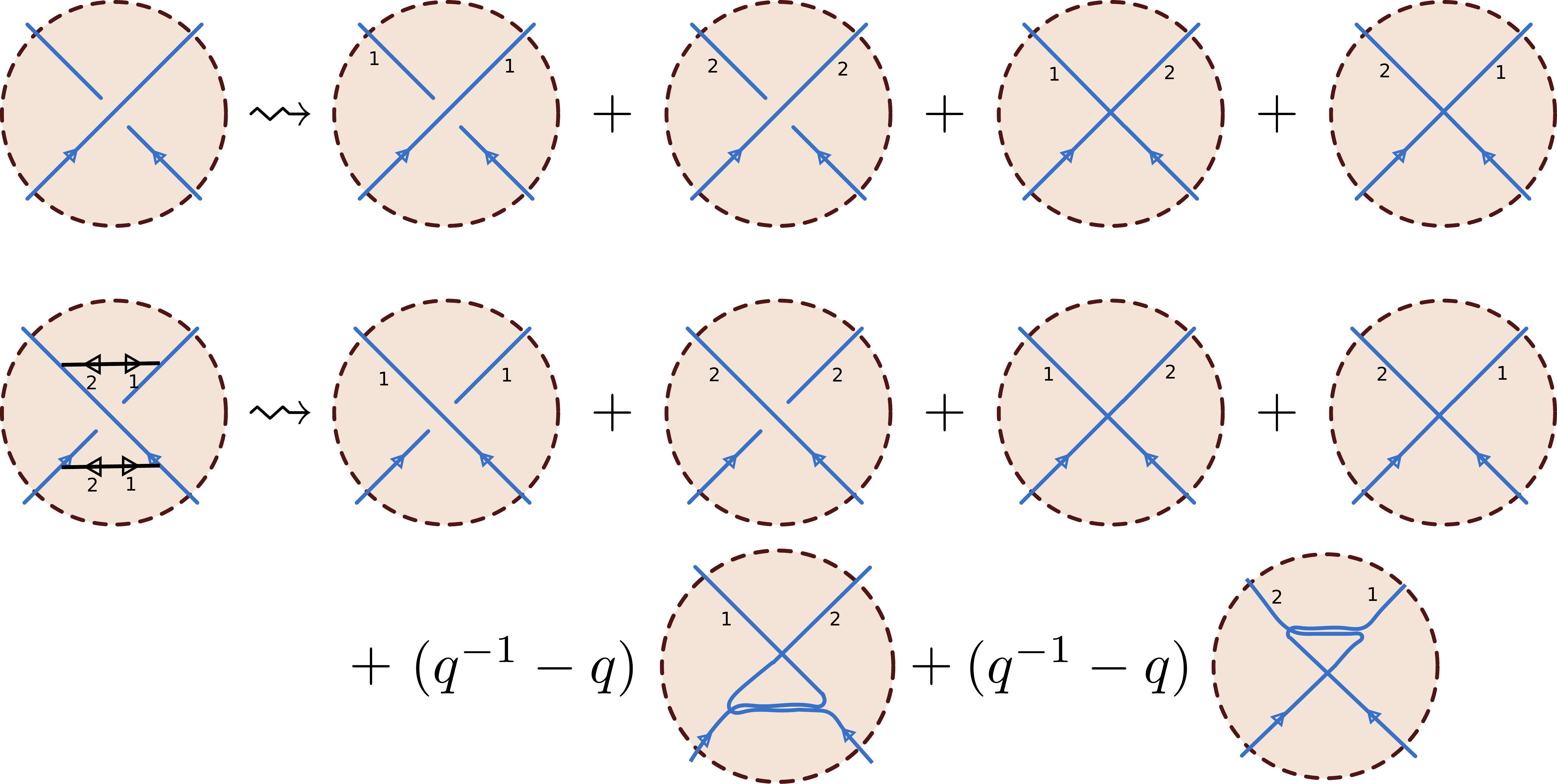}{0.25}{Applying $F$ to the LHS of skein relation (I) in $\Sk(M,\fgl(2))$.}

We show lifts of these two terms in \autoref{fig:gl2skeinI}. In the second term, in addition to the direct lifts, there are two lifts involving exchanges. 
Now subtracting these two sets of lifts, and using the skein relations
of $\Sk(\tM,\fgl(1))$ on the RHS, we get \autoref{fig:gl2skeinIRHS}.
The RHS of that figure is indeed the lift of the right side of
skein relation (I), as desired.

\insfigsvg{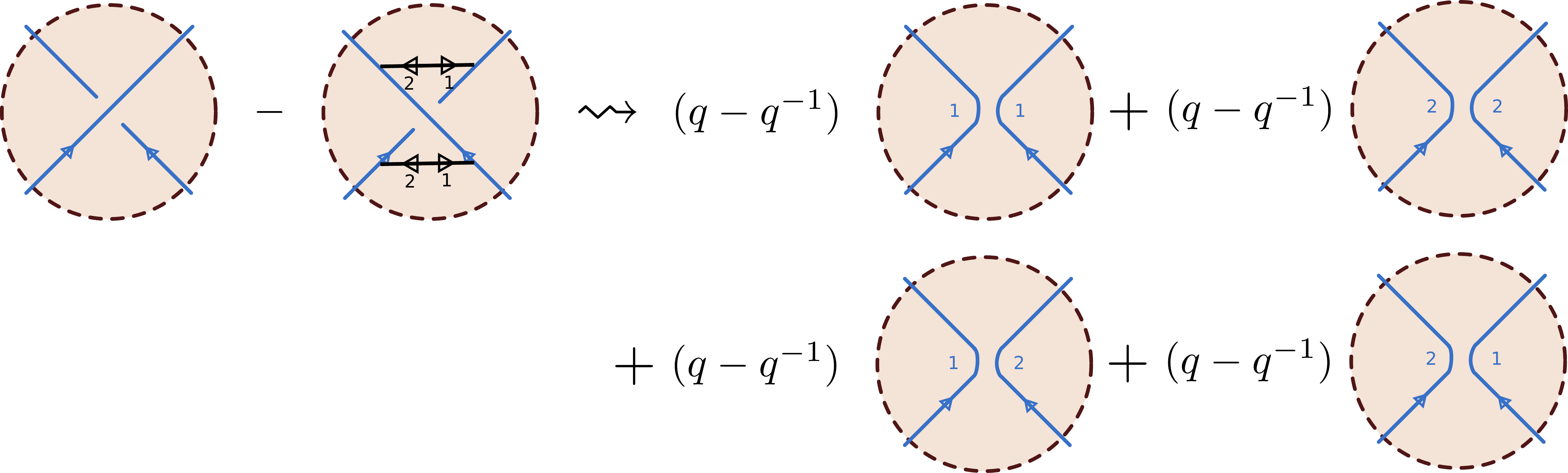}{0.25}{Applying $F$ to the LHS of skein relation (I) in $\Sk(M,\fgl(2))$.}

Next we consider skein relation (II) in $\Sk(M,\fgl(2))$ (the change-of-framing relation.) Again choose a generic direction for the foliation, 
and a simple monotonically decreasing height profile for the link strand. Due to the simple height profile the only lifts are direct lifts to sheet $1$ and sheet $2$. 
Each of these lifts comes with a framing fractor $q^{\half} \times q^{\half} = q$
from the two places where it is tangent to the foliation; combining this with the 
$q$ from the skein relations in $\Sk(\tM,\fgl(1))$ gives the desired factor $q^2$.
This is shown in \autoref{fig:gl2skeinII}.

\insfigsvg{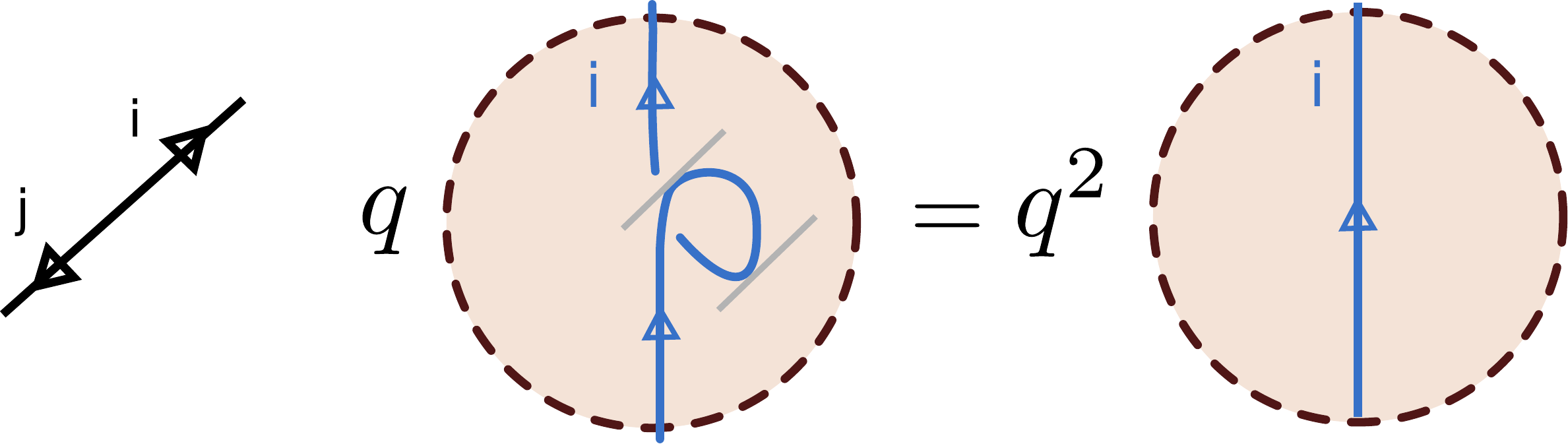}{0.3}{A direct lift of the open strand in skein relation (II), and its simplification in $\Sk(\tM,\fgl(1))$.}

Finally,
skein relation (III) in $\Sk(M,\fgl(2))$ follows from the fact
that $F$ assigns 
$q + q^{-1}$ to an unknot, which we have already checked
in \autoref{sec:unknotexample} (and again in \autoref{sec:unknotscritleaves}.)

\section{Framed wall-crossing}\label{sec:wall-crossing}

As we have discussed, the $q$-nonabelianization map for $M = C \times \R$
is defined
using a WKB foliation associated to a covering $\tC \to C$.
As we vary the covering $\tC \to C$ continuously, there are discrete
moments at which the topology of the WKB foliation can jump.
These jumps were called $\mathcal{K}$-walls in \cite{Gaiotto:2012rg}. 
Following the topology change, the $q$-nonabelianization map 
is expected to jump according to the {\it framed wall-crossing} 
formula \cite{Gaiotto:2008cd,MR2567745,Kontsevich:2008fj,Gaiotto:2009hg,Gaiotto:2010be,Dimofte:2009tm,Gaiotto:2012rg}.
In this section we verify that this indeed happens.

\subsection{The framed wall-crossing formula}

Let us first quickly recall the framed wall-crossing formula.

From the point of view of the 4d $\mathcal{N}=2$ theory $\fX[C, \fg]$,
the framed wall-crossing behavior is determined by the bulk BPS spectrum.
At any point in the Coulomb branch, the one-particle BPS Hilbert space is graded by 
the IR charge lattice $\Gamma$, $\mathcal{H}=\oplus_{\gamma\in\Gamma} \mathcal{H}_\gamma$. We factor out the center-of-mass degrees of freedom by writing
\begin{equation}
\mathcal{H}_\gamma=[({\mathbf 2},{\mathbf 1})\oplus({\mathbf 1},{\mathbf 2})]\otimes h_\gamma,
\end{equation}
and consider the {\it protected spin character} \cite{Gaiotto:2010be}: 
\begin{equation}
\Omega(\gamma,q):=\text{Tr}_{h_\gamma}(-q)^{2J_3} q^{2I_3}=\sum\limits_{n=1}^\infty \Omega_n(\gamma)q^n,
\end{equation}
where $J_3$, $I_3$ are Cartan generators of $\SU(2)_P$ and $\SU(2)_R$ respectively. The integers 
$\Omega_n(\gamma)$ are conveniently packaged into the {\it Kontsevich-Soibelman factor},
\begin{equation}
K(q;X_\gamma;\Omega_n(\gamma)):= \prod_{n=1}^\infty E_q \left( \sigma(\gamma)(-1)^nq^nX_\gamma \right)^{(-1)^n\Omega_n(\gamma)},
\end{equation}
where $E_q(z)$ is the quantum dilogarithm defined as
\begin{equation}
E_q(z):=\prod_{j=0}^\infty\Big(1+q^{2j+1}z\Big)^{-1},
\end{equation}
and $\sigma(\gamma) = \pm 1$ is a certain quadratic refinement defined in \cite{Gaiotto:2009hg}.
The $\cK$-walls are the (real codimension 1) loci 
in the Coulomb branch where, for some charge $\gamma$, 
the central charge $Z_\gamma \in \R_-$ and the Kontsevich-Soibelman factor
$K(q;X_\gamma;\Omega_n(\gamma)) \neq 1$.

Fix a framed link $L$ in $M$, and let 
$F_-([L])$ and $F_+([L])$ be the values of $F([L])$ before 
and after crossing a $\cK$-wall in the \ti{positive} direction,
i.e. the direction in which 
$\mathrm{Im}\ Z_\gamma$ goes from negative to positive.
The framed wall-crossing formula is \cite{Kontsevich:2008fj,Gaiotto:2010be}:
\begin{equation}\label{eq:WCF}
F_+([L]) = K F_-([L]) K^{-1}, \qquad K := K(q;X_\gamma;\Omega_i(\gamma)).
\end{equation}
In the following we will verify that our map $F$ indeed satisfies \eqref{eq:WCF},
in the two simplest situations:
\begin{itemize}
	\item A BPS hypermultiplet $\cK$-wall: then $K = E_q(-X_\gamma)$ 
	for some $\gamma$, which gives using \eqref{eq:quantumtorus}
	\begin{equation} \label{eq:hyper-concrete}
		K X_\mu K^{-1} = \begin{cases} X_\mu & \text{ for } \IP{\gamma,\mu} = 0, \\
		X_{\mu} \prod\limits_{j=1}^{\abs{\IP{\gamma,\mu}}} \big(1-q^{\text{sgn}(\IP{\gamma,\mu})(2j-1)}X_\gamma\big)^{\text{sgn}(\IP{\gamma,\mu})} & \text{ for } \IP{\gamma,\mu} \neq 0.
\end{cases}
	\end{equation}

	\item A BPS vector multiplet $\cK$-wall: then $K = E_q(-qX_\gamma) E_q(-q^{-1} X_\gamma)$ for some $\gamma$, which gives
	\begin{equation} \label{eq:vector-concrete}
		K X_\mu K^{-1} = \begin{cases} X_\mu & \text{ for } \IP{\gamma,\mu} = 0, \\
		X_{\mu} \prod\limits_{j=1}^{\IP{\gamma,\mu}} \big(1-q^{2j}X_\gamma\big) \big(1-q^{2j-2}X_\gamma\big)& \text{ for } \IP{\gamma,\mu} >0,\\
		X_{\mu} \prod\limits_{j=1}^{\IP{\mu,\gamma}} \big(1-q^{-2j}X_\gamma\big)^{-1} \big(1-q^{2-2j}X_\gamma\big)^{-1}& \text{ for } \IP{\gamma,\mu} <0.
\end{cases}
	\end{equation}

\end{itemize}

\subsection{Relative skein modules}

To compare $F_+$ to $F_-$ it will be convenient to observe that the 
definition of $F$ is local, in the following sense.

Let $M^\circ \subset M$ be a submanifold with boundary (and perhaps corners)
such that no leaf meets $\partial M$ transversely, i.e., the WKB 
foliation of $M$ by 1-manifolds 
restricts to a foliation of $\partial M$ by 1-manifolds.
Also fix a 0-chain $S$ in $\partial M^\circ$, 
given by a finite subset of $\partial M^\circ$
with each point labeled by $\pm 1$.
Then we can define a relative skein module $\Sk(M^\circ, S, \fgl(2))$,
by applying exactly the rules of \autoref{sec:gln-skein}, 
except that instead of oriented links $L$,
we use oriented tangles $L$, with boundary 
$\partial L = S$.
Also let $\tM^\circ \to M^\circ$ be the branched double cover of $M^\circ$
obtained by restricting $\tM \to M$.
Then we define a relative skein module $\Sk(M^\circ, S, \fgl(1))$
by applying again the rules of \autoref{sec:gl1-skein} to oriented tangles $\tL$,
where we allow $\partial \tL$ to be any 0-chain which projects to $S$. 
Our construction of $F$ applies directly in this relative situation,
to give a map 
\begin{equation}
F: \Sk(M^\circ, S, \fgl(2)) \to \Sk(\tM^\circ, S, \fgl(1)).
\end{equation}
Now we can formulate the crucial locality 
property of $F$. Suppose $M^\circ$ is obtained by
gluing $M^\circ_1$ and $M^\circ_2$ along part of their boundary,
and $L \subset M^\circ$ is a link, divided into $L_1 = L \cap M^\circ_1$ and 
$L_2 = L \cap M^\circ_2$.
Then we have
\begin{equation}
	F_{M^\circ}(L) = F_{M^\circ_1}(L_1) \cdot F_{M^\circ_2}(L_2)
\end{equation}
where the $\cdot$ on the right is defined by concatenation
of tangles.

The relative skein module $\Sk(\tM^\circ, S, \fgl(1))$ is a two-sided
module over the (absolute) skein algebra $\Sk(\tM^\circ, \fgl(1))$. 
For each relative
homology class $a$ we define an element
$X_a \in \Sk(\tM^\circ, S, \fgl(1))$, 
by the same rule we used to define $X_\gamma$
for ordinary homology classes $\gamma$
in \autoref{sec:quantumtorus}. Then we have
the analog of \eqref{eq:quantumtorus},
\begin{equation} \label{eq:relative-quantumtorus}
	X_\gamma X_a = (-q)^{\IP{\gamma,a}} X_{\gamma + a}.
\end{equation}

\subsection{The hypermultiplet \texorpdfstring{$\mathcal{K}$}{K}-wall}

We begin by considering a hypermultiplet $\mathcal{K}$-wall.

The transformation of the WKB foliation which occurs at such a $\cK$-wall
is described in \cite{Gaiotto:2012rg}.
The only nontrivial change occurs in the neighborhood
of a saddle connection on $C$, as shown in \autoref{fig:hyperKwall-foliation}.

\insfigsvg{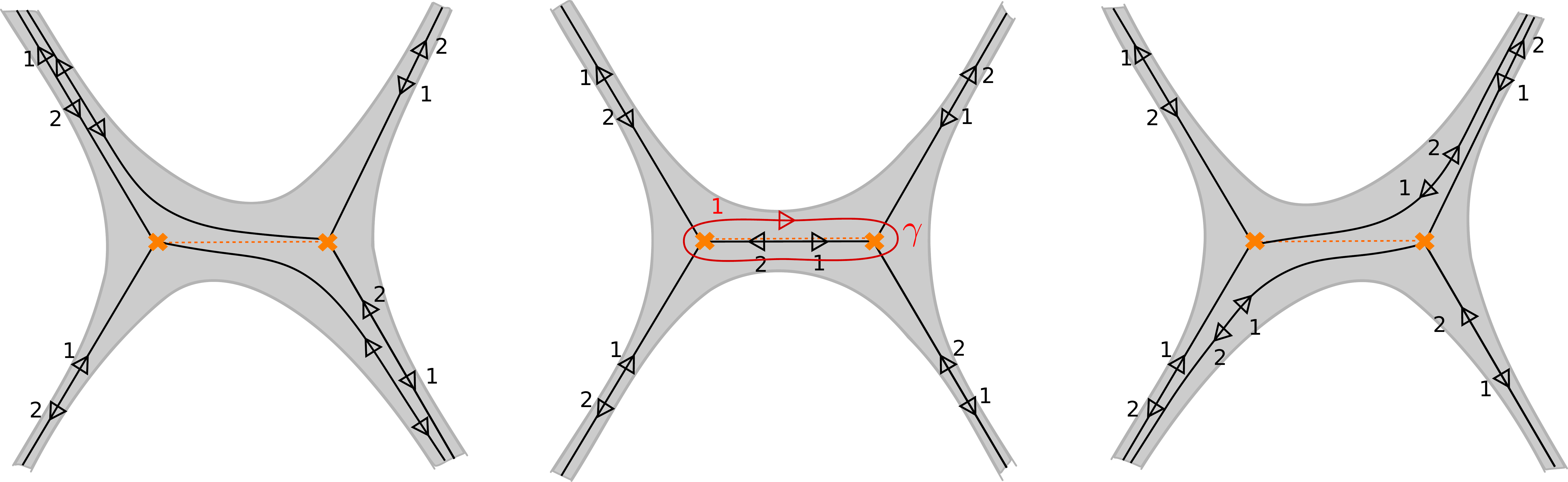}{0.2}{A region $C^\circ$ of $C$ bounded by generic leaves
of the WKB foliation, and the critical leaves
contained therein, before (left) and after (right) crossing a hypermultiplet $\mathcal{K}$-wall. The charge $\gamma$ of the
hypermultiplet is also shown.}

Let $C^\circ$ denote the region shown in \autoref{fig:hyperKwall-foliation},
and let $l_1, \dots, l_m$ be the components of the intersection $L \cap (C^\circ \times \R)$.
By an isotopy of $L$, we may assume that each
$l_n$ crosses $C^\circ$
in one of the simple ways shown in \autoref{fig:hyperKwall-crossings}. 

\insfigsvg{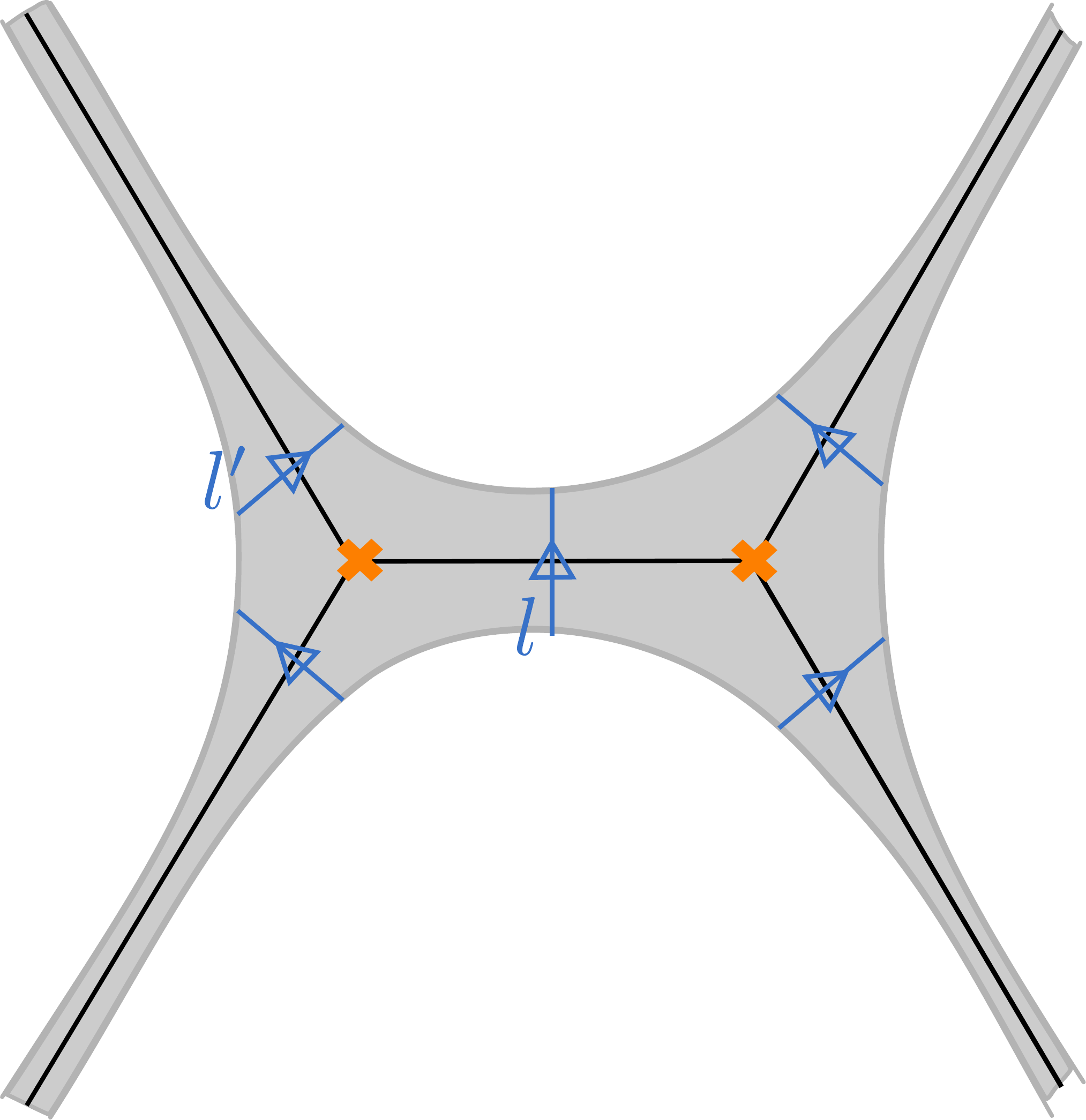}{0.2}{Five ways in which the projection
of $L$ could cross $C^\circ$; by an isotopy of $L$ we can arrange that
all crossings are of one of these five types.}

Moreover,
we may assume that each $l_n$ has $x^3$-coordinate 
monotonically increasing from $a_n$ to $b_n$,
with $b_n < a_{n+1}$.
Then we divide $M$ into the disjoint regions
$M^\circ_n = C^\circ \times [a_n,b_n]$
and the complementary region $M^\circ_0$.
We will prove that, in each region $M^\circ_n$,
we have
\begin{equation} \label{eq:wcf-local-1}
	F_{M^\circ_n,+}(l_n) = K F_{M^\circ_n,-}(l_n) K^{-1},
\end{equation}
where the conjugation acts concretely by \eqref{eq:hyper-concrete},
and in $M^\circ_0$ we have
\begin{equation} \label{eq:wcf-local-2}
	F_{M^\circ_0,+}(l_n) = F_{M^\circ_0,-}(l_n).
\end{equation}
Using the locality of $F$ this then implies the desired
formula \eqref{eq:WCF}.

First we consider the region $M^\circ_0$. In this region 
the topology of the WKB foliation
does not change as we cross the $\cK$-wall;
\eqref{eq:wcf-local-2} follows directly.

Next we consider a region $M^\circ_n$ containing an open strand 
$l_n$ isotopic
to the $l$ in \autoref{fig:hyperKwall-crossings}. $l$ has five lifts $\tl^1, \dots, \tl^5$ either before or after crossing the $\cK$-wall, as shown in \autoref{fig:hyperKwalllifts}. Using $p^i$ for the standard projection of $\tl^i$, $F_\pm([l])$ are:
\begin{align}
F_-([l])&=\sum\limits_{i=1}^4 q^{w_i}X_{p^i}+q^{w_2}X_{p^2+\gamma}=q^{w_1}X_{p^1}+q^{w_2}X_{p^2}(1-q^{-1}X_\gamma)+\sum\limits_{i=3,4}q^{w_i}X_{p^i}, \label{eq:fminus} \\
F_+([l])&=\sum\limits_{i=1}^4 q^{w_i}X_{p^i}+q^{w_1}X_{p^1+\gamma}=q^{w_1}X_{p^1}(1-qX_\gamma)+q^{w_2}X_{p^2}+\sum\limits_{i=3,4}q^{w_i}X_{p^i},
\label{eq:fplus}
\end{align} 
where $w_i$ is the winding of $p^i$ and we have used \eqref{eq:relative-quantumtorus} to rewrite $F_\pm([l])$.
Now \eqref{eq:fplus} is obtained from \eqref{eq:fminus}
by the substitutions $X_{p^1} \to X_{p^1}(1 - qX_\gamma)$,
$X_{p^2} \to X_{p^2} (1 - q^{-1} X_\gamma)^{-1}$, leaving
all other $X_a$ alone; this matches the expected
\eqref{eq:wcf-local-1}.

\insfigsvg{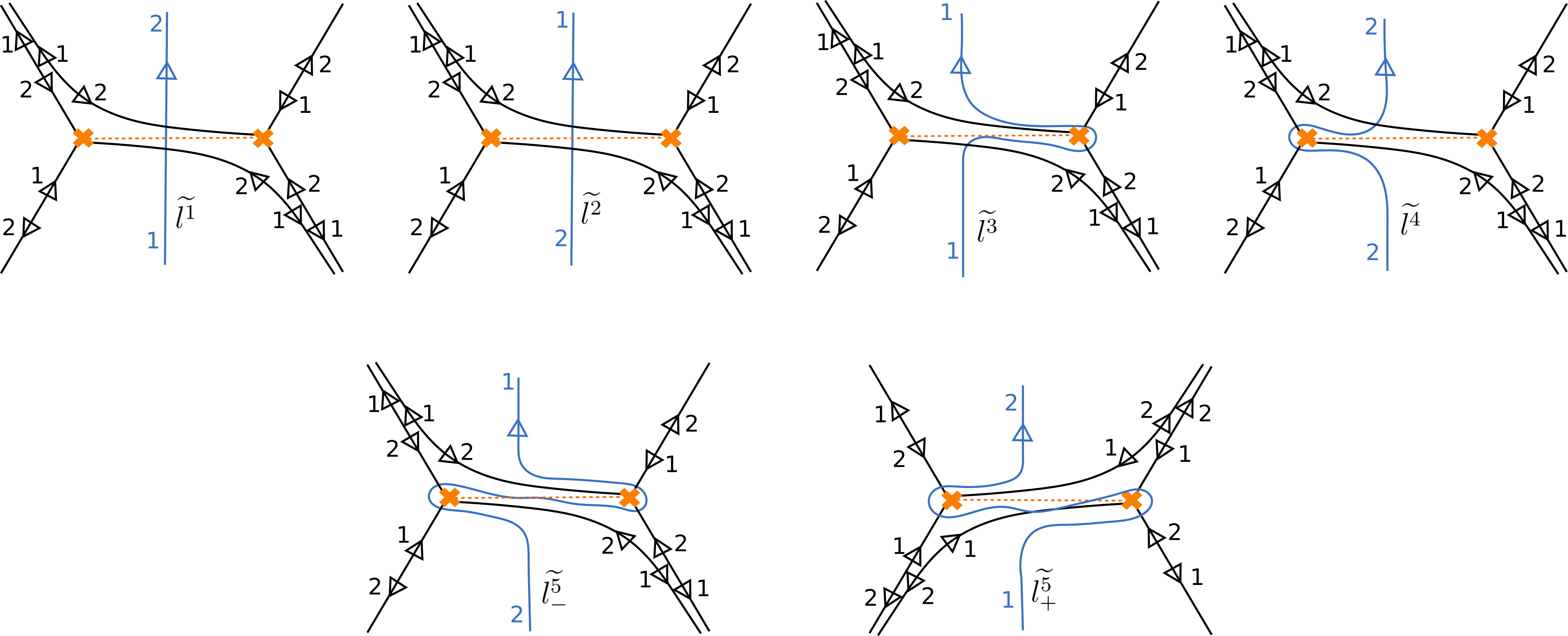}{0.23}{Top: four lifts $\tilde{l}^1, \dots, \tilde{l}^4$ 
	of the open strand $l$ which are present both
	before and after the flip. Bottom left: a lift $\tilde{l}_-^5$ which is present only
	before the flip. Bottom right: a lift $\tilde{l}_+^5$ which is present only
	after the flip.}

Finally we consider a region $M^\circ_n$ containing an open strand 
$l_n$ isotopic to the $l'$ in \autoref{fig:hyperKwall-crossings}.
There are two direct lifts of $l'$, which are the same on both sides of the $\cK$-wall; we denote their standard projections ${p^2}$, $p^3$.
On the $+$ side of the wall there is one lift containing a detour, 
with standard projection $p^1$.
On the $-$ side there are two lifts containing detours.
Altogether we find:
\begin{align*}
F_+([l']) &= q^{w_1}X_{p^1} +\sum\limits_{i=2,3}q^{w_i}X_{p^i}, \\
F_-([l']) &= q^{w_1}(X_{p^1}+X_{p^1+\gamma}) +\sum\limits_{i=2,3}q^{w_i}X_{p^i} = q^{w_1}X_{p^1}(1-q^{-1}X_\gamma)  +\sum\limits_{i=2,3}q^{w_i}X_{p^i}.
\end{align*}
This again agrees with the expected \eqref{eq:wcf-local-1},
using $\langle \gamma,p^1\rangle=-1$, $\langle \gamma,p^2 \rangle = 0$,
$\langle \gamma,p^3 \rangle = 0$.

\subsection{The vector multiplet \texorpdfstring{$\mathcal{K}$}{K}-wall}

Now we turn to the vector multiplet $\mathcal{K}$-wall,
where the WKB foliation develops an annulus of closed trajectories.
Here the transformation of the WKB foliation which occurs
at the wall is subtler to describe.
As described in \cite{Gaiotto:2009hg,Gaiotto:2012rg}, as we approach
the $\cK$-wall from either direction, the WKB foliation in the annulus
undergoes an
infinite sequence of flips, which accumulate at the $\cK$-wall.
The $F_\pm$ we are after in this case therefore have to be
understood as the \ti{limits} of $F$ as we approach the $\cK$-wall
from the two sides.
Moreover, the meaning of $F_\pm$ is a bit subtle: they involve
infinite sums of terms, so they really lie in some completion
of the skein module; we will optimistically gloss over this point
here, so what we are really giving here is a proof sketch rather than
a complete proof.

Similarly to the classical case discussed in \cite{Gaiotto:2012rg},
we can compute $F_\pm([l])$ by considering
the limiting behavior of the critical leaves of the 
foliation as we approach the wall.
A schematic picture of the limits is
shown in \autoref{fig:vectorKwall} (see also \cite{Gaiotto:2009hg,Gaiotto:2012rg} for more such pictures). The degenerated critical leaves
wind infinitely many times around the annulus.
Concretely speaking, to compute $F_\pm([l])$, we sum over direct lifts 
and detours along critical leaves in the usual way; the only tricky
point is that even for a compact $l$, $F_\pm([l])$ may involve
an infinite sum over detours, since $l$ may meet one of the
degenerated critical leaves infinitely many times.

\insfigsvg{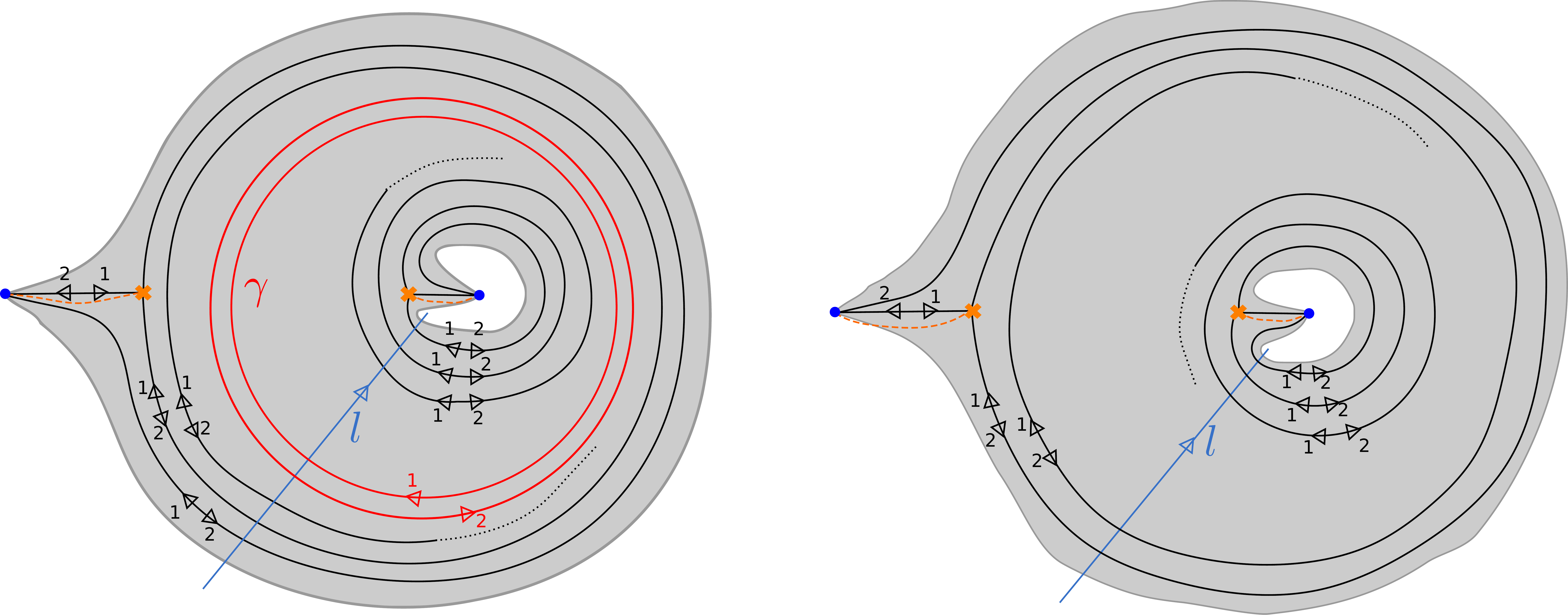}{0.18}{A region $C^\circ$ of $C$ bounded by generic leaves
of the WKB foliation, and degenerated versions of the WKB foliation
contained therein, immediately before (left) and after (right) crossing a vector multiplet $\mathcal{K}$-wall. The picture is a bit schematic;
the leaves which end in dotted lines should be understood as winding
infinitely many times around the annulus, 
and $l$ crosses each of these leaves infinitely many times. 
The charge $\gamma$ of the
vector multiplet is also shown on the left.}

Now we follow the same strategy we used for the hypermultiplet.
The key computation is to compare the two limits $F_\pm([l])$, where $l$ is an open strand crossing $C^\circ$ as shown in \autoref{fig:vectorKwall},
with its $x^3$-coordinate increasing monotonically.

We need some notation for various paths in $C^\circ$.
Let $a_{12}, a_{21}, b_{12}, b_{21}$ be the detour paths
shown in \autoref{fig:vectorKwallsoliton}.
Let $p^{1}$, $p^{2}$ be the standard projection of the direct lifts of
$l$ to sheet $1$, $2$ respectively. Finally, 
divide each $p^i$ into initial, middle and final segments $p^i_-$, 
$p^i_0$, $p^i_+$ respectively, in such a way that the following
paths are connected:
\begin{align*}
p^{11}&:=p_-^1+a_{12}+p_0^2+b_{21}+p^1_+,\quad
p^{22}:=p_-^2+a_{21}+p_0^1+b_{12}+p^2_+,\\
p^{12}&:=p_-^1+a_{12}+p_0^2+p_+ ^2,\quad
p'^{12}:=p_-^1+p_0^1+b_{12}+p_+ ^2,\\
p^{21}&:=p_-^2+a_{21}+p_0^1+p_+^1,\quad
p'^{21}:=p_-^2+p_0^2+b_{21}+p_+^1.
\end{align*}

\insfigsvg{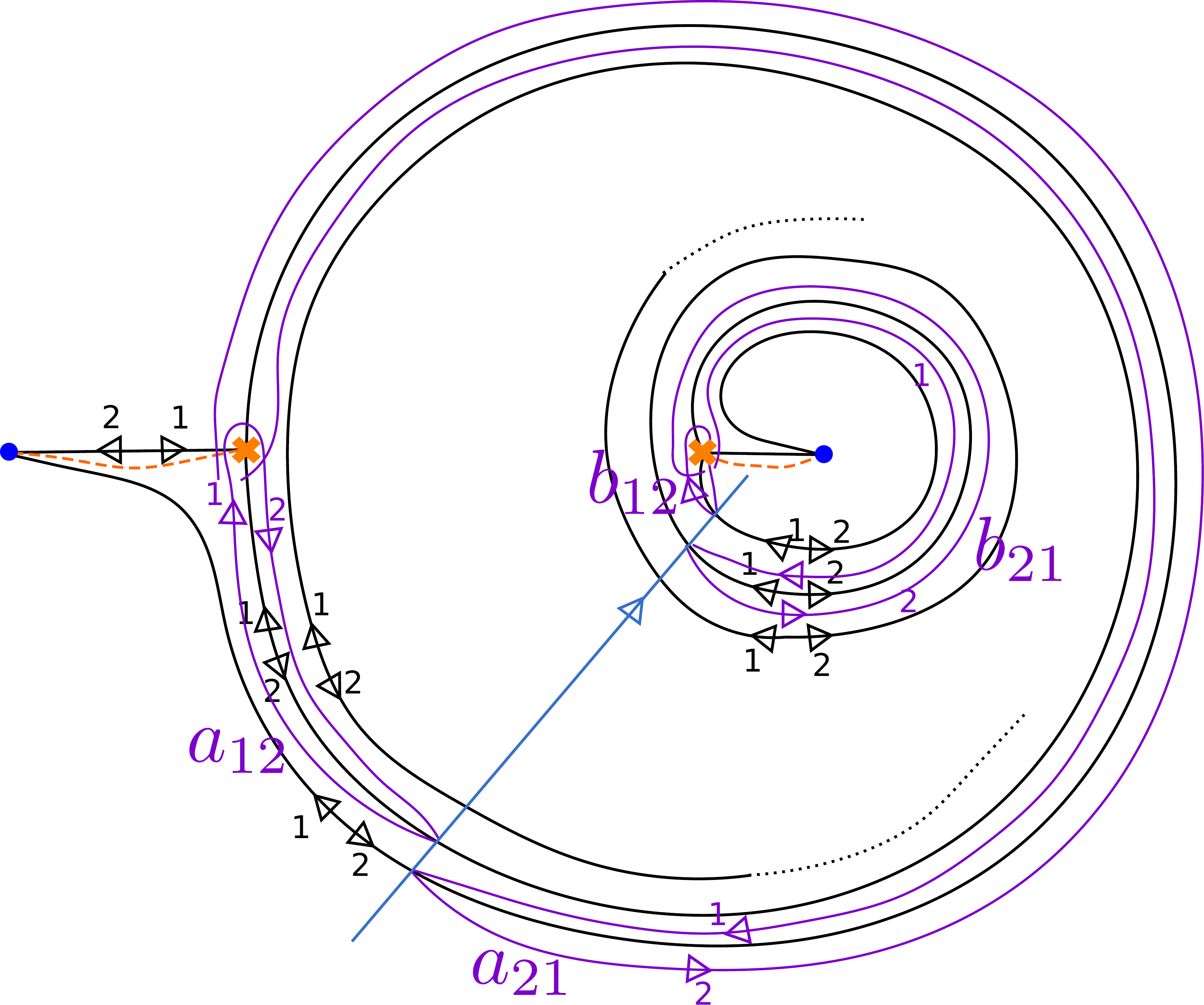}{0.20}{Basic constituents of open paths $\{p\}$ on $\tC$. The (minimal) detours $a_{12}, a_{21}, b_{12}, b_{21}$ are shown in purple.} 

\insfigsvg{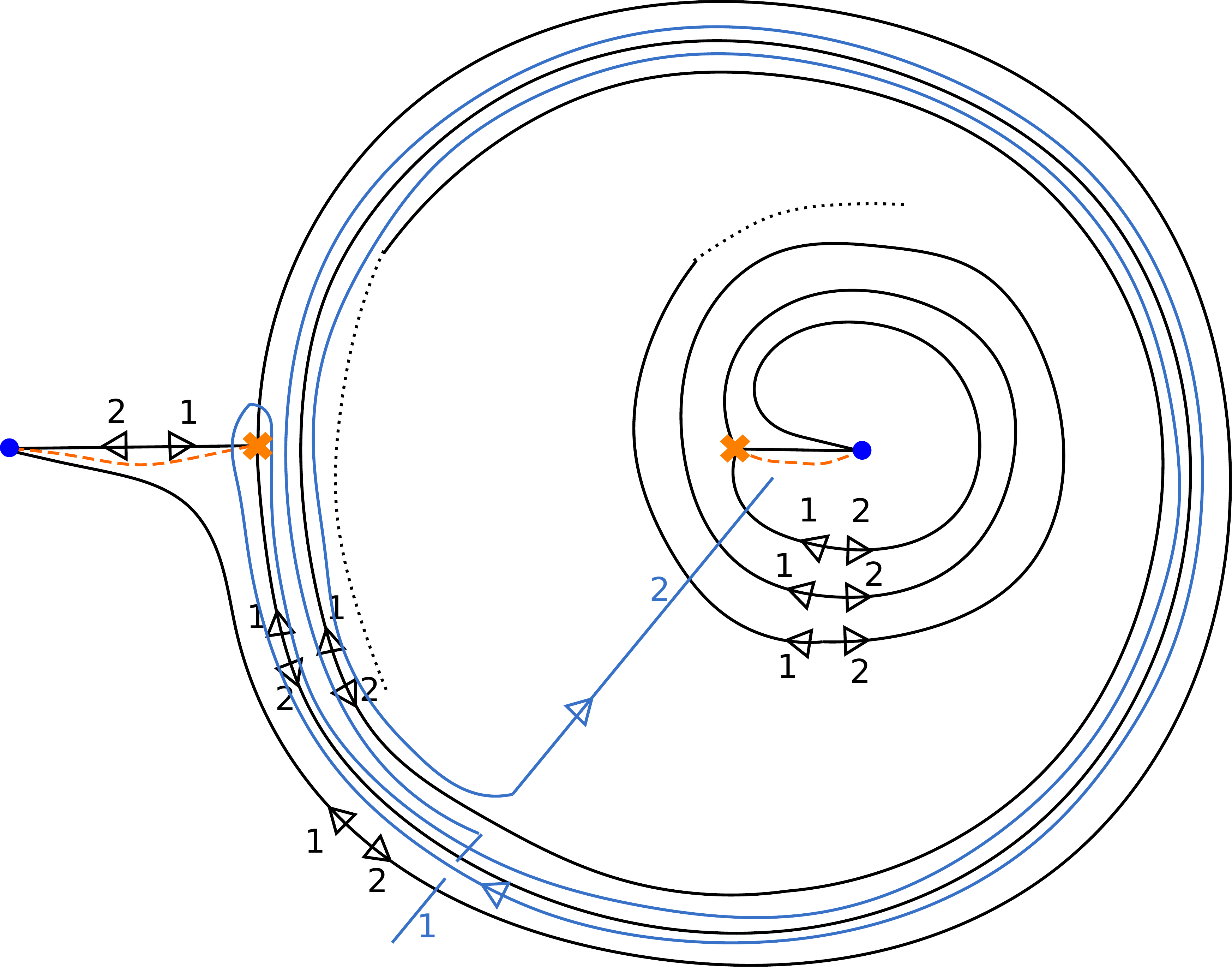}{0.20}{One of the lifts contributing to $F_-([l])^{(12)}$. This specific lift contributes $-q^{w^{12}-\frac{3}{2}}X_{p^{12}+\gamma}$.}
Let $F_\pm([l])^{(ij)}$  ($i,j\in\{1,2\}$) be the contribution to $F_\pm([l])$ from lifts that start on sheet $i$ and end on sheet $j$. For example, one lift of $l$ that contributes to $F_-([l])^{(12)}$ is shown in \autoref{fig:vectorKwall-lift}. Let $w^{ij}$ be the winding of the corresponding open strands in $\tC$.\footnote{Due to the monotonic height profile, the lift in the class $p+n \gamma$ has the same winding as the lift in the class $p$.}
A direct enumeration of all the allowed lifts gives:
\begin{align*}
F_-([l])^{(11)}&=q^{w^{11}}\Bigg(X_{p^1}+\sum\limits_{n_1,n_2=0}^\infty (-q)^{-n_1+n_2}X_{p^{11}+(n_1+n_2)\gamma}\Bigg)=q^{w^{11}}\Bigg(X_{p^1}+X_{p^{11}}\frac{1}{1-X_\gamma}\frac{1}{1-q^2X_\gamma}\Bigg),\\
F_+([l])^{(11)}&=q^{w^{11}}\Bigg(\sum\limits_{n_1,n_2=0}^\infty (-q)^{-n_1+n_2}X_{p^{1}+(n_1+n_2)\gamma}+X_{p^{11}}\Bigg)=q^{w^{11}}\Bigg(X_{p^{1}}\frac{1}{1-X_\gamma}\frac{1}{1-q^2X_\gamma}+X_{p^{11}}\Bigg),\\
F_-([l])^{(22)}&=q^{w^{22}}\Bigg(\sum\limits_{n_1,n_2=0}^\infty (-q)^{-n_1+n_2}X_{p^2+(n_1+n_2)\gamma}+X_{p^{22}}\Bigg)=q^{w^{22}}\Bigg(X_{p^2}\frac{1}{1-X_\gamma}\frac{1}{1-q^2X_\gamma}+X_{p^{22}}\Bigg),\\
F_+([l])^{(22)}&=q^{w^{22}}\Bigg(X_{p^2}+\sum\limits_{n_1,n_2=0}^\infty (-q)^{-n_1+n_2}X_{p^{22}+(n_1+n_2)\gamma}\Bigg)=q^{w^{22}}\Bigg(X_{p^2}+X_{p^{22}}\frac{1}{1-X_\gamma}\frac{1}{1-q^2X_\gamma}\Bigg),\\
F_-([l])^{(12)}&=q^{w^{12}-\frac{1}{2}}\Bigg(\sum\limits_{n_1,n_2=0}^\infty (-q)^{-n_1+n_2}X_{p^{12}+(n_1+n_2)\gamma}+X_{p'^{12}}\Bigg)=q^{w^{12}-\frac{1}{2}}\Bigg(X_{p^{12}}\frac{1}{1-X_\gamma}\frac{1}{1-q^2X_\gamma}+X_{p'^{12}}\Bigg),\\
F_+([l])^{(12)}&=q^{w^{12}-\frac{1}{2}}\Bigg(X_{p^{12}}+\sum\limits_{n_1,n_2=0}^\infty (-q)^{-n_1+n_2}X_{p'^{12}+(n_1+n_2)\gamma}\Bigg)=q^{w^{12}-\frac{1}{2}}\Bigg(X_{p^{12}}+X_{p'^{12}}\frac{1}{1-X_\gamma}\frac{1}{1-q^2X_\gamma}\Bigg),\\
F_-([l])^{(21)}&=q^{w^{21}+\frac{1}{2}}\Bigg(X_{p^{21}}+\sum\limits_{n_1,n_2=0}^\infty (-q)^{-n_1+n_2}X_{p'^{21}+(n_1+n_2)\gamma}\Bigg)=q^{w^{21}+\frac{1}{2}}\Bigg(X_{p^{21}}+X_{p'^{21}}\frac{1}{1-X_\gamma}\frac{1}{1-q^2X_\gamma}\Bigg),\\
F_+([l])^{(21)}&=q^{w^{21}+\frac{1}{2}}\Bigg(\sum\limits_{n_1,n_2=0}^\infty (-q)^{-n_1+n_2}X_{p^{21}+(n_1+n_2)\gamma}+X_{p'^{21}}\Bigg)=q^{w^{21}+\frac{1}{2}}\Bigg(X_{p^{21}}\frac{1}{1-X_\gamma}\frac{1}{1-q^2X_\gamma}+X_{p'^{21}}\Bigg).
\end{align*}
From these expressions one sees directly that 
$F_+([l]) = K F_-([l]) K^{-1}$, with $K$ given by 
\eqref{eq:vector-concrete}, as desired.

\section{Reduction to \texorpdfstring{$\fsl(2)$}{sl(2)}} \label{sec:sl2}

\subsection{The Kauffman bracket skein module}

In this final section we briefly discuss
the $\fsl(2)$ variant of the skein module,
which we denote $\Sk(M, \fsl(2))$; it is also
known as the Kauffman bracket skein module 
(see \cite{MR1723531} for a review).
In this version of the skein module we use
unoriented links and $\Z[(-q)^\half, (-q)^{-\half}]$ coefficients;
the skein relations are shown in \autoref{fig:sl2-skein-relations}.
Reflecting the relation $\fgl(2) = \fsl(2) \oplus \fgl(1)$
there is a map of skein modules
\begin{equation} \label{eq:skein-module-map}
	\Sk(M, \fgl(2)) \ \to \ \Sk(M, \fsl(2)) \otimes \Sk^{\half}(M, \fgl(1)),
\end{equation}
where by $\Sk^{\half}$ we mean the skein module with 
the replacement $q \mapsto -(-q)^{\half}$.
The map is given by the obvious operation, $[L] \mapsto [L] \otimes [L]$,
which one can check directly respects the skein relations.

\insfigsvg{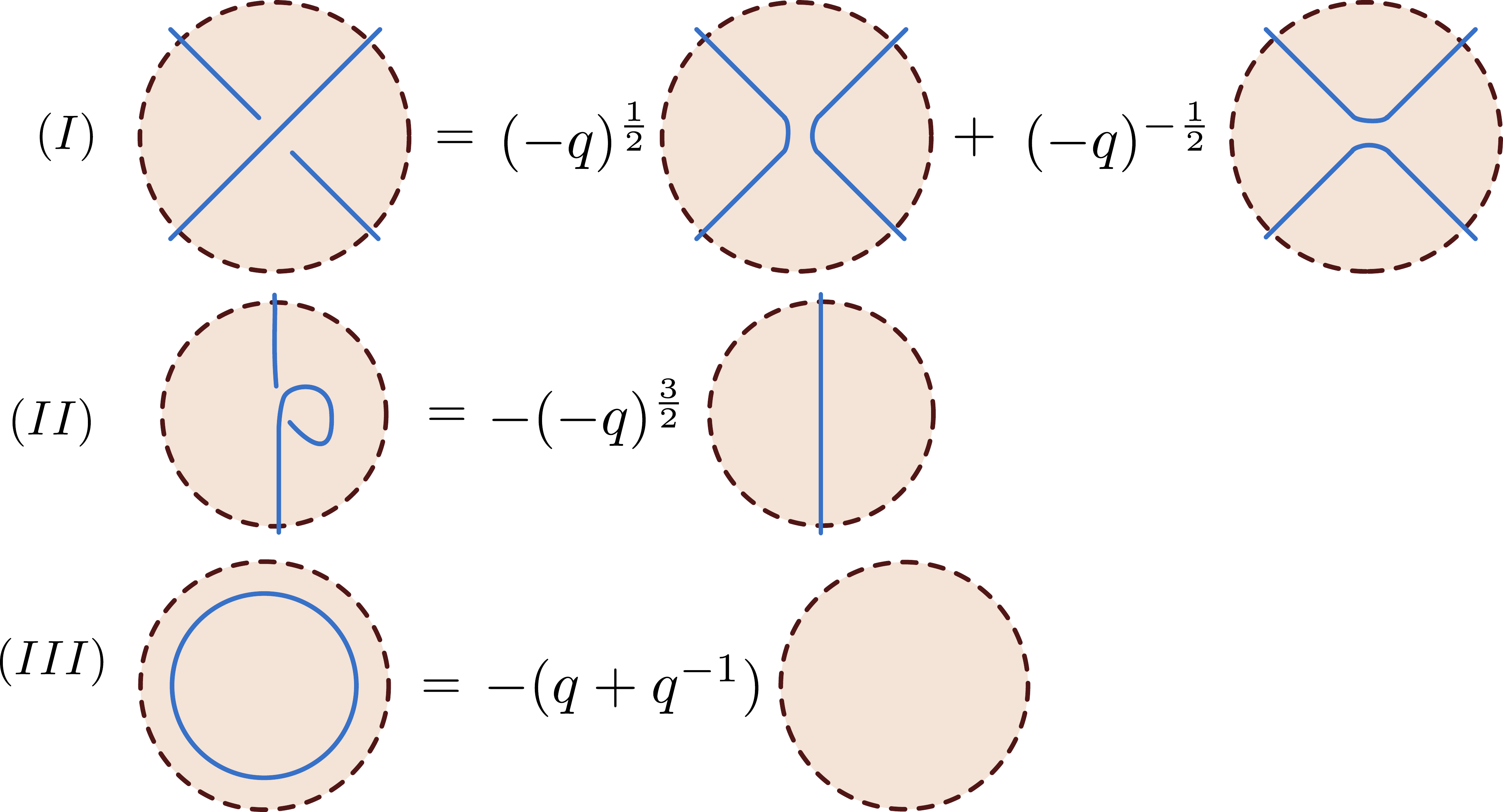}{0.25}{Skein relations defining
	$\Sk(M, \fsl(2))$.}

\subsection{Factorization of \texorpdfstring{$q$}{q}-nonabelianization, for \texorpdfstring{$M = C \times \R$}{M = C x R}}

Now suppose $M = C \times \R$.
Let $\Gamma = H_1(\tC,\Z)$, and let $\sigma$ be the involution
on $\Gamma$ induced by the deck transformation of $\tC$.
Then we have the map
\begin{align}
\begin{split}
	\Gamma &\to \half \Gamma^\odd \times \half \Gamma^\even \\
	\gamma &\mapsto \left(\half(\gamma - \sigma(\gamma)), \half (\gamma + \sigma(\gamma))\right),
\end{split}
\end{align}
which preserves the pairings and thus induces a map
\begin{equation}
	Q_\Gamma \to Q_{\half\Gamma^\odd} \otimes Q_{\half\Gamma^\even},
\end{equation}
where the quantum tori $Q_{\half\Gamma^\odd}$, $Q_{\half\Gamma^\even}$
are algebras over $\Z[(-q)^{\pm \frac14}]$, because the intersection
pairings on these lattices take values in $\frac14 \Z$.

We conjecture that the $q$-nonabelianization map $F$ can be factorized into odd and even parts, or
more precisely, that there is a commuting diagram
\begin{equation}
\begin{tikzcd}[column sep = large]
    \Sk(M,\fgl(2)) \arrow{r}{F} \arrow{d} & \Sk(\tM,\fgl(1)) \arrow{d} = Q_\Gamma \\
    \Sk(M, \fsl(2)) \otimes \Sk^{\half}(M, \fgl(1)) \arrow{r}{F^\odd \otimes F^\even} & Q_{\half\Gamma^\odd} \otimes Q_{\half\Gamma^\even},
  \end{tikzcd}
\end{equation}
where
\begin{align}
 	F^\even([L]) &= q^{\half w(L)} X_{\half \pi^{-1}(L)}, \\
    F^\odd([L]) &= q^{-\half w(L)} \rho (F([L])), \qquad \rho(X_\gamma) = X_{\half(\gamma-\sigma(\gamma))}, \label{eq:Fodd}
 \end{align}
and the map $F^\odd$ coincides with the quantum trace map of \cite{Bonahon2010}.
One approach to verifying this factorization would be to 
interpret $Q_{\half \Gamma^\odd}$ directly as a variant of 
the skein module which includes factors associated 
to non-local crossings. We will not undertake this verification 
here, but remark that a similar comparison
(between the approach of \cite{Gabella:2016zxu} 
to $q$-nonabelianization and the quantum trace of \cite{Bonahon2010})
was carried out in \cite{Kim:2018dux}.

\newpage

\bibliographystyle{utphys}

\bibliography{q-ab}

\end{document}